\documentclass[a4paper, 11pt]{article}
\pdfoutput=1 % if your are submitting a pdflatex (i.e., if you have
\usepackage{jheppub} % for details on the use of the package, please
\usepackage[T1]{fontenc} % if needed    
\usepackage{amsmath}  
\usepackage{amssymb}   
\usepackage{latexsym}
\usepackage{graphicx}

\usepackage{multirow}
\usepackage{ytableau}
\usepackage{float}
\usepackage{caption}
\usepackage{subcaption}
 \allowdisplaybreaks 
\usepackage{braket}
\usepackage{amssymb}
\usepackage{amsmath}
\usepackage{amsthm}
\usepackage{mathrsfs}
\usepackage{framed}
\usepackage{hyperref}
\usepackage{slashed}
\usepackage{array}

\hypersetup{colorlinks=true}
\hypersetup{linkcolor=violet}
\hypersetup{citecolor=violet}
\hypersetup{urlcolor=violet}
\numberwithin{equation}{section}

\usepackage{tikz, pgf}
\usetikzlibrary{shapes.misc}
\usetikzlibrary{shapes}
\usetikzlibrary{fit}
\usetikzlibrary{arrows}

\usetikzlibrary{decorations.pathmorphing}	% For Feynman Diagrams
\usetikzlibrary{decorations.markings}
 \usetikzlibrary{patterns}

\tikzset{
    sugra/.style={decorate, decoration={snake}, draw=black},
    scalarphi/.style={dashed,draw=black, postaction={decorate},
        },
    scalarchi/.style={draw=brown}, 
    hwbou/.style={draw=blue, postaction={decorate}, ultra thick
        },
    vector/.style={draw=blue,decorate, decoration={snake}, draw},
	provector/.style={decorate, decoration={snake,amplitude=2.5pt}, draw},
	antivector/.style={decorate, decoration={snake,amplitude=-2.5pt}, draw},
   	 fermion/.style={draw=cyan, postaction={decorate},
        decoration={markings,mark=at position .55 with {\arrow[draw=black]{>}}}},
    fermionbar/.style={draw=cyan, postaction={decorate},
        decoration={markings,mark=at position .55 with {\arrow[draw=black]{<}}}},
    fermionnoarrow/.style={draw=black},
    gluon/.style={decorate, draw=red,
        decoration={coil, amplitude=4pt, segment length=5pt}},
    scalar/.style={dashed,draw=black, postaction={decorate},
        decoration={markings,mark=at position .55 with {\arrow[draw=black]{>}}}},
    scalarbar/.style={dashed,draw=black, postaction={decorate},
        decoration={markings,mark=at position .55 with {\arrow[draw=black]{<}}}},
    electron/.style={draw=black, postaction={decorate},
        decoration={markings,mark=at position .55 with {\arrow[draw=black]{>}}}},
    scalarnoarrow/.style={dashed, draw=black},
    electron/.style={draw=black, postaction={decorate},
        decoration={markings, mark=at position .55 with {\arrow[draw=black]{>}}}},
	bigvector/.style={decorate, decoration={snake, amplitude=4pt}, draw},
    photon/.style={draw=violet, decorate, decoration={snake}, draw},
    higgs/.style={dashed, draw=black, postaction={decorate},
        },	
        goldstone/.style={draw=brown, postaction={decorate},
        },    
          ghost/.style={dashed, draw=magenta, postaction={decorate},
        decoration={markings, mark=at position .55 with {\arrow[draw=black]{>}}}
        },  
          antighost/.style={dashed, draw=magenta, postaction={decorate},
        decoration={markings, mark=at position .55 with {\arrow[draw=black]{<}}}
        }, 
            scalartwo/.style={dashed,draw=brown, postaction={decorate},
        decoration={markings,mark=at position .55 with {\arrow[draw=black]{>}}}},
    scalarbartwo/.style={dashed,draw=brown, postaction={decorate},
        decoration={markings,mark=at position .55 with {\arrow[draw=black]{<}}}}, 
    fermiontwo/.style={draw=purple, postaction={decorate},
        decoration={markings,mark=at position .55 with {\arrow[draw=black]{>}}}},
    fermionbartwo/.style={draw=purple, postaction={decorate},
        decoration={markings,mark=at position .55 with {\arrow[draw=black]{<}}}},    
        mphoton/.style={decorate, decoration={snake}, draw=violet},
        realscalar/.style={draw=black}, 
        fakerealscalar/.style={draw=white}, 
        realscalarone/.style={ draw=black},
    	realscalartwo/.style={draw=brown},    	    pseudoscalar/.style={draw=brown},
        mgluon/.style={decorate, draw=blue,
        	decoration={coil,amplitude=4pt, segment length=5pt}},
         weylfermion/.style={draw=orange, postaction={decorate},
        decoration={markings,mark=at position .55 with {\arrow[draw=black]{>}}}},
         weylfermionbar/.style={draw=orange, postaction={decorate},
        decoration={markings,mark=at position .55 with {\arrow[draw=black]{<}}}}, 
    majorana/.style={draw=cyan, postaction={decorate},
        decoration={markings,mark=at position .55 with {\arrow[draw=black]{><}}}},
    majoranabar/.style={draw=cyan, postaction={decorate},
        decoration={markings,mark=at position .55 with {\arrow[draw=black]{><}}}},    
   	wboson/.style={draw=blue,decorate, decoration={snake,amplitude=4pt}, draw},  
    zboson/.style={draw=violet, decorate, decoration={snake}, draw},   
    lepton/.style={draw=black, postaction={decorate},
        decoration={markings, mark=at position .55 with {\arrow[draw=black]{>}}}},
    leptonbar/.style={draw=black, postaction={decorate},
        decoration={markings, mark=at position .55 with {\arrow[draw=black]{<}}}}, 
    clepton/.style={draw=cyan, postaction={decorate},
        decoration={markings, mark=at position .55 with {\arrow[draw= black]{>}}}},
    cleptonbar/.style={draw=cyan, postaction={decorate},
        decoration={markings, mark=at position .55 with {\arrow[draw=black]{<}}}},   
   nlepton/.style={draw=orange, postaction={decorate},
        decoration={markings, mark=at position .55 with {\arrow[draw=black]{>}}}},
    nleptonbar/.style={draw=orange, postaction={decorate},
        decoration={markings, mark=at position .55 with {\arrow[draw=black]{<}}}},              
        graviton/.style={draw=blue, decorate, decoration={snake, amplitude=4pt}, draw},  
        spinj/.style={draw=red, decorate, decoration={snake, amplitude=4pt}, draw},  
        bgraviton/.style={draw=blue, decorate, decoration={snake, amplitude=4pt}, draw},  
        gravitino/.style={draw=red, postaction={decorate}, 
        decoration={snake,  markings, mark=at position .55 with {\arrow[draw=black]{><}}}},
    	gravitinobar/.style={draw=red, postaction={decorate},
        decoration={snake, markings, mark=at position .55 with {\arrow[draw=black]{><}}} },  
    phir/.style={draw=blue, postaction={decorate},},
   phil/.style={dashed,draw=blue,},
     phiav/.style={draw=cyan, postaction={decorate},},
   phidif/.style={dashed,draw=cyan,},  
    chir/.style={draw=red, postaction={decorate},},
   chil/.style={dashed,draw=red,},  
}

 \def\arnote#1{{\color{red} #1}}

\def\arnote#1{}

\usepackage{comment}

\def\arnote#1{{\color{red} #1}}

\def\iimg{ {\bf i}} 

\def\iimg{ {\bf i}}

\usepackage{subfiles}

\hbadness=\maxdimen
\vbadness=\maxdimen

\allowdisplaybreaks[1]

\usepackage{pifont}% http://ctan.org/pkg/pifont
\newcommand{\gegen}{\mathcal{G} }
\newcommand{\legendre}{P}
\newcommand{\jacobi}{\mathcal{J} }

\title{\boldmath On-shell Supersymmetry and higher-spin amplitudes}

\vspace{10mm}

\author[]{ Mahesh K.N. Balasubramanian}
\author[]{, Kushal Chakraborty}
\author[ ]{, Arnab Rudra}
\author[ ]{and  Arnab Priya Saha}
\author[]{\\ }
\vspace*{4.0ex} 
\vspace{10mm}
\affiliation[]{Indian Institute of Science Education and Research Bhopal,\\
 Bhopal Bypass Rd, Bhauri, Madhya Pradesh 462066, India.\\ }
\vspace*{4.0ex}  
\emailAdd{mahesh21@iiserb.ac.in}
\emailAdd{kushal16@iiserb.ac.in} 
\emailAdd{rudra@iiserb.ac.in } 
\emailAdd{apsaha@iiserb.ac.in.}  
\abstract{We use on-shell Supersymmetry to constrain the three-point function of two massless particles and one massive particle in 3+1 dimensions. We use this information to write down the tree-level four-point function of massless particles for $\mathcal{N}=1$, $2$ and $4$ theories. In particular, we derive the expressions for four-photon/gluon amplitudes with massive higher spin exchange in theories with $\mathcal{N}=4$ Supersymmetry in 3+1 dimensions. 
}

\begin{document}. 
\maketitle
%\raggedbottom    

\section{Introduction}
\label{sec:bcrsintro}
 
There is no known example of a consistent perturbative local quantum theory with massive particles with spin greater than two. Such a theory is commonly known as a higher spin theory. In the presence of a cosmological constant, it is possible to construct theories of massless higher spin particles (the so-called Vasiliev theories \cite{Vasiliev:1990en} ). In flat spacetime, the only known example of higher spin theories is string theory; but string theory is also non-local. A common feature of all these theories is that their spectrum contains infinitely many particles, and the spin of the particles is not bounded from above. In the case of string theory, the infinite spectrum arranges itself in a very regular pattern (also known as the Regge trajectories). 

Moreover, the spectrum and the three-point functions of the theory are controlled by two parameters: a dimensionful parameter $\alpha^\prime$ ($=\ell_s^2$; $\ell_s$ is the string length) and $\textrm{g}_s$ (string coupling constant; which is determined by the vacuum expectation value of a massless scalar). In any quantum theory of gravity, there is always a dimensionful parameter: the Planck mass $M_p$. The Planck mass $M_p$ is parametrically larger than $\alpha^\prime $ in perturbative string theory. The known examples of perturbative string theory requires supersymmetry and in 10 non-compact dimension, there are only five perturbative string theories.  The handfullness of the known examples and the high degree of uniqueness of string theory begs for an explanation in terms of general principles of physics: Causality, unitarity etc. In recent years, there has been some effort to better understand quantum theories of higher spin. Motivated by string theory, we consider theories of higher spin particles with a scale $\alpha$ parametrically smaller than the Planck scale $M_p$. In such theories, there are two sets of higher derivative corrections: one controlled by $\alpha$; these are usually known as the classical corrections. $ M_p$ controls the other set; these are called quantum corrections. Consider the set of theories where $M_p$ is infinity. So we are interested in classical theories with higher derivative corrections. We restrict our attention to tree-level amplitudes. In such theories, it has been argued that higher derivative correction must come with higher spin particles due to  the strong restrictions from causality \cite{Camanho:2014apa}; the spin of the higher spin particles has to be unbounded from above. Nevertheless, it is not clear what are the good $S$ matrices of such theories. One important property of a good $S$ matrix is unitarity; the presence of a pole corresponds to the exchange of physical particles. In \cite{Maity:2021obe, Arkani-Hamed:2022gsa} this approach has been used to check the unitarity of perturbative string theory. To understand the classical theory of higher spins, the authors of \cite{Chakraborty:2020rxf} classified the three-point functions of two massless and one massive particle in various spacetime dimensions. The expression for four-photon and four graviton amplitudes due to the exchange of higher spin particles (transforming under the completely symmetric traceless representation of the little group) were constructed in \cite{Balasubramanian:2021act}.

In this work, we incorporate supersymmetry and explore the higher spin theories in the presence of supersymmetry. The motivation behind incorporating supersymmetry is twofold: In the past, supersymmetry has provided us with tools to understand various physical phenomena by providing us with analytically controllable toy models. For example, the study of a supersymmetric black hole has been extremely fruitful in understanding the microscopic origin of black-hole entropy. The second motivation comes from string theory. String theory, currently the only known theory of higher spins, relies on supersymmetry for good infrared properties. Amplitudes with massless external legs due to massive exchange in a supersymmetric theory are useful to check the unitarity of various supersymmetric string theories.

The Poincare group is a symmetry group of any relativistic quantum field theory in flat spacetime. Coleman-Mandula \cite{Coleman:1967ad} showed that for an interacting theory, no non-trivial extension of the Poincare group is possible in $D+1$ ($D\geq 2$)  dimensions. This means that in an interacting theory, no symmetry generator can mix particles of different spins. Supersymmetry manages to bypass the Coleman-Mandula theorem in an interesting way. One of the crucial assumptions of Coleman-Mandula's analysis is that all the generators are representations of Lorentz group. Supersymmetry consists of fermionic generators (i.e. representation of the double cover of the Lorentz group). The action of supersymmetry generators can change spin. In this work, we want to see the consequence of supersymmetry on higher spin theories. The different number of supersymmetry generators characterise different supersymmetric theories. In $3+1$ dimensions, the minimum number of supersymmetry generators in any theory is 4. In a theory with Lagrangian description in $3+1$ dimensions, the allowed numbers of supersymmetry generators are 8, 16 and 32. A theory with 32 supersymmetry generators and massless particles necessarily contains graviton (and gravitino). In this work, we restrict to non-gravitational theories, and hence we will consider theories with $4$, $8$ and $16$ supersymmetries (also known as $\mathcal{N}=1$, $2$ and $4$ theories in $3+1 $ dimensions). We have primarily used the on-shell supersymmetry approach \cite{Herderschee:2019ofc, Herderschee:2019dmc,Engelbrecht:2022aao, Chiodaroli:2022ssi, Abhishek:2022nqv} using the spinor-helicity formalism to impose the constraints of supersymmetry. We use these results from on-shell supersymmetry to write the result in a Lorentz covariant method.  Here we summarize the key results of the paper before we move onto elaborate discussion

\paragraph{Main results} 

In this paper, we used on-shell supersymmetry to constrain the interaction of massless particles with the massive higher spin particles. The summary of the result is the following
\begin{enumerate}

	\item We have mainly used on-shell supersymmetry approach developed in \cite{Herderschee:2019ofc}. This approach depends on writing down various on-shell superfields and then imposing super-charge conservation through grassmann variables.   We write the massive on-shell supermultiplet for $\mathcal{N}=4$ supersymmetry in $3+1$ dim with arbitrary spin of the Clifford vacuum. The expressions for $\mathcal{N}=1$ and $\mathcal{N}=2$ supersymmetry in $3+1$ dimensions are available in the literature \cite{Herderschee:2019ofc, Liu:2020fgu}.

	\item Using the massless and massive super-fields\footnote{We use the phases `super-multiplet' and `superfield' interchangeably.} , we constraint the three-point function of two photons and one massive higher spin particle.

	We write down a map to convert the expressions from spinor helicity to Lorentz invariant expression. Then we use this map to obtain the three point amplitudes in a supersymmetric theory in terms of Lorentz invariant basis.  In this analysis, the new parity-violating three-point functions that appear in 3+1 dimensions \cite{Chakraborty:2020rxf} play very crucial role (see \eqref{photonreview44} to find the expression of these three-point functions).

	\item We have constructed the four-point functions using the knowledge of the three-point functions. We write down the explicit answers for four photon amplitude in $\mathcal{N}=1,2$ and $4$ theories and demonstrate how the expression of the amplitude simplifies.

The results for four point functions by gluing two parity preserving three point functions are available in the literature \cite{Balasubramanian:2021act}. 	In this work, we also compute the four-point function obtained from gluing two of the parity-violating three-point functions given in \cite{Chakraborty:2020rxf} as they become necessary in a supersymmetric theory.

	 \item In particular the four-photon amplitude in $\mathcal{N}=4$ theories is 
\begin{equation}
	t_8\mathcal{B}^4 \Bigg[  s^{\texttt{j}}\frac{\legendre_{\texttt{j} }(z)}{s-m_{\texttt{j}}^2}+\textrm{two more channels}\Bigg]
\qquad,\qquad
	\texttt{j}\in 2\mathbb{Z}^+
\end{equation}
Here $\mathcal{B}$ is the linearize field strength constructed out of polarization of the photon. $t_8$ is a rank 8 tensor. $t_8\mathcal{B}^4$ is the Lorentz scalar constructed out of $t_8$ tensor and four $\mathcal{B}$s. It is given by 
\begin{eqnarray}
	t_8\mathcal{B}^4=\frac{1}{2}\times \Bigg\{2^4 \textrm{tr}\left(\mathcal{B}_1\mathcal{B}_2\mathcal{B}_3\mathcal{B}_4\right)-2^2\Big( \textrm{tr}\left(\mathcal{B}_1\mathcal{B}_2\right)\textrm{tr}\left(\mathcal{B}_3\mathcal{B}_4\right)+\textrm{perm.}\Big)\Bigg\}
\end{eqnarray}
  $t_8\mathcal{B}^4$ is a unique tensor that is invariant under $\mathcal{N}=4$ supersymmetry \cite{Alday:2007hr}. The coefficient $t_8\mathcal{B}^4$ also turns out to be simple; it is just legendre polynomial. The expression is signicantly elegant compared to the non-supersymmetric theory \cite{Balasubramanian:2021act}.
    
\end{enumerate}
In our analysis, we focus on the exchange contributions and do not keep track of  any contribution to the contact terms. The contact terms do not contribute to residue at any pole. In our analysis, we have ignored any terms which do not contribute to residue at the poles. Please see the discussion below \eqref{nonephotonfourpoint14} for an explicit example of this issue.

\paragraph{Organisation of the paper} The content of the paper is organised as follows: In section \ref{sec:bcrsreviewnonsusy}, we review the known results of the three-point functions of two photons and one higher spin particle and the expressions of four-photon amplitudes due to higher spin exchange in a non-supersymmetric theory. The formalism of on-shell supersymmetry in the massive spinor-helicity formalism is summarised in section \ref{sec:bcrsonshellsusy}. The subsequent three sections contain the key results of the paper. They contain the analysis of the three-point and four-point functions in the presence of supersymmetry. The results for $\mathcal{N}=1,2$ and $4$ theories can be found in \ref{sec:bcrsnone}, \ref{sec:bcrsntwo} and \ref{sec:bcrsnfour} respectively. We end with conclusion and future directions in section \ref{sec:bcrsconclusion}. There are four appendices to help the readers: Appendix \ref{app:bcrsnotation} summarizes the notation and convention followed in this paper. Useful formulae in the spinor helicity formalism are summarized in \ref{app:bcrsshconevtions}. In appendix \ref{app:bcrstreelevelshamp}, we have given a derivation of the tree level amplitude of massless particles due to a massive exchange in the spinor helicity formalism. In the literature, the four-photon amplitudes are written in more than one basis. Appendix \ref{app:bcrspppptsturtcures} summaries those bases. It also has a discussion on dimensional analysis for amplitudes.

\section{Results from non-supersymmetric theories}
\label{sec:bcrsreviewnonsusy}

In this section, we review some of the results that already exist in the literature. Scattering amplitude computations are usually done using fields which are Lorentz tensors or spinors. For future purposes, we refer to this as Lorentz basis computation. In 3+1 dimensions, we also have spinor helicity formalism to compute scattering amplitudes. We use the spinor-helicity formalism to impose constrains from supersymmetry. As a result, we will be going back and forth between the spinor-helicity basis and Lorentz basis. Spinor helicity formalism is extremely efficient in 3+1 dimensions, whereas the Lorentz basis is more useful for discussing things in various dimensions. For example, $\mathcal{N}=4$ supersymmetry in $3+1$ dimension follow from $\mathcal{N}=1$ supersymmetry in $9+1$ dimensions. Even though we restrict to mostly 3+1 dimensions in this work, the Lorentz basis analysis will lay the foundation to consider supersymmetric theories in higher dimensions. 

\subsection{Photon amplitudes in spinor-helicity formalism}

We will be mostly using the massive spinor helicity formalism introduced in  \cite{Arkani-Hamed:2017jhn} (also \cite{Boels:2012if, Conde:2016vxs, Conde:2016izb}). Massless spinor helicity formalism relies on the fact that in $3+1$ dimensions, a null vector can be written as a product of two spinors. Massless spinor-helicity, along with the BCFW formalism, has been very successful in understanding the amplitudes of the massless particles. The formalism with the massive particles is relatively new. The key advantage of this formalism is that it deals with on-shell quantities in a little group covariant representation. As a result, it doesn't suffer from the ambiguity of the field redefinition. A time-like vector can be written as a sum of two null vectors; both the null vectors can be written as the product of two spinors. 
\begin{eqnarray}
p_{a\dot a }&=&-( \zeta_a\tilde \zeta_{\dot a}+\mu_a \tilde \mu_{\dot a})
%\nonumber\\
=-|\zeta]_a \langle \tilde \zeta|_{\dot a}-|\mu]_a  \langle\mu|_{\dot a} 	
\label{photonreview2}
\end{eqnarray}
such that 
\begin{eqnarray}
\langle \zeta \mu \rangle =m =[\tilde \zeta \tilde \mu ]	
\label{photonreview3}
\end{eqnarray}
We need two sets of spinors to write a time-like vector. The above expression is not unique. We can rotate the two spinors as a $SU(2)$ fundamental representation without changing the above equations. In order to make this point manifest consider $\lambda^I$ which is defined as follows 
\begin{equation}
	\lambda^I=(\zeta,\mu) 
\implies 	p_{a\dot a}= -\lambda_{a}^I\, \tilde \lambda_{\dot a I}	
\label{photonreview4}
\end{equation}
We can identify the $SU(2)$ as the little group for massive particles in $3+1$ dimensions \cite{Arkani-Hamed:2017jhn, Conde:2016vxs, Conde:2016izb}. And this allows us to write down amplitudes involving massive particles very easily. We start from three-point functions. For massless spinning particles we write amplitude in two ways: polarization stripped amplitude ($\mathcal{M}^{\mu_1\cdots \mu_n}$) and helicity amplitude ($A(\{h_i\})$). 
\begin{equation}
	\mathcal{M}^{\mu_1\cdots \mu_n}
\qquad,\qquad
A(\{h_i\})= \epsilon_{\mu_1}^{(h_1)}\cdots 	 \epsilon_{\mu_n}^{(h_n)}\mathcal{M}^{\mu_1\cdots \mu_n}
\label{photonreview5}
\end{equation}
For massless particles, we know that the helicity amplitudes are the simplest to study. Similarly for massive particle one can define polarization stripped amplitude and an analogue of helicity amplitude which has the following expression in 
\begin{equation}
A^{I_1\cdots I_{2\texttt{j}}}	= \lambda_{a_1}^{I_1}\cdots \lambda_{a_{2\texttt{j}}}^{I_{2\texttt{j}}}\mathcal{M}^{\{a_1\cdots a_{2\texttt{j}} \}}
\label{photonreview6}
\end{equation}
A large part of this paper deals with the three-point function of two massless particles and one massive particle. The three-point function of two massless and one massive particle in the spinor helicity formalism is unique and it is given by \cite{Arkani-Hamed:2017jhn} $g_{h_1,h_2,\texttt{j}}\,  \widehat{A}_3(h_1,h_2,\texttt{j}) $ where $\widehat{A}_3(h_1,h_2,\texttt{j})$ is given by 
\begin{equation}
	\widehat{A}_3(h_1,h_2,\texttt{j})=\frac{1}{m^{3\texttt{j}+h_1+h_2-1}}[12]^{\texttt{j}+h_1+h_2}\langle 13^{(I_1} \rangle \cdots \langle 13^{I_{\texttt{j}+h_2-h_1}} \rangle \langle 23^{\texttt{j}+h_2-h_1+1} \rangle\cdots \langle 23^{2\texttt{j})} \rangle 
\label{photonreview7}
\end{equation}
$g_{h_1,h_2,\texttt{j}}$ is a real-valued dimensionless coupling constant (see appendix \ref{subapp:bcrspppptsturtcurestwo} for dimensional analysis) .

\paragraph{Action of parity}
Let's now discuss the action of parity on this amplitude. Under parity transformation \cite{Herderschee:2019ofc}
\begin{equation}
P\quad:\quad	\Big([12],\langle 12\rangle,  \langle 13^I \rangle, \langle 23^{I} \rangle \Big)\longrightarrow 
\Big(-\langle 12\rangle, -[12], [ 13^I ], [ 23^{I} ] \Big)	
\label{photonreview11}
\end{equation}
So the action of parity transformation on the amplitude given in  \eqref{photonreview7} is 
\begin{equation}
(-1)^{\texttt{j}+h_1+h_2}\frac{1}{m^{3\texttt{j}+h_1+h_2-1}}\langle 12\rangle^{\texttt{j}+h_1+h_2}[ 13^{(I_1} ] \cdots [13^{I_{\texttt{j}+h_2-h_1}} ] [ 23^{\texttt{j}+h_2-h_1+1} ]\cdots [ 23^{2\texttt{j})} ] 
\label{photonreview12}
\end{equation}
Now we can use the following identities 
\begin{equation}
\langle 12\rangle[12]=-m^2
\qquad,\qquad	
 \langle 13^I \rangle=\frac{\langle 12\rangle}{m}[ 23^I ]
 \qquad,\qquad
 \langle 23^{I} \rangle= -\frac{\langle 12\rangle}{m}[13^I]
\label{photonreview13}
\end{equation}
to write \eqref{photonreview12} in the following way
\begin{equation}
\begin{split}
&	(-1)^{\texttt{j}+h_1+h_2}\frac{1}{m^{3\texttt{j}-h_1-h_2-1}}[12]^{\texttt{j}-h_1-h_2}\langle 13^{(I_1} \rangle \cdots \langle 13^{I_{\texttt{j}-h_2+h_1}} \rangle \langle 23^{\texttt{j}-h_2+h_1+1} \rangle\cdots \langle 23^{2\texttt{j})} \rangle 
\\
&=(-1)^{\texttt{j}+h_1+h_2}	\widehat{A}_3(-h_1,-h_2,\texttt{j})
\end{split}
\label{photonreview13}
\end{equation}
This shows that under parity transformation
\begin{equation}
P\qquad:\qquad \widehat{A}_3(h_1,h_2,\texttt{j})\longrightarrow (-1)^{\texttt{j}+h_1+h_2} \widehat{A}_3(-h_1,-h_2,\texttt{j})	
\label{photonreview14}
\end{equation}
Consider the case when $h_1+h_2=0$; we refer to this as minimal coupling\footnote{For massless particles, one definition of minimal coupling is to replace a partial derivative with a covariant derivative. Another definition is to consider the on-shell three-point function that has the least number of momentum/derivatives. These two definitions are the same. We thank Sourav Ballav for discussion on this point.}. If the massive particle has an even spin, then the amplitude is parity invariant, and otherwise, it is parity odd. 
 
\paragraph{Exchange symmetry} Consider the special case for the amplitude in \eqref{photonreview7} when $h_1=h_2(=h)$; In this case, there is an exchange symmetry between particles 1 and 2
\begin{equation}
1\leftrightarrow 2 \implies \widehat{A}_3(h,h,\texttt{j})	\rightarrow (-1)^{\texttt{j}+2h}\widehat{A}_3(h,h,\texttt{j})
\label{photonreview21}
\end{equation}
In absence of any other internal charges, the amplitude has to satisfy the constraints coming from spin-statistics.  In that case, $\texttt{j}$ must be even.

\paragraph{Four point function}

We glue two three-point functions using the propagators to get the four-point function. The propagator for massive particles in $3+1$ dimensions has very simple expressions
\begin{equation}
	\langle \Phi^{I_1\cdots I_{2\texttt{j}}}(-p)\, \Phi^{J_1\cdots J_{2\texttt{j}}}(p)\rangle
	=\frac{\epsilon^{I_1J_1}\cdots \epsilon^{I_{2\texttt{j}}J_{2\texttt{j}}} }{p^2+M^2-\iimg \varepsilon} \Bigg|_{\textrm{symmetric in } Is}
\label{photonreview31}
\end{equation}
If we glue the three-point amplitudes using the propagator, we get the massless four-point function due to a massive spin $\texttt{j}$ exchange. Let's restrict to the case of external photons/gluons\footnote{We deal with only the colour stripped amplitude.}. In the spinor-helicity basis, various $s$ channel amplitudes are given by (see appendix \ref{app:bcrstreelevelshamp})
\begin{equation}
\begin{split}
\mathcal{M}(1^+2^-3^+4^-)&=\frac{(g_{1,-1,\texttt{j}})^2}{2^{\texttt{j-4}}\, m^{\texttt{2j-2}}}    [13]^2\langle 24\rangle^2\,\frac{ s^\texttt{j-2} \widetilde{\mathcal{N}}_{\texttt{j};2,2}\, \jacobi^{(0,4)}_{\texttt{j}-2}\left(\frac{t-u}{s} \right)}{s-m^2}
\\
\mathcal{M}(1^+2^+3^+4^-)&=\frac{(g_{1,1,\texttt{j}})(g_{1,-1,\texttt{j}})}{2^{\texttt{j-4}}\,m^{\texttt{2j}}}    ([12]^2[13]^2\langle 14\rangle^2) 
\frac{s^{\texttt{j}-2}\widetilde{\mathcal{N}}_{\texttt{j};2,2}\, \jacobi^{(2,2)}_{\texttt{j}-2}\left(\frac{t-u}{s} \right) }{s-m^2} 
\\
\mathcal{M}(1^+2^+3^+4^+)&=\frac{(g_{1,1,\texttt{j}})^2}{2^{\texttt{j}}\,  m^{\texttt{2j+2}}}   ([12]^2[34]^2) \frac{s^{\texttt{j}}\widetilde{\mathcal{N}}_{\texttt{j};0,0}\,  \jacobi^{(0,0)}_{\texttt{j}}\left(\frac{t-u}{s} \right)}{s-m^2}
\\
\mathcal{M}(1^+2^+3^-4^-)&=\frac{(g_{1,1,\texttt{j}})  (g_{-1,-1,\texttt{j}})}{2^{\texttt{j}}\, m^{\texttt{2j+2}}}   ([12]^2\langle 34\rangle^2) \frac{s^{\texttt{j}}\widetilde{\mathcal{N}}_{\texttt{j};0,0}\,  \jacobi^{(0,0)}_{\texttt{j}}\left(\frac{t-u}{s} \right)}{s-m^2}
\end{split}	 
\label{photonreview41}
\end{equation}
where $\jacobi^{(\alpha,\beta)}_{\texttt{j}}(z) $ are Jacobi polynomials. It is given by 
\begin{equation}
	\mathcal{J}_{n}^{(\alpha, \beta)}(x) = \sum_{s=0}^{n}\frac{(n+\alpha)!(n+\beta)!}{s!(n+\alpha-s)!(\beta+s)!(n-s)!}\left(\frac{x-1}{2}\right)^{n-s}\left(\frac{x+1}{2}\right)^{s}
\end{equation}
 $\widetilde{\mathcal{N}}_{\texttt{j};h,h^\prime}$ is given by
\begin{equation}
	\widetilde{\mathcal{N}}_{\texttt{j};h,h^\prime}=	\mathcal{C}_{\texttt{j},h}\, \mathcal{C}_{\texttt{j},h^\prime }
\qquad;\qquad 
	 \mathcal{C}_{\texttt{j},h}=	2^{\frac{\texttt{j}}{2}-h} \sqrt{ \frac{ \Gamma (\texttt{j}-h+1)\, \Gamma (\texttt{j}+h+1)}{ \Gamma (2 \texttt{j}+1)} }
\label{photonreview46}
\end{equation}

\subsection{Photon amplitudes in the Lorentz basis}
The expressions in terms of the spinor helicity variables are very particular to the number of space-time dimensions (in this case $3+1$ dimensions). If one wants to compare scattering amplitudes in different space-time dimensions, then the expressions in terms of Lorentz tensors are better suited for the job. Here we give a quick introduction to the three-point function of two massless and one massive field \cite{Chakraborty:2020rxf} in the Lorentz basis and give the conversion to the spinor helicity basis in 3+1 dimensions. At first, we focus on the case when all three particles are bosonic. The three-point function of two-photon (labelled by 1 and 2) and one massive spin $\texttt{j}$ particle (with polarisation $(\epsilon_3)_{\mu_1\cdots \mu_\texttt{j} }$ ) is given by,  
\begin{equation}
\mathcal{A}_{\texttt{ppj}}=
	\frac{g_{\texttt{ppj}}^\texttt{(0)}}{m^{\texttt{j}-1}}\Big[\mathcal{W}_{(12)}^{\mu \nu }(\epsilon_3)_{\mu \nu \mu_1\cdots \mu_{\texttt{j}-2} }k_{12}^{\mu_1}\cdots k_{12}^{\mu_{\texttt{j}-2}}   \Big]
+	\frac{g_{\texttt{ppj}}^\texttt{(1)}}{m^{\texttt{j}+1}}\Big[\mathcal{W}_{(12)}^{\mu \nu }\eta_{\mu\nu}(\epsilon_3)_{\mu_1\cdots \mu_\texttt{j} }k_{12}^{\mu_1}\cdots k_{12}^{\mu_{\texttt{j}}}\Big]
\label{photonreview41}
\end{equation}
Here $g_{\texttt{ppj}}^\texttt{(0)}$ and $g_{\texttt{ppj}}^\texttt{(1)}$ are dimensionless coupling constants. From bosonic statistics we can conclude that both the three point functions are non-zero only for even $\texttt{j}$.

Here $\mathcal{W}_{\mu \nu}^{(i_1i_2)}$ is defined in the following way 
\begin{equation}
	\mathcal{W}_{\mu \nu}^{(i_1i_2)}= \eta^{\rho\sigma}\mathcal{B}_{\mu \rho }^{(i_1)}\, \mathcal{B}_{\sigma \nu }^{(i_2)}
\qquad,\qquad
	\mathcal{B}_{\mu \nu}^{(i)}= k_\mu^{(i)} \epsilon_\nu^{(i)} -k_\nu^{(i)} \epsilon_\mu^{(i)}	
\label{photonreview42}
\end{equation} 
$\mathcal{B}_{\mu \nu}^{(i)}$ is the linearized field strength of the photon. Using the substitution given in appendix \ref{app:bcrsshconevtions}, we can write the amplitudes in \eqref{photonreview41} in terms of spinor-helicity variables  
\begin{equation}
\begin{split}
\mathcal{A}_{\texttt{ppj}}(++\texttt{j})&=2^{\frac{\texttt{j}}{2}} \texttt{g}_{\texttt{ppj}}^\texttt{(1)}  \widehat{A}_3(+1,+1,\texttt{j})
\\	
\mathcal{A}_{\texttt{ppj}}(--\texttt{j})&=2^{\frac{\texttt{j}}{2}} \texttt{g}_{\texttt{ppj}}^\texttt{(1)} \widehat{A}_3(-1,-1,\texttt{j}) 
\\
\mathcal{A}_{\texttt{ppj}}(+-\texttt{j})&= 2^{\frac{\texttt{j}}{2}-2}\texttt{g}_{\texttt{ppj}}^\texttt{(0)} \widehat{A}_3(+1,-1,\texttt{j})
\\
\mathcal{A}_{\texttt{ppj}}(-+\texttt{j})&= 2^{\frac{\texttt{j}}{2}-2}\texttt{g}_{\texttt{ppj}}^\texttt{(0)} \widehat{A}_3(-1,+1,\texttt{j})
\end{split}	
\label{photonreview43}
\end{equation}
$\widehat{A}_3$ is defined in \eqref{photonreview7}. The two structures given in \eqref{photonreview41} are allowed in any number of spacetime dimensions. In $3+1$ dimensions there are two other structures \cite{Chakraborty:2020rxf}. 
\begin{equation}
\begin{split}
\mathcal{A}_{\texttt{ppj}} &= \frac{  g_{\texttt{ppj}}^\texttt{(2)}}{2\,m^{\texttt{j}+1}}\Bigg(\bigg(\epsilon^{\mu_1\mu_2\mu_3\mu_4}\mathcal{B}^{(1)}_{\mu_1\mu_2}k_{2\,\mu_3}\epsilon_{\texttt{j}\,\mu_4}\bigg)(\mathcal{B}^{(2)}_{\nu_1\nu_2}\epsilon^{\nu_1}_\texttt{j}k^{\nu_2}_1)-(1\leftrightarrow 2)\Bigg)(\epsilon_\texttt{j}\cdot k_{12})^{\texttt{j}-2}\enspace,\enspace\text{for \texttt{j}}\in 2\mathbb{Z}+1
\\
\mathcal{A}_{\texttt{ppj}} &= \frac{ g_{\texttt{ppj}}^\texttt{(3)}}{2m^{\texttt{j}+1}}\bigg(\epsilon^{\mu_1\mu_2\mu_3\mu_3}\mathcal{B}^{(1)}_{\mu_1\mu_2}\mathcal{B}^{(2)}_{\mu_3\mu_4}\bigg)\big(\epsilon_{\texttt{j}}\cdot k_{12}\big)^{\texttt{j}}\enspace,\enspace\text{for \texttt{j}}\in 2\mathbb{Z}
\end{split}
\label{photonreview44}
\end{equation} 
Both of these structures are parity-violating. In order to convert them to spinor-helicity we need to use the following formula
\begin{equation}
\epsilon_{\mu\nu\rho\sigma} \rightarrow \epsilon_{a\dot{a}b\dot{b}c\dot{c}d\dot{d}} = -\iimg \big(\epsilon_{ad}\epsilon_{bc}\epsilon_{\dot{a}\dot{c}}\epsilon_{\dot{b}\dot{d}}-\epsilon_{ac}\epsilon_{bd}\epsilon_{\dot{a}\dot{d}}\epsilon_{\dot{b}\dot{c}}\big)
\label{photonreview45}
\end{equation}
Their expression of the amplitudes in \eqref{photonreview44} in terms of spinor helicity is 
\begin{equation}
\begin{split}
\mathcal{A}_{\texttt{ppj}}(++\texttt{j})&=2^{\frac{\texttt{j}}{2}} \texttt{g}_{\texttt{ppj}}^\texttt{(3)}  \widehat{A}_3(+1,+1,\texttt{j})
\\	
\mathcal{A}_{\texttt{ppj}}(--\texttt{j})&=-2^{\frac{\texttt{j}}{2}} \texttt{g}_{\texttt{ppj}}^\texttt{(3)} \widehat{A}_3(-1,-1,\texttt{j}) 
\\
\mathcal{A}_{\texttt{ppj}}(+-\texttt{j})&= 2^{\frac{\texttt{j}}{2}-2}\texttt{g}_{\texttt{ppj}}^\texttt{(2)} \widehat{A}_3(+1,-1,\texttt{j})
\\
\mathcal{A}_{\texttt{ppj}}(-+\texttt{j})&= 2^{\frac{\texttt{j}}{2}-2}\texttt{g}_{\texttt{ppj}}^\texttt{(2)} \widehat{A}_3(-1,+1,\texttt{j})
\end{split}	
\label{photonreview56}
\end{equation}
In $3+1$ dimensions, the Hodge dual of a 2 form is another 2 form. We define
\begin{equation}
	\widetilde{\mathcal{B}}_{\mu \nu}= \frac{\iimg}{2}\epsilon_{\mu \nu \rho \sigma}\mathcal{B}^{\rho\sigma}
\label{photonreview51}
\end{equation}
Then one can form the following two combinations 
\begin{equation}
	{\mathcal{B}}_{\mu \nu}+\widetilde{\mathcal{B}}_{\mu \nu}
\qquad,\qquad
{\mathcal{B}}_{\mu \nu}-\widetilde{\mathcal{B}}_{\mu \nu}	
\label{photonreview52}
\end{equation}
The first one is non-zero only for positive helicity of the photon and the second one only to negative helicity (of the photon). Further-more we can check that 
\begin{equation}
	\widetilde{\mathcal{B}}_{\mu \nu}\widetilde{\mathcal{B}}^{\mu \nu}={\mathcal{B}}_{\mu \nu}{\mathcal{B}}^{\mu \nu}
\label{photonreview53}
\end{equation}
At this point it is convenient to define the following combinations 
\begin{equation}
	\widehat{\mathcal{B}}^{\pm}_{\mu \nu}=\frac{1}{2}\left( {\mathcal{B}}_{\mu \nu}\pm \widetilde{\mathcal{B}}_{\mu \nu}\right)
\label{photonreview54}
\end{equation}
Using these information, we can form non-minimal three point functions which only couples to positive helicities 
\begin{equation}
	\frac{1}{m^{\texttt{j}-1}}\Big[\eta^{\mu \sigma}\eta^{\nu \rho}\widehat{\mathcal{B}}^{+(1)}_{\mu \nu}\widehat{\mathcal{B}}^{+(2)}_{\rho \sigma}  \Big](\epsilon_3)_{ \mu_1\cdots \mu_{\texttt{j} } }k_{12}^{\mu_1}\cdots k_{12}^{\mu_{\texttt{j} }}
\label{photonreview55}
\end{equation}
For negative helicities, we need to substitute $\widehat{\mathcal{B}}^{+(i)}$ by $\widehat{\mathcal{B}}^{-(i)}$.  Later we will see that this three point function plays an important role in supersymmetric theory. 

Let's now consider the three-point function of two massless fermions and one massive bosonic higher spin particle. The three-point function of two massless spins 1/2 particles is 
\begin{equation}
\mathcal{A}_{\texttt{ffj}}=		\Big[\frac{g_{\textrm{ffj}}^{(0)}}{m^{\texttt{j}-1}}(\bar u_2\gamma^{\mu_1} u_1) 
		+\frac{g_{\textrm{ffj}}^{(1)}}{m^{\texttt{j}}}(\bar u_2  u_1) k_{12}^{\mu_1}\Big](\epsilon_3)_{\mu_1\cdots \mu_\texttt{j} }k_{12}^{\mu_2}\cdots k_{12}^{\mu_{\texttt{j}}}
\label{photonreview64}
\end{equation}
The second term has more derivatives than the first term. In the spinor helicity language, this takes the following form 
\begin{equation} 
\begin{split}
\mathcal{A}_{\texttt{ffj}}\Big(-\frac{1}{2},+\frac{1}{2},\texttt{j}\Big) &= 2^{\frac{\texttt{j}}{2}-1} g^{(0)}_{\texttt{ffj}} \widehat{A}_{\texttt{ffj}}(-\frac{1}{2},+\frac{1}{2},\texttt{j}\Big) \\
\mathcal{A}_{\texttt{ffj}}\Big(+\frac{1}{2},-\frac{1}{2},\texttt{j}\Big) &= - 2^{\frac{\texttt{j}}{2}-1} g^{(0)}_{\texttt{ffj}} \widehat{A}_{\texttt{ffj}}(+\frac{1}{2},-\frac{1}{2},\texttt{j}\Big)
\\
\mathcal{A}_{\texttt{ffj}}\Big(+\frac{1}{2},+\frac{1}{2},\texttt{j}\Big) &=   2^{\frac{\texttt{j}}{2} }g^{(1)}_{\texttt{ffj}} \widehat{A}_{\texttt{ffj}}\Big(+\frac{1}{2},+\frac{1}{2},\texttt{j}\Big)
\\
\mathcal{A}_{\texttt{ffj}}\Big(-\frac{1}{2},-\frac{1}{2},\texttt{j}\Big) &= - 2^{\frac{\texttt{j}}{2}} g^{(1)}_{\texttt{ffj}}  \widehat{A}_{\texttt{ffj}}\Big(-\frac{1}{2},-\frac{1}{2},\texttt{j}\Big)
\end{split}
\label{photonreview65}
\end{equation}
Here we summarise the relation between the coupling constant in the spinor-helicity language and the Lorentz co-variant description 
\begin{equation}
\begin{split}
&g_{+1+1\texttt{j}}=2^{\frac{\texttt{j}}{2}} \texttt{g}_{\texttt{ppj}}^\texttt{(1)}  
\qquad,\qquad	
g_{-1-1\texttt{j}}=2^{\frac{\texttt{j}}{2}} \texttt{g}_{\texttt{ppj}}^\texttt{(1)} 
\qquad,\qquad	
g_{+1-1\texttt{j}}= 2^{\frac{\texttt{j}}{2}-2}\texttt{g}_{\texttt{ppj}}^\texttt{(0)}  
\qquad,\qquad
g_{-1+1\texttt{j}}= 2^{\frac{\texttt{j}}{2}-2}\texttt{g}_{\texttt{ppj}}^\texttt{(0)} 
\\
&g_{-\frac{1}{2}+\frac{1}{2}\texttt{j}}=2^{\frac{\texttt{j}}{2}-1} g^{(0)}_{\texttt{ffj}} 
\qquad,\qquad	
g_{\frac{1}{2}-\frac{1}{2}\texttt{j}}=-2^{\frac{\texttt{j}}{2}-1} g^{(0)}_{\texttt{ffj}} 
\qquad,\qquad	
g_{+\frac{1}{2}+\frac{1}{2}\texttt{j}}=2^{\frac{\texttt{j}}{2}} g^{(1)}_{\texttt{ffj}} 
\qquad,\qquad	
g_{-\frac{1}{2}-\frac{1}{2}\texttt{j}}=-2^{\frac{\texttt{j}}{2} } g^{(1)}_{\texttt{ffj}} 
\end{split}	
\label{photonreview71}
\end{equation}
For future purposes we also define the following three point amplitudes, 
\begin{equation}
\begin{split}
\widehat{\mathcal{A}}^{(0)}_{\texttt{ppj}}&=
	\frac{1 }{m^{\texttt{j}-1}}\Big[\mathcal{W}_{(12)}^{\mu \nu }(\epsilon_3)_{\mu \nu \mu_1\cdots \mu_{\texttt{j}-2} }k_{12}^{\mu_1}\cdots k_{12}^{\mu_{\texttt{j}-2}} \Big]	
\qquad,\qquad\texttt{j}\in 2\mathbb{Z}	
\\
&= \frac{1}{2\,m^{\texttt{j}+1}}\Bigg(\bigg(\epsilon^{\mu_1\mu_2\mu_3\mu_4}\mathcal{B}^{(1)}_{\mu_1\mu_2}k_{2\,\mu_3}\epsilon_{\texttt{j}\,\mu_4}\bigg)(\mathcal{B}^{(2)}_{\nu_1\nu_2}\epsilon^{\nu_1}_\texttt{j}k^{\nu_2}_1)-(1\leftrightarrow 2)\Bigg)(\epsilon_\texttt{j}\cdot k_{12})^{\texttt{j}-2} 
\qquad,\qquad
\texttt{j}\in 2\mathbb{Z}+1	
\\
\widehat{\mathcal{A}}^{(1+)}_{\texttt{ppj}}&=
\frac{1}{m^{\texttt{j}-1}}\Big[\eta^{\mu \sigma}\eta^{\nu \rho}\widehat{\mathcal{B}}^{+(1)}_{\mu \nu}\widehat{\mathcal{B}}^{+(2)}_{\rho \sigma}  \Big](\epsilon_3)_{ \mu_1\cdots \mu_{\texttt{j} } }k_{12}^{\mu_1}\cdots k_{12}^{\mu_{\texttt{j} }}
\qquad,\qquad\texttt{j}\in 2\mathbb{Z}
\\
\widehat{\mathcal{A}}^{(1-)}_{\texttt{ppj}}&=
\frac{1}{m^{\texttt{j}-1}}\Big[\eta^{\mu \sigma}\eta^{\nu \rho}\widehat{\mathcal{B}}^{-(1)}_{\mu \nu}\widehat{\mathcal{B}}^{-(2)}_{\rho \sigma}  \Big](\epsilon_3)_{ \mu_1\cdots \mu_{\texttt{j} } }k_{12}^{\mu_1}\cdots k_{12}^{\mu_{\texttt{j} }}
\qquad,\qquad\texttt{j}\in 2\mathbb{Z}
\end{split}	
\label{photonreview73}
\end{equation}

\subsubsection{Four photon amplitude}
\label{sec:fourphotonamplitudebasis}

This section computes the four-point function of external photons in a non-supersymmetric theory. Four-point scattering amplitude is a function of external polarization/momenta as well as the Mandelstam variables. A very convenient way to write down scattering amplitudes is 
\begin{equation}
	\sum_{\alpha}T_\alpha(\epsilon_i,k_i)\,  F_\alpha(s,t,u)
\label{photonreview101}
\end{equation} 
$T_\alpha(\epsilon_i,k_i)$ are called tensor factors; they capture the dependence on the polarisation entirely. For external massless particles, every tensor factor is gauge-invariant. $\alpha$ runs over the linearly independent tensor factors. $F_\alpha(s,t,u)$s are called form factors; they are functions of the Mandelstam variables only. We restrict to the tree level amplitudes. As a result, they can have at most simple poles in the Mandelstam variables. $F_\alpha(s,t,u)$ depends on the mass and spin of the particles in theory. If a theory has more than massive spinning particle then the form factor is given by  
\begin{equation}
	F_\alpha(s,t,u) = \sum_{\texttt{j}} F_\alpha(s,t,u|\texttt{j})
\label{photonreview102}
\end{equation}
$\texttt{j}$ is the spin of the exchange.
For example, the expression for the form factor due to symmetric traceless exchange can be found in \cite{Balasubramanian:2021act}. This paper wants to find the tree-level amplitudes in a supersymmetric theory. 

Let us now consider four-photon amplitude. We start by defining following Lorentz invariant and  Gauge invariant structures of polarization and momenta. 
\begin{equation}
\begin{split}
	\mathcal{W}_{i_1i_2}&= \mathcal{B}^{\mu_1\nu_1}_{(i_1)} \mathcal{B}^{\mu_2\nu_2}_{(i_2)}\eta_{\nu_1\mu_2}\eta_{\nu_2\mu_1}	
\\
\mathcal{W}_{i_1i_2i_3}&=	\mathcal{B}^{\mu_1\nu_1}_{(i_1)} \mathcal{B}^{\mu_2\nu_2}_{(i_2)}\mathcal{B}^{\mu_3\nu_3}_{(i_3)}\eta_{\nu_1\mu_2}\eta_{\nu_2\mu_3}	\eta_{\nu_3\mu_1}	
\\
\mathcal{W}_{i_1i_2i_3i_4}&=	\mathcal{B}^{\mu_1\nu_1}_{(i_1)} \mathcal{B}^{\mu_2\nu_2}_{(i_2)}\mathcal{B}^{\mu_3\nu_3}_{(i_3)}\mathcal{B}^{\mu_4\nu_4}_{(i_4)}\eta_{\nu_1\mu_2}\eta_{\nu_2\mu_3}	\eta_{\nu_3\mu_4}\eta_{\nu_4\mu_1}
\\
\mathcal{Y}_i&=k_{(i+1)\mu }\, \mathcal{B}^{\mu\nu}_{(i)}\, k_{(i-1)\nu }
\end{split}
\label{photonreview104}
\end{equation}
From this, we also define 
\begin{equation}
\begin{split}
	\mathcal{Z}_{i_1i_2i_3i_4}&=\mathcal{W}_{i_1i_2i_3i_4}-\frac{1}{4}\Big(\mathcal{W}_{i_1i_2 }\mathcal{W}_{ i_3i_4}+ \mathcal{W}_{i_1 i_3 }\mathcal{W}_{ i_2 i_4}+ \mathcal{W}_{i_1i_4}\mathcal{W}_{i_2i_3} \Big)
\\
\mathcal{V}&=\mathcal{Y}_1 \mathcal{W}_{234}+\mathcal{Y}_2\mathcal{W}_{341}+\mathcal{Y}_3\mathcal{Y}_{412}+\mathcal{Y}_4\mathcal{W}_{123} 	
\end{split}
\label{photonreview105}
\end{equation}
Then a Tensor factor basis for writing down parity invariant four-photon amplitude is 
\begin{equation}
\begin{split}
	&\mathcal{T}_1= \mathcal{Z}_{1234} +\frac{1}{2}\mathcal{W}_{12} \mathcal{W}_{34} \qquad,\qquad \mathcal{T}_2=\mathcal{Z}_{1324} +\frac{1}{2}\mathcal{W}_{13} \mathcal{W}_{24} \qquad,\qquad \mathcal{T}_3= \mathcal{Z}_{1423} +\frac{1}{2}\mathcal{W}_{14} \mathcal{W}_{23} 	
\\ 
	&\mathcal{T}_4= \mathcal{Z}_{1234} 	
\qquad,\qquad \mathcal{T}_5=\mathcal{Z}_{1324} 	
\qquad,\qquad \mathcal{T}_6=\mathcal{Z}_{1243} 		
\qquad,\qquad \mathcal{T}_7= \mathcal{V} 		
\end{split}	
\label{photonreview107}
\end{equation}  
Every tensor factor has certain symmetries, and that will put some constraint on the corresponding form factor. For example, 
$\mathcal{V} $ is invariant under all possible exchanges and hence 
\begin{equation}
	\mathcal{F}_7(s,t,u)=	\mathcal{F}_7(t,s,u)
=	\mathcal{F}_7(u,t,s)
\label{photonreview108}
\end{equation}
The angular distribution of these basis elements are given in table \ref{tab:joycebasisangdist}. 
\begin{table}[h]
\begin{center}
    
\begin{tabular}{ ||p{2cm}||p{1.8cm}|p{1.8cm}|p{1.8cm}|p{1.8cm}| p{1.8cm}||  }
 \hline
 \hline
 \multicolumn{6}{|c|}{{\bf Angular distribution of the tensor structures} } \\
 \hline
 \hline
$T$ structure & $(++++)$  $(----)$& $(+++-)$  $(++-+)$ $(+-++)$ $(-+++)$  & $(++--)$  $(--++)$& $(+-+-)$  $(-+-+)$ & $(+--+)$  $(-++-)$    \\
 \hline
  \hline
 &&&&&
 \\
 $ \mathcal{T}_{1}$   & $0$   &$ 0$  & $s^2$ & $0$ &$0$  \\
 &&&&&\\
 \hline
  &&&&&
 \\
 $ \mathcal{T}_{2}$   & $0$   &$ 0$  &  $0$& $s^2\cos^4 (\frac{\theta}{2}) $ &  $0$ \\
 &&&&&\\
 \hline
  &&&&&
 \\
 $ \mathcal{T}_{3}$   & $0$   &$ 0 $  & $0$&  $0$ &   $s^2\sin^4 (\frac{\theta}{2})  $\\
 &&&&&\\
 \hline
  &&&&&
 \\
 $ \mathcal{T}_{4}$   &$ -s^2 $  & $0$ &  $0$ & $0$ &$0$ \\
 &&&&&\\
 \hline
  &&&&&
 \\
  $ \mathcal{T}_{5}$   &$-s^2\cos^4 (\frac{\theta}{2}) $ & $0$ &  $0$    &  $0$ & $0$ \\
 &&&&&\\
 \hline 
  &&&&&
 \\
 $ \mathcal{T}_{6}$  &$-s^2\sin^4 (\frac{\theta}{2})  $   & $0$  &  $0$ & $0$& $0$ \\
 &&&&&\\
 \hline
  &&&&&
 \\
 $ \mathcal{T}_{7}$   & $\frac{1}{4} s^3\sin^2\theta $   &$-\frac{1}{16} s^3\sin^2\theta $   & $0$  &  $0$& $0$ \\
 &&&&&\\
 \hline
 \hline
\end{tabular}

\caption{Angular distribution of the tensor structures}
\label{tab:joycebasisangdist}
\end{center}

\end{table}
The key advantage of this basis is that for $(+++-)$, $(++--)$, $(+-+-)$ and $(+--+)$ configurations, only one of the tensor structures is non-zero. Note that $\mathcal{T}_{4}$, $\mathcal{T}_{5}$ and $\mathcal{T}_{6}$ are non-zero only for $(++++)$ configuration. This fact indicates that there is a choice of $\mathcal{F}_4$, $\mathcal{F}_5$ and $\mathcal{F}_6$ such that the combination (given in \eqref{photonreview101}) has zero angular distribution for any choice of helicity of external state. Consider an function of the following form (at this point, we are agnostic about the origin of this amplitude; the reason will become clear soon)
\begin{equation}
	\mathcal{T}_{4}\mathcal{F}_4+
	\mathcal{T}_{5}\mathcal{F}_5+
	\mathcal{T}_{6}\mathcal{F}_6
\label{photonreview111}
\end{equation}
By construction, this is zero for $(+++-)$, $(++--)$, $(+-+-)$ and $(+--+)$ configurations. Let's now consider $(++++)$ configuration. In this case, the amplitude becomes 
\begin{equation}
	-s^2\Bigg[\mathcal{F}_4(s,\cos\theta)
	+\cos^4 \left(\frac{\theta}{2}\right) \mathcal{F}_5(s,\cos\theta)
	+\sin^4 \left(\frac{\theta}{2}\right) \mathcal{F}_6(s,\cos\theta)
	\Bigg]
\label{photonreview112}
\end{equation}
There are choices of $\mathcal{F}_4$, $\mathcal{F}_5$ and $\mathcal{F}_6$ for which this combination is zero and hence the function is zero for all possible helicity configurations. This strongly suggests that the function can be set to zero; not all choices of $\mathcal{F}_4$, $\mathcal{F}_5$ and $\mathcal{F}_6$ give rise to different amplitudes. The authors of \cite{Chowdhury:2019kaq} arrived at this statement from a different consideration and have made this more precise. We present their analysis in appendix \ref{subapp:bcrspppptsturtcuresone}. The discussion on the basis for Tensor factors in the spinor helicity at the level of four-point function can be found in \cite{DeAngelis:2022qco}; in this paper we always use Lorentz basis for the four point and hence do not go into that discussion. 

\subsubsection{Tree-level answer in a non-supersymmetric theory \cite{Balasubramanian:2021act}} 
We restrict to the case when both the three-point that we glue to get the four-point function are the same. Let's consider the non-minimal couplings first. There are two such three point functions \cite{Chakraborty:2020rxf}.  In both of these cases, the spin of the massive particle must be even. The four point function for these cases are given by  
\begin{enumerate}
	\item Parity preserving  non-minimal coupling (given in \eqref{photonreview41}): The non-zero form factors are given by 
\begin{equation}
	\mathcal{F}_1(s,t,u)=-\mathcal{F}_4(s,t,u)=\widetilde{\mathcal{N}}_{{\texttt{j},0,0} } \frac{(g_{\texttt{ppj}}^\texttt{(1)})^2}{m^{2\texttt{j}+2}} s^{\texttt{j} }\frac{\legendre_\texttt{j}(z)}{s-m^2}
\label{photonreview131}
\end{equation}
$\legendre_\texttt{j}$ is the Legendre polynomial and the above answer for $s$-channel  exchange. This was computed in \cite{Balasubramanian:2021act}, but here we present the answer in choice of basis for the tensor factors

	\item Parity violating  non-minimal coupling (given in \eqref{photonreview44}):  The non-zero form factors are given by 
\begin{equation}
	\mathcal{F}_1(s,t,u)=\mathcal{F}_4(s,t,u)=2\widetilde{\mathcal{N}}_{{\texttt{j},0,0} }\frac{(g_{\texttt{ppj}}^\texttt{(3)})^2}{m^{2\texttt{j}+2}}   s^{\texttt{j} } \frac{\legendre_\texttt{j}(z)}{s-m^2}
\label{photonreview132}
\end{equation}
\end{enumerate}
Note that $\mathcal{F}_1(s,t,u)$ is the same in both cases (up to a factor of $2$) and $\mathcal{F}_4(s,t,u)$ flips a sign (up to a factor of $2$) . This means that there is a combination of these two amplitudes such that it is non-zero only for $(++--)$ (or $(++++)$ configurations). The relevance of this comment will become clear when we discuss supersymmetric cases. 
\\
Let's now consider minimal coupling. Again there are two cases  
\begin{enumerate}
	\item Parity preserving  minimal coupling (given in \eqref{photonreview41}): This is consistent with spin statistics only for even $\texttt{j}$
\begin{equation}
\mathcal{F}_{3}(s,-z)=
\mathcal{F}_{2}(s,z)=	2\frac{(g_{\texttt{ppj}}^\texttt{(0)})^2}{  m^{2\texttt{j}-2}} s^{\texttt{j}-2}\,  \widetilde{\mathcal{N}}_{{\texttt{j},2,2} }	 \, \frac{\jacobi_{\texttt{j}-2}^{(0,4)}(z) }{s-m^2}
\label{photonreview136}
\end{equation}
$\jacobi^{(\alpha,\beta)}_{\texttt{j}}(z) $ are Jacobi polynomials. This result for the general dimension can be found in \cite{Balasubramanian:2021act} where the answer was written on a different basis. We have taken that answer, converted the basis given in \eqref{photonreview107} and then removed the redundancy that is discussed above and in appendix \ref{subapp:bcrspppptsturtcuresone}. 

	\item Parity violating minimal coupling (given in the second eqn of \eqref{photonreview44}): This is consistent with spin statistics only for odd $\texttt{j}$. The non-zero form factors are given by 
\begin{equation}
\mathcal{F}_{3}(s,-z)=
\mathcal{F}_{2}(s,z)=	2\frac{(g_{\texttt{ppj}}^\texttt{(2)})^2}{m^{2\texttt{j}+2} }
s^{\texttt{j}}\,  \widetilde{\mathcal{N}}_{{\texttt{j},2,2} }	  
\frac{\jacobi_{\texttt{j}-2}^{(0,4)}(z)}{s-m^2}
\label{photonreview137}
\end{equation}
\end{enumerate}

\paragraph{Angular distribution} The residue for the $s$ channel in the center of mass frame has a very simple expression
\begin{equation}
	A(1^{h_1},2^{h_2},3^{h_3},4^{h_4}; \texttt{j})=(g_{h_1,h_2,\texttt{j}})(g_{h_3,h_4,\texttt{j}})\, \widetilde{\mathcal{N}}_{\texttt{j};h,h^\prime} s^\texttt{j}\,  d^{(\texttt{j})}_{h,h^\prime}(\theta)
\label{photonreview32}
\end{equation} 
where $h=h_1-h_2$, $h^\prime=h_3-h_4$. $d^{(\texttt{j})}_{h,h^\prime}(\theta)$ is the Legendre polynomial.

\section{On-shell supersymmetry}
\label{sec:bcrsonshellsusy}

Supersymmetry is the only non-trivial extension of the Poincare symmetry for any interacting theory living in 3+1 dimensions or above. Apart from the Poincare generators, Supersymmetry algebra includes fermionic generators, which mutually anti-commute to give momentum. The supersymmetry algebra in the language of spinor helicity variables takes the following form
\begin{equation}
	\Big\{\mathcal{Q}_a^A,\mathcal{Q}^\dagger_{B\dot a}\Big\}=-2{\delta^A}_B p_{a\dot a}
\qquad,\qquad	
\Big\{\mathcal{Q}_a^A,\mathcal{Q}^B_{b}\Big\}=2Z^{AB} \epsilon_{a b}
\qquad,\qquad	
\Big\{\mathcal{Q}_{A\dot a},\mathcal{Q}_{B\dot b}\Big\}=2Z_{AB} \epsilon_{\dot a \dot b}
\label{onshellsusyreview1}
\end{equation}
$a,b, \dot a, \dot b= 1,2$ are spinor indices, $A,B$ are indices for the $R$-symmetry group. $Z_{AB}$ are called central charges. Central charges play an important role in understanding representations of the SUSY algebra. However, for this work, we are interested in the massive ``long"-multiplets, and hence it doesn't play any role in our analysis. From now on, we set it to zero. 

An irreducible representation of super-Poincare (Supersymmetry) algebra is known as a supermultiplet. A supermultiplet consists of more than one Poincare multiplets ($\equiv$ particle). The mass of all the particles in a supermultiplet is the same. For any super-symmetry multiplet the number of bosonic d.o.f is the same as the number of fermionic d.o.f. The content of the supermultiplet is distinctly different depending on the mass being zero or non-zero \footnote{To be precise, the structure of the supermultiplet depends on the mass and the value of the central charge. However, in this paper, we are considering states with zero central charges.}. 

Poincare symmetry is the symmetry of Minkowski spacetime. A useful way to realize Supersymmetry is to consider a manifold where Supersymmetry becomes the isometry of the manifold; Such a construction is known as superspace \cite{Salam:1976ib}. Superspace contains extra Grassmann valued coordinates along with Minkowski coordinates. The number of Grassmann valued coordinates depends on the amount of Supersymmetry. For $\mathcal{N}=n$ we introduce $n$ Grassmann co-ordinates
\begin{equation}
	\eta^A \qquad,\qquad A=1,\cdots, n
\label{onshellsusyreview2}
\end{equation}
In the absence of central charge, the $R$ symmetry group is $SU(\mathcal{N})$; $\eta^A $s transform as a fundamental representation of $SU(\mathcal{N})$. In terms of the spinor helicity variables, the super-charge has the following expressions in the $\eta$ basis 
\begin{equation}
	\mathcal{Q}_{a}^A=\sum_{i} \mathcal{Q}_{ia}^A 
\qquad ,\qquad
\mathcal{Q}_{ia}^A =	
\left\{	
\begin{matrix}
	i=\textrm{massless}\qquad,\qquad\sqrt{2}|i]\frac{\partial}{\partial \eta_{i}^A}
\\	
\\	
	i=\textrm{massive}\qquad,\qquad\sqrt{2}|i_I]\frac{\partial}{\partial \eta_{i,I}^A}
\end{matrix}	
\right.
\label{onshellsusyreview3}
\end{equation}
$\mathcal{Q}^\dagger$s are given by 
\begin{equation}
(\mathcal{Q}^\dagger)_{i}^{aA} =	
\left\{	
\begin{matrix}
	i=\textrm{massless}\qquad,\qquad-\sqrt{2}|i\rangle^a\eta_{i}^A
\\	
\\	
	i=\textrm{massive}\qquad,\qquad-\sqrt{2}|i^I\rangle^a \eta_{i,I}^A
\end{matrix}	
\right.
\label{onshellsusyreview4}
\end{equation}
$\mathcal{Q}$s and $\mathcal{Q}^\dagger$s provide a representation for the supersymmetry algebra in terms of the superspace co-ordinates and derivatives.  

We know that every irreducible unitary representation can be realised in terms of fields which are a function of the Minkowski coordinates. There is an analogous way to represent the irreducible unitary representation of the super-Poincare group in terms of superfields. In this paper, we work with on-shell superfield, which is a function of momentum and the grassmann variable 
\begin{equation}
	\Phi_\texttt{j}(k^\mu, \eta^A)
\label{onshellsusyreview5}
\end{equation}
On-shell condition and hence the expression of on-shell superfields are better suited in the momentum space. Since $\eta^A$ is Grassmann valued, the Taylor expansion truncates after a few terms. The coefficients of the Taylor expansion can be written as a linear sum of Poincare representations. Superfields and their expansions for a different amount of Supersymmetry can be found in the following sections.

\paragraph{Parity and Supersymmetry}
Parity flips the spatial part of the momenta
\begin{equation}
	\mathcal{P}\qquad:\qquad k_\mu \longrightarrow {\mathcal{P}_\mu}^\nu k_\nu 
\qquad,\qquad	{\mathcal{P}_\mu}^\nu=\textrm{diag}(1,-1,-1,-1)
\label{onshellsusyreview11}
\end{equation}
Under Parity, the spinor helicity variable for massless particles transforms in the following way 
\begin{equation}
	\mathcal{P}\qquad:\qquad \begin{matrix}
		|k\rangle
		\\
		\langle k|
		\\
		|k]
		\\
		[k|
	\end{matrix}
\quad	
\longrightarrow	
\quad	
\left\{
\begin{matrix}
		-e^{\iimg \varphi}|k]
		\\
		e^{\iimg \varphi} [k|
		\\
		e^{-\iimg \varphi}|k\rangle
		\\
		-e^{-\iimg \varphi}\langle k|
	\end{matrix}
\right.	
\label{onshellsusyreview12}
\end{equation}
\begin{equation}
	\mathcal{P}\qquad:\qquad \begin{matrix}
		|k^I\rangle
		\\
		\langle k^I|
		\\
		|k^I]
		\\
		[k^I|
	\end{matrix}
\quad	
\longrightarrow	
\quad	
\left\{
\begin{matrix}
		|k^I]
		\\
		-[k^I|
		\\
		|k^I\rangle
		\\
		-\langle k^I|
	\end{matrix}
\right.	
\label{onshellsusyreview13}
\end{equation}
\begin{equation}
\mathcal{P}\eta_I^A\mathcal{P}^{-1}=\iimg \eta_{I,A}^\dagger  
\label{onshellsusyreview14}
\end{equation}
Let's now consider the action of Parity on the massless superfield $\Sigma^{(h)}$. We start by recalling the action of Parity on any field (in the momentum space)
\begin{equation}
	\mathcal{P}\Phi^{\mu_1\cdots \mu_n}(k^\mu)\mathcal{P}^{-1}
=	{\mathcal{P}^{\mu_1}}_{\nu_1}\cdots{\mathcal{P}^{\mu_n}}_{\nu_n}  \Phi^{\nu_1\cdots {\nu_n}}({\mathcal{P}^\mu}_\nu k^\nu)
\label{onshellsusyreview15}
\end{equation}
Using equation \eqref{onshellsusyreview14} we get the following action on a massless superfield 
\begin{equation}
	\mathcal{P}\Sigma^{(h)}(k^\mu, \eta^A)\mathcal{P}^{-1}=\zeta^{(h)} \Sigma^{(-h)}({\mathcal{P}^\mu}_\nu k^\nu, \iimg \eta_A^\dagger )
\label{onshellsusyreview21}
\end{equation}
$\zeta^{(h)}$ is a phase. $h\rightarrow -h $ is due to the fact that helicity flips under Parity. Since parity squares to identity we get $\zeta^{(h)}\zeta^{(-h)}=1$. A massless superfield is called Parity self-conjugate if
\begin{equation}
	\Sigma^{(h)} ( \eta^A)\propto \widetilde{\Sigma}^{(-h)} ( \eta^A)
\label{onshellsusyreview22}
\end{equation} 
$\widetilde{\Sigma}^{(-h)}(\eta^A)$ is Grassmann Fourier transform of $\Sigma^{(-h)}( \eta_A^\dagger )$. Similarly, we can find action of Parity on massive superfield
\begin{equation}
	\mathcal{P}\Phi^{I_1\cdots I_{\texttt{2j}}}_{\texttt{j}}(k^\mu, \eta_I^A)\mathcal{P}^{-1}
=	\zeta_{(\texttt{j})}  \Phi^{I_1\cdots I_{\texttt{2j}}}_{\texttt{j}}({\mathcal{P}^\mu}_\nu  k^\nu,  \iimg \eta_{IA}^\dagger)
\label{onshellsusyreview23}
\end{equation}
$\zeta_{(\texttt{j})}  $ is a phase; since parity squares to identity $\zeta_{(\texttt{j})}=\pm 1$\footnote{From general considerations one would except 
\begin{equation}
	\mathcal{P}\Phi^{I_1\cdots I_{\texttt{2j}}}_{\texttt{j}}(k^\mu, \eta_I^A)\mathcal{P}^{-1}
=	{\zeta^{I_1\cdots I_{\texttt{2j}}}}_{{K_1\cdots K_{\texttt{2j}}}}  \Phi^{K_1\cdots K_{\texttt{2j}}}_{\texttt{j}}({\mathcal{P}^\mu}_\nu k^\nu,  \iimg \eta_{IA}^\dagger)
\nonumber
\label{onshellsusyreview24}
\end{equation}
But since Parity commutes with little group generators, we get 
\begin{equation}
	{\zeta^{I_1\cdots I_{\texttt{2j}}}}_{{K_1\cdots K_{\texttt{2j}}}}= 
	\zeta_{\texttt{j}}{\delta^{I_1}}_{K_1}\cdots {\delta^{I_\texttt{2j}}}_{K_\texttt{2j}}
\nonumber	
\label{onshellsusyreview25}
\end{equation}
}. The implication of these relations for the component fields will be analysed in the later sections. 

\subsection{A quick introduction to Super-amplitude}

A $n$-point amplitude is a Poincare scalar constructed out of $n$ fields(=Poincare representations). One can analogously define superamplitude, a super-Poincare invariant quantity constructed out of $n$- supermultiplets. In \cite{Herderschee:2019ofc}, it is shown that supersymmetric invariant three-point amplitudes can be uniquely fixed. These amplitudes are annihilated by the supercharges $\mathcal{Q}$ and $\mathcal{Q}^{\dagger}$. A natural way to achieve this is by incorporating supercharge conserving delta functions. In the $\eta$ basis, the delta function  in an $n$-point superamplitude is given by
\begin{equation}
	\delta^{(2\mathcal{N})}\left(\mathcal{Q}^{\dagger}\right) = \prod_{A=1}^{\mathcal{N}}\left(\sum_{i<j =1}^{n}\langle i^{I}j^{J}\rangle\eta_{iI}^{A}\eta_{jJ}^{A} + \frac{1}{2}\sum_{i=1}^{n}m_{i}\eta_{iI}^{A}\eta_{i}^{IA}\right).
\label{onshellsusyreview31}
\end{equation}
Here $\mathcal{N}$ is the amount of supersymmetry. 

In the $\eta$ basis a three-point superamplitude with $M = \{1,2,3\}$ number of massive legs can be written as product of supercharge conserving delta function and a polynomial in Grassmann variables, $F(\eta_{I})$. The degree of the polynomial satisfies an upper bound of $\mathcal{N}(M-1)$. For $M=1$ $F$ is a degree zero polynomial in $\eta_{I}$. The superamplitude with two massless and one massive superfields can then be given in terms of any of the component amplitudes, like
\begin{eqnarray}
	\mathcal{A}_{3}\left(\Sigma^{(h_{1})}_{1}, \Sigma^{(h_{2})}_{2}, \Omega_{3}^{(\texttt{j})}\right) & = & \frac{\mathfrak{g}_{h_{1},h_{2},\texttt{j}}}{m^{\mathcal{N}}}\widehat{A}_{3}\left(\varphi_{1}^{h_{1}}, \varphi_{2}^{h_{2}}, \tilde{\phi}_{3}^{(\texttt{j})}\right)\delta^{(2\mathcal{N})}\left(\mathcal{Q}^{\dagger}\right) \\
	& = & \frac{\mathfrak{g}_{h_{1},h_{2},\texttt{j}}}{m^{3\texttt{j}+h_{1}+h_{2} +\mathcal{N}-1}}\delta^{(2\mathcal{N})}\left(\mathcal{Q}^{\dagger}_{(3)}\right)[12]^{\texttt{j}+h_{1}+h_{2}}\langle 1\mathbf{3}\rangle^{\odot \texttt{j}+h_{2}-h_{1}}\langle 2\mathbf{3}\rangle^{\odot \texttt{j}+h_{1}-h_{2}} .
\label{onshellsusyreview32}
\nonumber
\end{eqnarray}
$\widehat{A}_3$ is given in \eqref{photonreview7}. For the purpose of this paper we will be interested in three-point amplitudes with two massless and one massive higher spin states. Four-point amplitudes with all external massless states are obtained by gluing two such three-point amplitudes with the massive states being exchanged in the internal channel of the four-point amplitudes.

\paragraph{Four photon amplitude in supersymmetric theory} We have already introduce form factor and tensor factor in section \ref{sec:fourphotonamplitudebasis}. In a supersymmetric theory, we will find that the form factor depends on the amount of Supersymmetry. To make this part manifest, we write the above expression as 
\begin{equation}
	\sum_{\alpha}T_\alpha(\epsilon_i,k_i)\,  F_\alpha(s,t,u|\mathcal{N},\texttt{j})
\label{photonreview103}
\end{equation} 
$\mathcal{N}$ is the amount of Supersymmetry, and $\texttt{j}$ is the spin of the superfield which defined as the spin of Clifford vacuum. 
We will see that as the number of Supersymmetry increases, the number of independent form factors reduces. As a by-product, this also provides the partial wave in supersymmetric theories.

\section{$\mathcal{N}=1$ supersymmetry} 
\label{sec:bcrsnone}

Our notation is that we use $g$ for coupling between three Poincare representations and $\mathfrak{g}$ for three super-Poincare representations. To put it differently, 
\begin{equation}
\begin{split}
	\mathfrak{g}&= \textrm{coupling of super-amplitude}
\\
g&= \textrm{coupling of ordinary amplitude}
\end{split}
\end{equation}
In this work, we focus on the three-point function of two massless (super-)fields and massive (super-)field. The coupling constant for those (super-) amplitudes is ($\mathfrak{g}_{h_1h_2\texttt{j}}$) $g_{h_1h_2\texttt{j}}$. $h_1$ and $h_2$ denote the helicity of the massless (super-)fields, and $\texttt{j}$ denotes the spin of the massive (super-)field.

\subsection{Superfields}

We start with the simplest supersymmetric theory in $3+1$ dimensions. The massless supermultiplet in $\mathcal{N}=1$ theories has one bosonic field and one fermionic field
\begin{equation}
	\Sigma^{(h)}= \varphi^{(h)}+\eta\,  \varphi^{(h-\frac{1}{2})}
\label{nonesuperfield0.1}
\end{equation} 
$h$ is the helicity of the particles; For massless superfields, the helicity of the Clifford vacuum defines the helicity of the superfield. The massless multiplet is not CPT conjugate. $h=1,-\frac{1}{2}$ multiplets are known as the vector/photon multiplets 
\begin{equation}
	\Sigma^{(1)}= p^{(1)}+\eta\,  f^{(\frac{1}{2})}
\qquad,\qquad
\Sigma^{(-\frac{1}{2})}= f^{(-\frac{1}{2})}+\eta\,  p^{(-1)}	
\label{nonesuperfield0.2}
\end{equation}

A $\mathcal{N}=1$  massive supermultiplet has $4(2\texttt{j}+1)$ states; here $\texttt{j}$ is the spin of the Clifford vacuum.  An on-shell massive supermultiplet is given by  
\cite{Herderschee:2019ofc} 
\begin{equation}
		\Phi_{\texttt{j}}^{(I_1\cdots I_\texttt{2j})}= 
		\alpha_1\phi_{\texttt{j}}^{(I_1\cdots I_\texttt{2j})}+\alpha_2
		\eta_J\psi_{\texttt{j}+\frac{1}{2}}^{(JI_1\cdots I_\texttt{2j})}+
		\alpha_3 \eta^{(I_1} \psi_{\texttt{j}-\frac{1}{2}}^{I_2\cdots I_\texttt{2j})} 
		+\alpha_4\frac{1}{2}\eta_J\eta^J{\widetilde  \phi}_{\texttt{j}}^{(I_1\cdots I_\texttt{2j})}
\label{nonesuperfield1}		
\end{equation}
 $\alpha$ s are real from the reality condition of the superfield. Here we spell out the strategy to determine these coefficients. In lates sections, we write down only the expressions for the superfields. 
To determine the coefficients, we demand that the Supersymmetry invariant quadratic term made of out superfields must have the canonical normalization in terms of the component fields. Now the supersymmetry invariant inner product of fields can be obtained from the product of the superfields. 
\begin{equation}
\begin{split}
	&\int d^2\eta \Bigg[ \Phi_{\texttt{j}}^{(I_1\cdots I_\texttt{2j})}(\Phi_{\texttt{j}})_{(I_1\cdots I_\texttt{2j})}\Bigg]
\\
&= 	\phi_{\texttt{j}}^{(I_1\cdots I_\texttt{2j})}\Big({\widetilde  \phi}_{\texttt{j}}\Big)_{(I_1\cdots I_\texttt{2j})}
+\psi_{\texttt{j}+\frac{1}{2}}^{(I_1\cdots I_\texttt{2j+1})}
\Big(\psi_{\texttt{j}+\frac{1}{2}}\Big)_{(I_1\cdots I_\texttt{2j+1})}
+\psi_{\texttt{j}-\frac{1}{2}}^{(I_1\cdots I_\texttt{2j-1})}
\Big(\psi_{\texttt{j}-\frac{1}{2}}\Big)_{(I_1\cdots I_\texttt{2j-1})}	
\end{split}
\label{nonesuperfield2}		
\end{equation}
If we take the expression in \eqref{nonesuperfield1} and put it in the first line of \eqref{nonesuperfield2} we get
\begin{equation}
\begin{split}
\alpha_1\alpha_4\, 	\phi_{\texttt{j}}^{(I_1\cdots I_\texttt{2j})}\Big({\widetilde  \phi}_{\texttt{j}}\Big)_{(I_1\cdots I_\texttt{2j})}
+(\alpha_2)^2\psi_{\texttt{j}+\frac{1}{2}}^{(I_1\cdots I_\texttt{2j+1})}
\Big(\psi_{\texttt{j}+\frac{1}{2}}\Big)_{(I_1\cdots I_\texttt{2j+1})}
+(\alpha_3)^2
\frac{2\texttt{j}+1} {2\texttt{j}}
\psi_{\texttt{j}-\frac{1}{2}}^{(I_1\cdots I_\texttt{2j-1})}
\Big(\psi_{\texttt{j}-\frac{1}{2}}\Big)_{(I_1\cdots I_\texttt{2j-1})}	
\end{split}
\label{nonesuperfield2a}		
\end{equation}	
Comparing this expression with the second line of \eqref{nonesuperfield2} we obtain
\begin{equation}
	\alpha_1\alpha_4=1\qquad,\qquad 
	(\alpha_2)^2=1\qquad,\qquad 
	(\alpha_3)^2
\frac{2\texttt{j}+1}{2\texttt{j}} =1
\label{nonesuperfield2b}		
\end{equation}
This fixes $\alpha_2$ and $\alpha_3$. This doesn't uniquely fix $\alpha_1$ and $\alpha_4$. We will see later that the parity transformation becomes simple if we choose $\alpha_1=\alpha_4=1$. Note that the supersymmetry invariant inner product has no diagonal term involving two $\phi_{\texttt{j}}$s or two ${\widetilde  \phi}_{\texttt{j}}$s. As a consequence, the propagator is also off-diagonal in this basis. This is different from \cite{Herderschee:2019ofc, Liu:2020fgu} and this technique was used in \cite{Engelbrecht:2022aao} \footnote{Even though we agree on the procedure, we disagree on some numbers. For example, for $\mathcal{N}=1, \texttt{j}=\frac{3}{2}$ the expression is available in eqn 2.12 of their paper. We disagree on the coefficient of the second term.}. The key advantage of this criteria is that we get a canonically normalized propagator in this way. So, the $\mathcal{N}=1$ superfield is 
\begin{equation}
		\Phi_{\texttt{j}}^{(I_1\cdots I_\texttt{2j})}= 
		\phi_{\texttt{j}}^{(I_1\cdots I_\texttt{2j})}+
		\eta_J\psi_{\texttt{j}+\frac{1}{2}}^{(JI_1\cdots I_\texttt{2j})}+
		\sqrt{\frac{2\texttt{j}}{2\texttt{j}+1}} \eta^{(I_1} \psi_{\texttt{j}-\frac{1}{2}}^{I_2\cdots I_\texttt{2j})} 
		+\frac{1}{2}\eta_J\eta^J{\widetilde  \phi}_{\texttt{j}}^{(I_1\cdots I_\texttt{2j})}
\label{nonesuperfield3}		
\end{equation}
In our convention
\begin{equation}
	\eta^{(I_1} \psi_{\texttt{j}-\frac{1}{2}}^{I_2\cdots I_\texttt{2j})} =\frac{1}{2\texttt{j}}\Bigg(\sum_{k=1}^{2\texttt{j}} \eta^{I_k} \psi_{\texttt{j}-\frac{1}{2}}^{(I_1\cdots I_{k-1}I_{k+1}\cdots I_\texttt{2j})}\Bigg)
\label{nonesuperfield4}
\end{equation}

The number in front of the third term is simply a Clebsh-Gordon coefficient $ \langle j, \frac{1}{2};j,-\frac{1}{2}|j-\frac{1}{2},j-\frac{1}{2}\rangle $\footnote{We are using the notation $ \langle j_1, j_2;m_1,m_2|J,M\rangle $}. And it is not surprising because the second and third term of \eqref{nonesuperfield3} is simply
\begin{equation}
	\frac{1}{2}\otimes \texttt{j}= \left(\texttt{j}+\frac{1}{2}\right)\oplus  \left(\texttt{j}-\frac{1}{2}\right)
\label{nonesuperfield6}
\end{equation}
And the criteria in \eqref{nonesuperfield2} is simply demanding the orthonormality of the states in terms of the component amplitudes. From this expansion, we can extract various components in the following way
\begin{equation}
\begin{split}
\phi_{\texttt{j}}^{(I_1\cdots I_\texttt{2j})}&=\Phi^{(I_1\cdots I_\texttt{2j})}\Big|_{\eta_J=0}
\\
\psi_{\texttt{j}-\frac{1}{2}}^{(I_1\cdots I_\texttt{2j-1})}&=
\sqrt{\frac{(2\texttt{j})}{(2\texttt{j}+1)}}\partial_{I_{\texttt{2j}}}\Phi^{(I_1\cdots I_\texttt{2j})}\Big|_{\eta_J=0}
\\
\psi_{\texttt{j}+\frac{1}{2}}^{(I_1\cdots I_\texttt{2j+1})}&=
\partial^{(\texttt{2j+1}}\Phi^{I_1\cdots I_\texttt{2j})}\Big|_{\eta_J=0}
\\
\widetilde{\phi}_{\texttt{j}}^{(I_1\cdots I_\texttt{2j})}&=
\frac{1}{2}\partial^J\partial_J\Phi^{(I_1\cdots I_\texttt{2j})}\Big|_{\eta_J=0}
\end{split}	
\label{nonesuperfield8}
\end{equation} 
where $\partial_I$ and $\partial^I$ are defined as
\begin{equation}
	\partial_I =\frac{\partial}{\partial \eta^I}
\qquad,\qquad	
\partial^I =\frac{\partial}{\partial \eta_I}
\label{nonesuperfield21}
\end{equation}

\paragraph{Parity}
There is no parity self-conjugate massless superfield in $\mathcal{N}=1$ theories. So we discuss the parity transformation of the massive superfield. From \eqref{onshellsusyreview23} we know that a massive superfield transforms in the following way 
\begin{equation}
	\Phi^{(I_1\cdots I_\texttt{2j})}(\eta) \longrightarrow \zeta\,  \widetilde{\Phi}^{(I_1\cdots I_\texttt{2j})}(\iimg\eta)
\end{equation}
$\zeta$ is $\pm 1$. $\widetilde{\Phi}^{(I_1\cdots I_\texttt{2j})}(\eta)$ is defined as
\begin{equation}
\begin{split}
\widetilde{\Phi}^{(I_1\cdots I_\texttt{2j})}(\iimg \eta)&\equiv	\int d^2\eta^\dagger e^{\eta\eta^\dagger}\Phi^{(I_1\cdots I_\texttt{2j})}(\eta^\dagger)
\\
&= 	
-\Bigg[{\widetilde  \phi}_{\texttt{j}}^{(I_1\cdots I_\texttt{2j})}+
		\iimg\eta_J\psi_{\texttt{j}+\frac{1}{2}}^{(JI_1\cdots I_\texttt{2j})}+
		\iimg\sqrt{\frac{2\texttt{j}}{2\texttt{j}+1}} \eta^{(I_1} \psi_{\texttt{j}-\frac{1}{2}}^{I_2\cdots I_\texttt{2j})} 
		+\frac{1}{2}\eta_J\eta^J\phi_{\texttt{j}}^{(I_1\cdots I_\texttt{2j})}\Bigg]
\end{split}
\end{equation}
then the parity transformation of the component fields are 
\begin{equation}
	P\qquad:\qquad 
	(\phi_{\texttt{j}},\psi_{\texttt{j}+\frac{1}{2}} , \psi_{\texttt{j}-\frac{1}{2}},{\widetilde  \phi}_{\texttt{j}})\longrightarrow 
	\zeta (-{\widetilde  \phi}_{\texttt{j}} ,\iimg \psi_{\texttt{j}+\frac{1}{2}} ,\iimg  \psi_{\texttt{j}-\frac{1}{2}}, -\phi_{\texttt{j}})
\end{equation}

\subsection{Three point function}
Let's consider the three-point function of two massless superfields and one massive superfield. This is non-zero only if $h_1+h_2+\texttt{j}\in \mathbb{Z}$. The three point super-amplitude is given by 
\begin{equation}
	\mathcal{A}(\Sigma_1^{(h_1)}\Sigma_2^{(h_2)}\Phi_\texttt{j})\Big|^{(I_1\cdots I_{2\texttt{j}})}
	=
\frac{\mathfrak{g}_{h_1, h_2,\texttt{j}}}{m^{3\texttt{j} +h_1+h_2 }}\delta^{(2)}(\mathcal{Q}_{(3)}^\dagger)
	\Big([12]^{\texttt{j}+ h_1+h_2} 
	\langle 13^{(I_1} \rangle \cdots \langle 13^{I_{\texttt{j}+h_2 -h_1}} \rangle \langle 23^{\texttt{j}-\frac{1}{2}} \rangle\cdots \langle 23^{2\texttt{j})} \rangle 
\Big)	
\label{nonethreepoint1}
\end{equation}
where the delta function is given by 
\begin{equation}
	\delta^{(2)}(\mathcal{Q}_{(3)}^\dagger)=   \langle 12\rangle \eta_1\eta_2 + \langle 13^J\rangle \eta_1\eta_{3,J} + \langle 23^J\rangle \eta_2\eta_{3,J} + \frac{m_3}{2} \eta_{3,J}\eta_{3}^{J}
\label{nonethreepoint2}
\end{equation}
The role of the delta function is to ensure that the amplitude is supersymmetry invariant. 
$\mathfrak{g}_{h_1, h_2,\texttt{j}}$ is the dimensionless coupling constant appearing in the three-point super-amplitude. We will see below those coupling constants of the three-point amplitudes of component fields are determined in terms of $\mathfrak{g}_{h_1, h_2,\texttt{j}}$. From this, we can extract the component amplitudes. 
\begin{equation}
	A\Big(\varphi_1^{(h_1)}\varphi_2^{(h_2-\frac{1}{2})}\psi_{\texttt{j}-\frac{1}{2}}\Big)\Big|^{(I_1\cdots I_{2\texttt{j}-1})}
= \sqrt{\frac{(2\texttt{j})}{(2\texttt{j}+1)}}
\partial_{(2)} \left( \partial_{(3)}\right)_{I_{2\texttt{j}}} 
\Big[
	\mathcal{A}(\Sigma_1^{(h_1)}\Sigma_2^{(h_2)}\Phi_\texttt{j})\Big]^{(I_1\cdots I_{2\texttt{j}})}
\label{nonethreepoint3}
\end{equation} 
From \eqref{nonethreepoint2}, we can see that action of the derivative operators is the same as multiplying by $\langle 23_{I_{2\texttt{j}}}\rangle $. Now we know that 
\begin{equation}
	\langle 23_I\rangle \langle 13^I\rangle =m\langle 21\rangle =\frac{m^3}{[12]}
\qquad,\qquad
	\langle 23_I\rangle \langle 23^I\rangle =0 	
\label{nonethreepoint4}
\end{equation}
So the three-point function turns out to be 
\begin{equation}
	\frac{(\texttt{j}+h_2-h_1)}{\sqrt{2\texttt{j}(2\texttt{j}+1)}}
	\frac{\mathfrak{g}_{h_1, h_2,\texttt{j}}}{m^{3\texttt{j} +h_1+h_2-3}}
	\Big([12]^{\texttt{j}+h_1+h_2-1} 
	\langle 13^{(I_1} \rangle \cdots \langle 13^{I_{\texttt{j}+h_2 -h_1-1}} \rangle \langle 23^{\texttt{j}+h_2-h_1} \rangle\cdots \langle 23^{I_{2\texttt{j-1})}} \rangle 
\Big)
\label{nonethreepoint5}
\end{equation}
This gives 
\begin{equation}
	g_{h_1,h_2-\frac{1}{2},\texttt{j}-\frac{1}{2}}=\mathfrak{g}_{h_1, h_2,\texttt{j}}\frac{ (\texttt{j}+h_2-h_1)}{\sqrt{2\texttt{j}(2\texttt{j}+1)}}
\label{nonethreepoint6}
\end{equation}
Now we want to compute 
\begin{equation}
	A\Big(\varphi_1^{(h_1)}\varphi_2^{(h_2-\frac{1}{2})}\psi_{\texttt{j}+\frac{1}{2}}\Big)\Big|^{(I_1\cdots I_{2\texttt{j+1}})}
=
\partial_{(2)}\partial_{(3)}^{(I_{2\texttt{j+1}}} \Big[
	\mathcal{A}(\Sigma_1^{(h_1)}\Sigma_2^{(h_2)}\Phi_\texttt{j})\Big]^{I_1\cdots I_{2\texttt{j}})}
\label{nonethreepoint7}
\end{equation} 
It is straightforward to find the action of the derivatives, and from that, we get
\begin{equation}
	\frac{\mathfrak{g}_{h_1, h_2,\texttt{j}}}{m^{3\texttt{j} +h_1+h_2}}
	\Big([12]^{\texttt{j}+h_1+h_2} 
	\langle 13^{(I_1} \rangle \cdots \langle 13^{I_{\texttt{j}+h_2 -h_1}} \rangle \langle 23^{\texttt{j}+h_2 -h_1+1} \rangle\cdots \langle 23^{2\texttt{j}+1)} \rangle 
\Big)
\label{nonethreepoint8}
\end{equation}
From this, we obtain that the coupling constant is given by 
\begin{equation}
	g_{h_1,(h_2-\frac{1}{2}),(\texttt{j}+\frac{1}{2})}=\mathfrak{g}_{h_1, h_2,\texttt{j}} 
\label{nonethreepoint9}
\end{equation}
Using the results given in \eqref{nonethreepoint6} and in \eqref{nonethreepoint9} we can compute the following ratio
\begin{equation}
	\frac{g_{h_1,(h_2-\frac{1}{2}),(\texttt{j}-\frac{1}{2})}g_{h_3,(h_4-\frac{1}{2}),(\texttt{j}-\frac{1}{2})}}{g_{h_1,(h_2-\frac{1}{2}),(\texttt{j}+\frac{1}{2})}g_{h_3,(h_4-\frac{1}{2}),(\texttt{j}+\frac{1}{2})}}
= \frac{(\texttt{j}+h_2-h_1)(\texttt{j}+h_4-h_3)}{2\texttt{j}(2\texttt{j}+1)}	
\label{nonethreepoint21}
\end{equation}
This ratio can also be determined from the expansion of the four-point superamplitude \cite{Liu:2020fgu}. 

In the same way, we can determine the other two coupling constants. We are not presenting the details here. They are given by  
\begin{equation}
	g_{h_1,h_2, \texttt{j}}=\mathfrak{g}_{h_1, h_2,\texttt{j}}
\qquad,\qquad	
g_{h_1-\frac{1}{2},h_2-\frac{1}{2},{\texttt{j}}}=-\mathfrak{g}_{h_1, h_2,\texttt{j}}
\label{nonethreepoint31}
\end{equation}
The spin $\texttt{j}$ particle appearing with the first coupling constant is ${\widetilde  \phi}_{\texttt{j}}$ and the particle appearing with the second constant is ${  \phi}_{\texttt{j}}$.

\subsection{Three point function of two photons}

Let's consider the amplitude between two-photon multiplets and one massive spin $\texttt{j}$ multiplet. There are two possible three-point functions 
\begin{equation}
A(\Sigma^{(+1)},\Sigma^{(+1)},\Phi_{\texttt{j}}) 
\quad,\quad	
A(\Sigma^{(+1)},\Sigma^{(-\frac{1}{2})},\Phi_{\texttt{j}}) 
\label{nonephotonthreepoint1}
\end{equation}
We can use parity to determine $A(\Sigma^{(-\frac{1}{2})},\Sigma^{(-\frac{1}{2})},\Phi_{\texttt{j}}) $ and $A(\Sigma^{(-\frac{1}{2})},\Sigma^{(+1)},\Phi_{\texttt{j}}) $ from the above two using Parity.  

For $\mathcal{N}=1$ Supersymmetry, there are two separate cases based on whether the superfield has an integer spin or half-integer spin. We first consider the case when the superfield has a half-integer spin. In the absence of Supersymmetry, there are twelve independent coupling constants
\begin{equation}
	g^\texttt{(0,1)}_{\texttt{pp(j}-\frac{1}{2})}
\quad,\quad	
g^\texttt{(0,1)}_{\texttt{pp(j}+\frac{1}{2})}
\quad,\quad	
	g^\texttt{(0,1)}_{\texttt{ff(j}-\frac{1}{2})}
\quad,\quad	
g^\texttt{(0,1)}_{\texttt{ff(j}+\frac{1}{2})}	
\quad,\quad	
	g^\texttt{(0,1)}_{\texttt{fpj}}
	\quad,\quad	
	g^\texttt{(0,1)}_{\texttt{fp}{\texttt{j}}}
\label{nonephotonthreepoint2}
\end{equation} 
The super-scripts are there to denote the minimal/non-minimalness of the amplitude; we use $\texttt{(0)}$ for minimal amplitudes and $\texttt{(1)}$ for non-minimal amplitudes.  We have not written down  $g^\texttt{(0,1)}_{\texttt{ppj}}$ and $g^\texttt{(0,1)}_{\texttt{ffj}}$, since they are zero
from conservation of angular momentum  ($\texttt{j}\in \mathbb{Z}+\frac{1}{2}$.)

From the consideration of spin-statistics, we get the following constraints 
\begin{equation}
	\texttt{j}\in \mathbb{Z}+\frac{1}{2} \implies A(\Sigma^{(+1)},\Sigma^{(+1)},\Phi_{\texttt{j}}) = A(\Sigma^{(-\frac{1}{2})},\Sigma^{(-\frac{1}{2})},\Phi_{\texttt{j}})= 0
\label{nonephotonthreepoint3}
\end{equation} 
and this immediately gives that 
\begin{equation}
	g^\texttt{(1)}_{\texttt{pp(j}-\frac{1}{2})}=0=g^\texttt{(1)}_{\texttt{pp(j}+\frac{1}{2})}
\qquad,\qquad	
g_{\texttt{fp} \texttt{j}  }^\texttt{(1)}=0 
\qquad,\qquad	
	g^\texttt{(1)}_{\texttt{ff(j}-\frac{1}{2})}=0=g^\texttt{(1)}_{\texttt{ff(j}+\frac{1}{2})}
\label{nonephotonthreepoint4}
\end{equation}
$\mathcal{N}=1$ Supersymmetry sets six coupling constants to zero. 
In particular, if the superfield has a half-integer spin then, then the photon has only minimal coupling. Let's now consider the case when $\texttt{j}$ is half-integer. In this case, the non-zero super-amplitude is given by 
\begin{equation}
	A(\Sigma^{(+1)},\Sigma^{(-\frac{1}{2})},\Phi_{\texttt{j}}) =\frac{\mathfrak{g}_{\texttt{ppj}}^{(0)}}{m^{3\texttt{j}+1/2}}\delta^{(2)}(\mathcal{Q}^\dagger)\Big([12]^{\texttt{j}+2}
	\langle 13^{(I_1} \rangle \cdots \langle 13^{I_{\texttt{j}}} \rangle \langle 23^{\texttt{j}} \rangle\cdots \langle 23^{2\texttt{j})} \rangle 
\Big)
\label{nonephotonthreepoint5}
\end{equation}
This is simply a special case of \eqref{nonethreepoint1} with $h_1=1$, $h_2=-\frac{1}{2}$. Then we get the following relation 
\begin{equation}
\begin{split}
&g^\texttt{(0)}_{\texttt{pp(j}-\frac{1}{2})}=\frac{\mathfrak{g}_{\texttt{ppj}}^{(0)} (\texttt{j}-\frac{3}{2})}{\sqrt{\texttt{j}(2\texttt{j}+1)}}
\qquad,\qquad
g^\texttt{(0)}_{\texttt{pp(j}+\frac{1}{2})}= \mathfrak{g}_{\texttt{ppj}}^{(0)}
\end{split}
\label{nonephotonthreepoint6}
\end{equation}
Now we write down the answer in the Lorentz covariant basis. We found that if $ \texttt{j} \in 2\mathbb{Z} $, only the photons can only couple to the supermultiplet non-minimally, through the bottom and top components. So the three point amplitude is given by   
\begin{equation}
\widehat{\mathcal{A}}_{j} = \Big(\widehat{\mathcal{A}}^{(1+)}_{\texttt{j}}(+,+,\phi_{\texttt{j}}) + \widehat{\mathcal{A}}^{(1-)}_{\texttt{j}}(-,-,\widetilde{\phi}_{\texttt{j}})\Big)\enspace,\enspace\text{$ \texttt{j} \in 2\mathbb{Z} $}
\end{equation}
$\widehat{\mathcal{A}}^{1+}$ and $\widehat{\mathcal{A}}^{1+}$ are defined in \eqref{photonreview73}. 
Now we move to the case where the superfield has integer spin. In the absence of Supersymmetry, there are twelve coupling constants in this case. 
\begin{equation}
 \texttt{j}\in \mathbb{Z}\implies
A(\Sigma^{(+1)},\Sigma^{(-\frac{1}{2})},\Phi_{\texttt{j}}) = A(\Sigma^{( -\frac{1}{2})},\Sigma^{(+1)},\Phi_{\texttt{j}})= 0
\label{nonephotonthreepoint7}
\end{equation}
In this case, the minimal couplings are absent. 
\begin{equation}
	 g_{\texttt{ppj}}^\texttt{(0)}=0 
\label{nonephotonthreepoint8}
\end{equation}
In this case, the nonzero amplitude is 
\begin{equation}
	A(\Sigma^{(+1)},\Sigma^{(+1)},\Phi_{\texttt{j}}) 
	=\frac{\mathfrak{g}_{\texttt{ppj}}^{(1)}}{m^{3\texttt{j}+2}}\delta^{(2)}(\mathcal{Q}^\dagger)\Big([12]^{\texttt{j}+2}
	\langle 13^{(I_1} \rangle \cdots \langle 13^{I_{\texttt{j}}} \rangle \langle 23^{\texttt{j}} \rangle\cdots \langle 23^{2\texttt{j})} \rangle 
\Big)	
\label{nonephotonthreepoint11}
\end{equation}
This superamplitude, after using \eqref{photonreview71}, gives the following relation between the coupling constant of the component amplitude
\begin{equation}
\begin{split}
	&g_{\texttt{ffj}}^{(1)}= \mathfrak{g}_{\texttt{ppj}}^{(1)}  
	\qquad,\qquad
	g_{\texttt{pp}{\texttt{j}}}^{(1)}=\mathfrak{g}_{\texttt{ppj}}^{(1)}	 
\end{split}
\label{nonephotonthreepoint12}
\end{equation}
In the Lorentz covariant basis, the three point function of two photons and one higher spin particle in a non-supersymmetric theory is given in \eqref{photonreview73}. The result in $\mathcal{N}=1$ theories are given by 
\begin{equation}
\widehat{\mathcal{A}}_{\texttt{j}}{}^{(0)}\Big|_{\mathcal{N}=1} =
\bigg(\widehat{\mathcal{A}}_{\texttt{j}+\frac{1}{2}}{}^{(0)} + \sqrt{2}\sqrt{\frac{2 \left(\texttt{j} \left(\texttt{j}-\frac{3}{2}\right) \left(\texttt{j}-\frac{3}{2}\right) \right)}{\left(4 \texttt{j}^2\right) (2 \texttt{j}+1)}}\widehat{\mathcal{A}}_{\texttt{j}-\frac{1}{2}}{}^{(0)}\bigg)\enspace,\enspace \texttt{j}+\frac{1}{2}\in \mathbb{Z}\\
\label{nonethreepoint41}
\end{equation}
We omitted the superscript $\texttt{ppj}$ for sake of brevity.   
\subsection{Four photon amplitude}

Let's consider the tree level four gluon amplitude in $\mathcal{N}=1$ supersymmetric theories. We note down the key feature here. 
\begin{itemize}
	\item From supercharge conservation, it follows that any higher spin either has nonzero minimal coupling or has non-minimal but never both. Amplitudes with three plus and one minus (three minus with one plus) are non-zero only when a massive field couples both minimally and non-minimally to two photons. This means in a supersymmetric theory that amplitudes with three plus and one minus (three minus with one plus) are also zero
\begin{equation}
\mathcal{N}=1\implies	\mathcal{M}(1^+2^+3^+4^-)=0=\mathcal{M}(1^-2^-3^-4^+)
\label{nonephotonfourpoint1}
\end{equation}
In a non-super-symmetric theory, this amplitude is nonzero only when the higher spin particle has both minimal and non-minimal couplings to two-photon. 

We can use table \ref{tab:joycebasisangdist} to conclude that in the basis given in \eqref{photonreview107}, $\mathcal{F}_7$ is zero.

	\item Let's now consider all plus amplitude $\mathcal{M}(1^+2^+3^+4^+)$ (all minus can be obtained by parity transformation. This gets a contribution only when there is nonzero non-minimal coupling. Hence it is zero when the superfield has a half-integer spin. It turns out that it is also zero when the superfield has an integer spin. We note that in this case, $\phi_{\texttt{j}}$ couples only when both the photons have positive helicity (see \eqref{nonethreepoint31} and the discussion below it) and $\widetilde{\phi}_{\texttt{j}}$ couples only when both the photons have negative helicity. From super-charge conservation it follows that $\phi_{\texttt{j}}-\phi_{\texttt{j}}$ propagator and $\widetilde{\phi}_{\texttt{j}}-\widetilde{\phi}_{\texttt{j}}$ propagator is zero. Only non-zero propagator is $\phi_{\texttt{j}}-\widetilde{\phi}_{\texttt{j}}$. From this consideration, it follows that 
\begin{equation}
\mathcal{N}=1\implies	\mathcal{M}(1^+2^+3^+4^+)=0=\mathcal{M}(1^-2^-3^-4^-)
\label{nonephotonfourpoint2}
\end{equation}
From table \ref{tab:joycebasisangdist} we can conclude that $\mathcal{F}_4=0$ is zero.

	\item Then, only nonzero amplitudes are when two photons have positive helicity, and two have negative helicity. To be precise, consider $1^+2^+3^-4^-$; The other cases with two helicities can be obtained from $2\leftrightarrow 3$, $2\leftrightarrow 4$ exchange. . The tensor factor in this case is 
\begin{equation}
\mathcal{M}(1^+2^+3^-4^-)=	[12]^2\langle 34\rangle^2 \mathcal{F}_1(s,t)
\label{nonephotonfourpoint3}
\end{equation}
From table \eqref{tab:joycebasisangdist}, we know that only $\mathcal{T}_1$ is non-zero for this configuration. The subscript $2$ denotes this fact. It can have $s$, $t$ channel pole due to planarity. In a non-supersymmetric theory, the $s$ channel pole is present due to non-minimal coupling to the exchanged particle, and the $t$ channel pole is present due to minimal coupling to the exchanged particles. In a supersymmetric theory, the situation depends on the spin of the supermultiplet in the following way:

\begin{enumerate}
	\item 	For supermultiplets with integer spin, only non-minimal couplings are nonzero; hence $t$ channel pole is absent 
\begin{equation}
\lim_{t\to m^2}	(t-m^2)\mathcal{F}_1\left(s,t,u|\mathcal{N}=1,\texttt{j}\in \mathbb{Z}\right)
=0
\label{nonephotonfourpoint4}
\end{equation}
Then the only nonzero contribution follows from the $s$ channel exchange
\begin{equation}
	\mathcal{F}_1\left(s,t,u|\mathcal{N}=1,\texttt{j}\in \mathbb{Z}\right)
	=\frac{(g_{\texttt{pp(j)}}^{(1)})^2}{m^{2\texttt{j}+2}}\frac{s^{\texttt{j}}\widetilde{\mathcal{N}}_{\texttt{j};0,0}\,  \jacobi^{(0,0)}_{\texttt{j}} \left(\frac{t-u}{s}\right) }{s-m^2}
\label{nonephotonfourpoint5}
\end{equation}

	\item If the supermultiplet has a half-integer spin, only the minimal couplings are nonzero, and hence $s$ channel pole is absent
\begin{equation}
\lim_{s\to m^2}	(s-m^2)	\mathcal{F}_1\left(s,t,u|\mathcal{N}=1,\texttt{j}\in \mathbb{Z}+\frac{1}{2}\right)=0
\label{nonephotonfourpoint11}
\end{equation}
Let the spin of the super-multiplet be $\texttt{j}+\frac{1}{2}$ where $\texttt{j}$ is a non-negative integer. There are $t$ channel poles due to exchange of particle with spin $\texttt{j}+1$ and $\texttt{j}$. 
\begin{equation}
	\frac{(g_{\texttt{pp(j+1)}}^{(0)})^2}{m^{2\texttt{j}-2}}
	\, t^{\texttt{j}-2} \Bigg[\frac{t}{m^2}  \widetilde{\mathcal{N}}_{\texttt{j+1};2,2}\jacobi_{\texttt{j}-1}^{(0,4)}\left(\frac{s-u}{t}\right)
+\frac{(g_{\texttt{pp(j)}}^{(0)})^2}{(g_{\texttt{pp(j+1)}}^{(0)})^2}
\widetilde{\mathcal{N}}_{\texttt{j};2,2}\jacobi_{\texttt{j}-2}^{(0,4)}\left(\frac{s-u}{t}\right)
\Bigg]	
\label{nonephotonfourpoint12}
\end{equation}
We know the ratio of the three-point function is 
\begin{equation}
	\frac{(g_{\texttt{pp(j)}}^{(0)})^2}{(g_{\texttt{pp(j+1)}}^{(0)})^2}
	= \frac{(\texttt{j}-1)^2}{(2\texttt{j}+1)(\texttt{j}+1)}
\label{nonephotonfourpoint13}
\end{equation}
We can use \eqref{basicidentity} to simplify the answer 
\begin{equation}
\mathcal{F}_1\left(s,t,u|\mathcal{N}=1,\texttt{j}\in \mathbb{Z}+\frac{1}{2}\right)
	= 	\frac{(g_{\texttt{pp(j+1)}}^{(0)})^2}{m^{2\texttt{j}}}\frac{t^{\texttt{j}+1}\widetilde{\mathcal{N}}_{\texttt{j}+\frac{1}{2};\frac{3}{2},\frac{3}{2} }\jacobi_{\texttt{j}-1}^{(0,3)}\left(\frac{s-u}{t}\right)}{t-m^2}
\label{nonephotonfourpoint14}
\end{equation}
Please note that to simplify the answer, we have set any term proportional $t-m^2$ to be zero because any such term does not contribute to the residue at the pole, and hence they correspond to a contact term. In this paper, we are not focusing on the four-point contact terms; hence, we have ignored such contributions.

\end{enumerate}

\end{itemize}
In a theory with $\mathcal{N}=1$ supersymmetry, the form factors are three dimensional; only $\mathcal{F}_1$, $\mathcal{F}_2$ and $\mathcal{F}_3$ are non-zero. The final answer from the above analysis is that for a supermultiplet with integer spin 
\begin{equation}
	\mathcal{F}\left(s,t,u|\mathcal{N}=1,\texttt{j}\in \mathbb{Z}\right)
\equiv 
\begin{pmatrix}
	\mathcal{F}_1
\\
\\
\mathcal{F}_2
\\
\\
\mathcal{F}_3	
\end{pmatrix}
=
\frac{(g_{\texttt{pp(j)}}^{(1)})^2}{m^{2\texttt{j}+2}}\widetilde{\mathcal{N}}_{\texttt{j};0,0}\,
\begin{pmatrix}
\frac{s^{\texttt{j}}  \jacobi^{(0,0)}_{\texttt{j}} \left(\frac{t-u}{s}\right) }{s-m^2}
\\
\\
\frac{u^{\texttt{j}}   \jacobi^{(0,0)}_{\texttt{j}} \left(\frac{t-s}{t}\right) }{u-m^2}	
\\
\\
\frac{t^{\texttt{j}}  \jacobi^{(0,0)}_{\texttt{j}} \left(\frac{s-u}{t}\right) }{t-m^2}
\end{pmatrix}	
\label{nonephotonfourpoint15}
\end{equation}
For supermultiplet with half-integer spin 
\begin{equation}
	\mathcal{F}\left(s,t,u|\mathcal{N}=1,\texttt{j}\in \mathbb{Z}+\frac{1}{2}\right)
\equiv 
\begin{pmatrix}
	\mathcal{F}_1
\\
\\
\mathcal{F}_2
\\
\\
\mathcal{F}_3	
\end{pmatrix}
=
\frac{(g_{\texttt{pp(j+1)}}^{(0)})^2}
{m^{2\texttt{j}}}\widetilde{\mathcal{N}}_{\texttt{j}+\frac{1}{2};\frac{3}{2},\frac{3}{2} }
\begin{pmatrix}
\frac{t^{\texttt{j}-1}\jacobi_{\texttt{j}-1}^{(0,3)}\left(\frac{s-u}{t}\right)}{t-m^2}
+\frac{u^{\texttt{j}-1}\jacobi_{\texttt{j}-1}^{(0,3)}\left(\frac{s-t}{u}\right)}{u-m^2}
\\
\\
\frac{s^{\texttt{j}-1}\jacobi_{\texttt{j}-1}^{(0,3)}\left(\frac{u-t}{s}\right)}{s-m^2}+
\frac{t^{\texttt{j}-1}\jacobi_{\texttt{j}-1}^{(0,3)}\left(\frac{s-u}{t}\right)}{t-m^2}

\\
\\
\frac{u^{\texttt{j}-1}\jacobi_{\texttt{j}-1}^{(0,3)}\left(\frac{s-t}{u}\right)}{u-m^2}
+\frac{s^{\texttt{j}-1}\jacobi_{\texttt{j}-1}^{(0,3)}\left(\frac{u-t}{s}\right)}{s-m^2}
\end{pmatrix}	
\label{nonephotonfourpoint21}
\end{equation}
If it is gluon instead of photon and we consider colour ordering 1,2,3,4, then all the $u$ channel poles are not there. In that case, the four photon amplitude simplifies to 
\begin{equation}
\frac{(g_{\texttt{pp(j+1)}}^{(0)})^2}{m^{2\texttt{j}}}\widetilde{\mathcal{N}}_{\texttt{j}+\frac{1}{2};\frac{3}{2},\frac{3}{2} }
\begin{pmatrix}
\frac{t^{\texttt{j}-1}\jacobi_{\texttt{j}-1}^{(0,3)}\left(\frac{s-u}{t}\right)}{t-m^2}
\\
\\
\frac{s^{\texttt{j}-1}\jacobi_{\texttt{j}-1}^{(0,3)}\left(\frac{u-t}{s}\right)}{s-m^2}+
\frac{t^{\texttt{j}-1}\jacobi_{\texttt{j}-1}^{(0,3)}\left(\frac{s-u}{t}\right)}{t-m^2}

\\
\\
\frac{s^{\texttt{j}-1}\jacobi_{\texttt{j}-1}^{(0,3)}\left(\frac{u-t}{s}\right)}{s-m^2}
\end{pmatrix}	
\label{nonephotonfourpoint22}
\end{equation}

\section{$\mathcal{N}=2$ supersymmetry} 
\label{sec:bcrsntwo}

The massless $\mathcal{N}=2$ consists of two  $\mathcal{N}=1$ supermultiplets. In this case, there are two Grassmann valued superspace co-ordinate $\eta^A$ ($A=1,2$). The massless multiplet with helicity $h$ is given by 
\begin{equation}
		\Sigma^{(h)}= \varphi^{(h)}+\eta^A\,  \varphi_A^{(h-\frac{1}{2})}+\frac{1}{2}\epsilon_{AB}\eta^A\eta^B\varphi^{(h-1)}
\label{ntwosuperfield1}
\end{equation} 
The massless multiplet is CPT self-conjugate iff $h=\frac{1}{2}$ (also known as $\mathcal{N}=2$ hypermultiplet). Otherwise, we have to add the CPT conjugate multiplet. There only three massless multiplet: hypermultiplet ($h=1/2$), vector multiplet ($h=1$ and it's CPT conjugate $h=0$) and gravity multiplet ($h=2$ and it's CPT conjugate $h=-1$). The graviton multiplet $(h=2,-1)$ is given by 
\begin{equation}
\Sigma^{(2)}= g^{(2)}+\eta^A\,  r_A^{(\frac{3}{2})}+\frac{1}{2}\epsilon_{AB}\eta^A\eta^B \zeta^{(1)}
\quad,\quad
\Sigma^{(-1)}= \zeta^{(-1)}+\eta^A\,  r_A^{(-\frac{3}{2})}+\frac{1}{2}\epsilon_{AB}\eta^A\eta^B g^{(-2)}	
\end{equation}
$\zeta$ is called the graviphoton. The photon multiplet is given by 
\begin{equation}
\Sigma^{(1)}= p^{(1)}+\eta^A\,  f_A^{(\frac{1}{2})}+\frac{1}{2}\epsilon_{AB}\eta^A\eta^B \phi^{(0)}
\quad,\quad
\Sigma^{(0)}= \phi^{(0)}+\eta^A\,  f_A^{(-\frac{1}{2})}+\frac{1}{2}\epsilon_{AB}\eta^A\eta^B p^{(-1)}	
\end{equation}
The hypermultiplet has the following expansion
\begin{equation}
	\Sigma^{(\frac{1}{2})}= \psi^{(\frac{1}{2})}+\eta^A\,  \phi_A^{(0)}+\frac{1}{2}\epsilon_{AB}\eta^A\eta^B \psi^{(-\frac{1}{2})}
\end{equation}
The massive $\mathcal{N}=2$ multiplets have $16(\texttt{2j}+1)$ states; they are self-conjugate under CPT symmetry. The components of the spin $\texttt{j}$ superfields are given by 
\begin{equation}
\begin{split}
\Phi_{\texttt{j}}^{(I_1\cdots I_\texttt{2j})}=& 		\phi_{\texttt{j}}^{(I_1\cdots I_\texttt{2j})}
+
\eta^A_J\psi_{A}^{J(I_1\cdots I_\texttt{2j})}
		+\eta^{A}_{J_1}\eta_{J_2}^{B}\xi_{AB}^{J_1J_2(I_1\cdots I_\texttt{2j})}
%+
%\\
%&
+ \eta^{A}_{J_1}\eta_{J_2}^{B}\eta^{C}_{J_3} \widetilde{\psi }_{ABC}^{J_1J_2J_3(I_1\cdots I_\texttt{2j})} 
%\\
%& 
+	\eta^4 {\widetilde  \phi}_{\texttt{j}}^{(I_1\cdots I_\texttt{2j})}
\end{split}	
\label{ntwosuperfield2}
\end{equation}
For the purpose of clarity, we present the superfield expansions using Young Tableaux of $SU(2)_R$ and $SU(2)_{\textrm{little group}}$. The complete decomposition in terms of $R$-symmetry can little group indices can be found in Table \ref{tab:ntwofieldcontents}. We started with the Clifford Vacuum  given below 
\begin{center}
		\begin{ytableau}
			1 & 2 & 3 & 4 & \dots  & 2\texttt{j} 
		\end{ytableau} \\	
\end{center}
The supercharges carry both little group indices and the $R$ symmetry indices. The green boxes in the table coming from ``additiona of angular momentum" due to the action of the supercharges (alternatively from the Grassmann expansion). The Yellow boxes tell us the representation under $R$ symmetry group, which is $SU(2)$ for $\mathcal{N}=2$. 

\begin{table}[h]
\begin{center}
\tiny 
	\begin{tabular}{|c|c|c|}
		\hline 
		Level & $R$-symmetry & Spin \\
		\hline
		&&  \\
		0 & $\odot$ & \ytableausetup
		{mathmode, boxsize=2em}
		\begin{ytableau}
			1 & 2 & 3 & 4 & \dots  & 2\texttt{j} 
		\end{ytableau} \\
		&& \\
		\hline
		&&\\
		\multirow{2}{*}{1} &  \multirow{2}{*}{ 
			\ydiagram[*(yellow)]{1}} & 
		\begin{ytableau}
			1 & 2 & 3 & 4 & \dots  & 2\texttt{j} & *(green)
		\end{ytableau} \\
		& & 
		\begin{ytableau}
			1 & 2 & 3 & 4 & \dots  & \scriptstyle 2\texttt{j}-1 
		\end{ytableau}  \\
		\hline
		&&\\
		\multirow{4}{*}{2} & \multirow{3}{*}{$\odot$} & 
		\begin{ytableau}
			1 & 2 & 3 & 4 & \dots  & 2\texttt{j} & *(green) & *(green) 
		\end{ytableau} \\
		& & 
		\begin{ytableau}
			1 & 2 & 3 & 4 & \dots  & \scriptstyle 2\texttt{j}-1 & *(green)  
		\end{ytableau} \\
		& & 
		\begin{ytableau}
			1 & 2 & 3 & 4 & \dots  & \scriptstyle 2\texttt{j}-2 
		\end{ytableau} \\
		\cline{2-3}
		& & \\
		&
		\ydiagram[*(yellow)]{2} & 
		\begin{ytableau}
			1 & 2 & 3 & 4 &\dots  & \scriptstyle 2\texttt{j}-1 & *(green)  
		\end{ytableau} \\
		\hline
		&&\\
		\multirow{2}{*}{3} &  \multirow{2}{*}{
			\ydiagram[*(yellow)]{1} } & 
		\begin{ytableau}
			1 & 2 & 3 & 4 & \dots  & 2\texttt{j} & *(green)
		\end{ytableau} \\
		& & 
		\begin{ytableau}
			1 & 2 & 3 & 4 & \dots  &  \scriptstyle 2\texttt{j}-1 
		\end{ytableau}  \\
		\hline
		&&\\
		4 & $\odot$ & 
		\begin{ytableau}
			1 & 2 & 3 & 4 & \dots  & 2\texttt{j} 
		\end{ytableau} \\
		&&\\
		\hline
	\end{tabular}
\caption{Field content of massive $\mathcal{N}=2$ multiplet }
\label{tab:ntwofieldcontents}	
\end{center}
	\end{table}

$	\eta^4$ is the short-form of 
\begin{equation}
	\eta^4=	\frac{\epsilon^{J_1J_2}}{2!} \frac{	\epsilon^{J_3J_4}}{2!}\frac{\epsilon_{AB}\epsilon_{CD}}{2!2!}\eta_{J_1}^{A}\eta_{J_2}^C\eta_{J_3}^{B}\eta_{J_4}^{D}
\label{ntwosuperfield3}
\end{equation}
The expression of massive $\mathcal{N}=2$ multiplet can also be found in \cite{Liu:2020fgu}. However, we used different normalization criteria. The decomposition of the third term in terms of irrep is given by 
\begin{equation}
\begin{split}
\eta^{A}_{J_1}\eta_{J_2}^{B}\xi_{AB}^{J_1J_2(I_1\cdots I_\texttt{2j})}=& 	
		\epsilon^{J_1J_2}\eta^{(A}_{J_1}\eta_{J_2}^{B)}\xi_{(AB),\texttt{j}}^{(I_1\cdots I_\texttt{2j})}+\frac{1}{2}\epsilon_{AB}\Bigg( \eta^{A}_{J_1}\eta_{J_2}^{B}\xi_{\texttt{j}+1}^{(J_1J_2I_1\cdots I_\texttt{2j})}
\\&		
		+\sqrt{\frac{\texttt{j}}{\texttt{j}+1}}
		\eta^{A}_{J}\eta^{B(I_1}\xi_{\texttt{j}}^{JI_2\cdots I_\texttt{2j})}
		+\sqrt{\frac{2\texttt{j}-1}{2\texttt{j}+1}}
		 \eta^{A(I_1}\eta^{|B|I_2}\xi_{\texttt{j}-1}^{I_3\cdots I_\texttt{2j})}
		\Bigg)
\end{split}	
\label{ntwosuperfield4}
\end{equation}
All the terms in the second line transform under the rank 2 anti-symmetric representation of the $R$-symmetry group. In the case of $SU(2)$, this implies that they transform as the trivial representation. Again we can see that various coefficients are simply Clebsh-Gordon coefficients.
\begin{equation}
	1\otimes  \texttt{j}=(\texttt{j}+1)\, \oplus \texttt{j} \oplus (\texttt{j}-1)
\label{ntwosuperfield5}
\end{equation}
Scattering amplitudes are invariant under the $R$ symmetry group. If we consider a tree-level three-point function with two external photons or gravitons (which are neutral under $R$ symmetry), then the third particle also has to be $R$ symmetry scalar. As a result, we restrict our discussions to the higher spin particles, which transform trivially under the $R$ symmetry group. The field content of the $R$ symmetry singlet sector in the $\mathcal{N}=2$ massive multiplet is 
\begin{equation}
\begin{split}
\texttt{j}+1\qquad&:\qquad \xi_{\texttt{j}+1}
\\	
\texttt{j}\qquad&:\qquad \phi_{\texttt{j}}\quad ,\quad \xi_{\texttt{j}}\quad ,\quad{\widetilde  \phi}_{\texttt{j}}
\\	
\texttt{j}-1\qquad&:\qquad \xi_{\texttt{j}+1}
\label{ntwosuperfield5.1}
\end{split}	
\end{equation}
Various components can be obtained by projections
\begin{equation}
\begin{split}
\phi_{\texttt{j}}^{(I_1\cdots I_\texttt{2j})}
&= 
\Phi_{\texttt{j}}^{(I_1\cdots I_\texttt{2j})}\Big|_{\eta \rightarrow 0 }
\\
\xi_{\texttt{j}+1}^{(I_1\cdots I_\texttt{2j+2})}&=	\epsilon^{AB}\partial_A^{(I_1}\partial_B^{I_2}\Phi_{\texttt{j}}^{I_3\cdots I_\texttt{2j+2})}\Big|_{\eta \rightarrow 0 }
\\
\xi_{\texttt{j}}^{(I_1\cdots I_\texttt{2j})}&=	-
\sqrt{\frac{2\texttt{j}}{\texttt{j}+1}}
\epsilon^{AB}\partial_{A,J}
\partial_{B}^{(I_1}\Phi_{\texttt{j}}^{(JI_2\cdots I_\texttt{2j})}\Big|_{\eta \rightarrow 0 }
\\
\xi_{\texttt{j}-1}^{(I_1\cdots I_\texttt{2j-2})}&=	
-\sqrt{\frac{2\texttt{j}-1}{2\texttt{j}+1}}
\epsilon^{AB}\partial_{A,J_1}\partial_{B, J_2}\Phi_{\texttt{j}}^{(J_1J_2I_1\cdots I_\texttt{2j-2})}\Big|_{\eta \rightarrow 0 }
\\
{\widetilde  \phi}_{\texttt{j}}^{(I_1\cdots I_\texttt{2j})}
&= 
 \epsilon_{J_1J_2}   \epsilon_{J_3J_4}  \epsilon^{AB}\epsilon^{CD} \frac{\partial^4}{\partial\eta_{J_1}^{A}\partial\eta_{J_2}^C\partial\eta_{J_3}^{B}\partial\eta_{J_4}^{D}}
\Phi_{\texttt{j}}^{(I_1\cdots I_\texttt{2j})}\Big|_{\eta \rightarrow 0 }
\end{split}	
\label{ntwosuperfield6}
\end{equation}

\paragraph{Parity}
Now we consider the action of parity on the supermultiplet. We start with the massless multiplet. Parity flips the helicity of the particle. Consider a supermultiplet and the action of parity on it 
\begin{equation}
	P\qquad:\qquad 
	\Big(\varphi^{h_1},\varphi^{h_1-\frac{1}{2}}, \varphi^{h_1-1} \Big)
\longrightarrow
	\Big(\varphi^{-h_1},\varphi^{-h_1+\frac{1}{2}}, \varphi^{-h_1+1} \Big)
\end{equation}
The multiplet that is invariant under parity iff $h=\frac{1}{2}$ (i.e  the hypermultiplet).

Any massive supermultiplet is parity self-conjugate. However, supersymmetry transformation fixes the parity transformation rules for various component superfields. For a spin $\texttt{j}$ supermultiplet, we get the following transformation for the component fields
\begin{equation}
\begin{split}
P\quad:\quad 
\left\{
\begin{matrix}
 \xi_{\texttt{j-1}} 
\rightarrow 
- \xi_{\texttt{j-1}} 
\\
\\
\Big(\psi_{A,\texttt{j}-\frac{1}{2}}\, ,\,  \widetilde{\psi}^A_{\texttt{j}-\frac{1}{2}}\Big)
\rightarrow 
\Big(-\iimg \widetilde{\psi}^A_{\texttt{j}-\frac{1}{2}}\, \iimg 
\psi_{A,\texttt{j}-\frac{1}{2}}  \Big)
\\
\\
\Big(\phi_{\texttt{j}}\, ,\,  \widetilde{\phi}_{\texttt{j}}\,,\,  \xi_{(AB),\texttt{j}}\Big)
\rightarrow 
\Big( \widetilde{\phi}_{\texttt{j}}\,,\, \phi_{\texttt{j}}\, ,\,   \xi_{(AB),\texttt{j}}\Big)	
\\
\\
\Big(\psi_{A,\texttt{j}+\frac{1}{2}}\, ,\,  \widetilde{\psi}^A_{\texttt{j}+\frac{1}{2}}\Big)
\rightarrow 
\Big(-\iimg \widetilde{\psi}^A_{\texttt{j}+\frac{1}{2}},\, \iimg 
\psi_{A,\texttt{j}+\frac{1}{2}} \Big)
\\
\\
 \xi_{\texttt{j+1}} 
\rightarrow 
- \xi_{\texttt{j+1}} 
\end{matrix}
\right.
\end{split}	
\end{equation}
Note that for all fields, parity squares to $1$.

\subsection{Three point functions}
Let's consider the three-point function of two massless superfields and one massive superfield $\mathcal{A}(\Sigma_1^{(h_1)}\Sigma_2^{(h_2)}\Phi_{\texttt{j}})\Big|^{(I_1\cdots I_{\texttt{2j}})}$ is given by 
\begin{equation}
\begin{split}
%&	
%\\	
%&	=
\frac{\mathfrak{g}_{h_1,h_2,\texttt{j}}}{m^{3\texttt{j}+h_1+h_2+1}}\delta^{(4)}(\mathcal{Q}_{(3)}^\dagger) 
	\Big([12]^{\texttt{j}+ h_1+h_2} 
	\langle 13^{(I_1} \rangle \cdots \langle 13^{I_{\texttt{j}+h_2 -h_1}} \rangle \langle 23^{\texttt{j}+h_2 -h_1+1} \rangle\cdots \langle 23^{I_{2\texttt{j}})} \rangle 
\Big)	
\end{split}
\label{ntwothreepoint1}
\end{equation} 
This is non-zero if $h_1+h_2+\texttt{j}\in \mathbb{Z}$. 
The delta function for the three-point function in the case of $\mathcal{N}=2$ theories is given by 
\begin{equation}
	\delta^{(4)}(\mathcal{Q}_{(3)}^\dagger)=  \prod_{A=1}^2\Bigg( \langle 12\rangle \eta_1^A\eta_2^A + \langle 13^J\rangle \eta_1^A\eta_{3,J}^A + \langle 23^J\rangle \eta_2^A\eta_{3,J}^A + \frac{m_3}{2} \eta_{3,J}^A\eta_{3}^{A,J}\Bigg)
\label{ntwothreepoint2}
\end{equation}
From this super-amplitude, we can extract various component amplitudes using the projectors defined in Eqn \eqref{ntwosuperfield6}.  For example, $A\Big(\varphi_1^{(h_1)}\varphi_2^{(h_2-1)}\xi_{\texttt{j}+1}\Big)\Big|^{(I_1\cdots I_{\texttt{2j+2}})}$ is given by 
\begin{equation}
\begin{split}
%	&
%\\
%&=  
	\frac{\mathfrak{g}_{h_1, h_2,\texttt{j}}}{m^{3\texttt{j} +h_1+h_2+1}}
	\Big([12]^{\texttt{j}+h_1+h_2} 
	\langle 13^{(I_1} \rangle \cdots \langle 13^{I_{\texttt{j}+h_2 -h_1}} \rangle \langle 23^{I_{\texttt{j}+h_2 -h_1+1}} \rangle\cdots \langle 23^{I_{2\texttt{j}+2})} \rangle 
\Big)
\end{split}
\label{ntwothreepoint3}
\end{equation}
From this, we obtain that the coupling constant is given by 
\begin{equation}
	g_{h_1,h_2-1,\texttt{j}+1}=\mathfrak{g}_{h_1, h_2,\texttt{j}} 
\label{ntwothreepoint4}
\end{equation}
Similarly we can compute $A\Big(\varphi_1^{(h_1)}\varphi_2^{(h_2-1)}\xi_{\texttt{j}}\Big)\Big|^{(I_1\cdots I_{\texttt{2j}})}$.  
\begin{equation}
	\frac{\mathfrak{g}_{h_1,h_2,\texttt{j}}}{m^{3\texttt{j}+h_1+h_2+1}}
	\Big([12]^{\texttt{j}+ h_1+h_2} 
	\langle 13^{(J} \rangle \langle 13^{I_1}\rangle \cdots \langle 13^{I_{\texttt{j}+h_2 -h_1}} \rangle \langle 23^{\texttt{j}+h_2 -h_1+1} \rangle\cdots \langle 23^{I_{2\texttt{j-1}}} \rangle 
\Big)\left( \sqrt{\frac{\texttt{j}}{\texttt{j}+1}}\langle 23^{I_{\texttt{2j})}}\rangle \langle 23_{J}\rangle 
\right)	
\label{ntwothreepoint4.1}
\end{equation}
The contraction of the the $J$ indices give a factor of $\frac{(\texttt{j}+h_2-h_1)m^3}{\texttt{2j}[12]}$. From this we get the coupling constant $g_{h_1,h_2-1,\texttt{j}}$ to be 
\begin{equation}
	g_{h_1,h_2-1,\texttt{j}}=\mathfrak{g}_{h_1, h_2,\texttt{j}}\, (\texttt{j}+h_2-h_1)\sqrt{\frac{2}{(2\texttt{j})(2\texttt{j}+2) }}
\label{ntwothreepoint10}
\end{equation}
and $A\Big(\varphi_1^{(h_1)}\varphi_2^{(h_2-1)}\xi_{\texttt{j}-1}\Big)\Big|^{(I_1\cdots I_{\texttt{2j-2}})}$ to obtain

\begin{equation}
	\frac{\mathfrak{g}_{h_1,h_2,\texttt{j}}}{m^{3\texttt{j}+h_1+h_2+1}}
	\Big([12]^{\texttt{j}+ h_1+h_2} 
	\langle 13^{(I_1} \rangle \cdots \langle 13^{I_{\texttt{j}+h_2 -h_1}} \rangle \langle 23^{\texttt{j}+h_2 -h_1+1} \rangle\cdots \langle 23^{I_{2\texttt{j}})} \rangle 
\Big)\left( \sqrt{\frac{2\texttt{j}-1}{2\texttt{j}+1}}
\langle 23_{I_{\texttt{2j-1}}}\rangle \langle 23_{\texttt{2j}}\rangle 
\right)	
\label{ntwothreepoint10.1}
\end{equation}
In this case, we have a fair of contracted indices, and those contractions give  a factor of 

$\frac{(\texttt{j}+h_2-h_1)(\texttt{j}+h_2-h_1-1) }{(\texttt{2j})^2[12]^2}$. And thus, we get the following relation for the coupling constant. 
\begin{equation}
	g_{h_1,h_2-1,\texttt{j}-1}=\mathfrak{g}_{h_1, h_2,\texttt{j}}\,  (\texttt{j}+h_2-h_1)(\texttt{j}+h_2-h_1-1)\sqrt{\frac{1}{(2\texttt{j}-1) (2\texttt{j})^2(2\texttt{j}+1) }}
\label{ntwothreepoint15}
\end{equation}
In the next section, we need a ratio of various coupling constants. From \eqref{ntwothreepoint4}, \eqref{ntwothreepoint10} and \eqref{ntwothreepoint15}, we get the following ratio of coupling constants  
\begin{equation}
\begin{split}
	\frac{g_{h_1,h_2-1,\texttt{j}-1}g_{h_3,h_4-1,\texttt{j}-1}}{g_{h_1, h_2-1, \texttt{j}+1}g_{h_3,h_4-1,\texttt{j}+1}}
&= \frac{(\texttt{j}+h_2-h_1)(\texttt{j}+h_4-h_3)(\texttt{j}+h_2-h_1-1)(\texttt{j}+h_4-h_3-1)}{(2\texttt{j}+1)(2\texttt{j})^2(2\texttt{j}-1)}	
\\	
	\frac{g_{h_1, h_2-1,\texttt{j}}g_{h_3, h_4-1,\texttt{j}}}{g_{h_1,h_2-1, \texttt{j}+1}g_{h_3, h_4-1,\texttt{j}+1}}
&=   \frac{2(\texttt{j}+h_2-h_1)(\texttt{j}+h_4-h_3)}{(2\texttt{j})(2\texttt{j}+2)}	
\end{split}
\label{ntwothreepoint20}
\end{equation} 
We have computed the coupling constants when all the external states are $R$ charge scalar. There is only two other three-point function of $R$-symmetry scalar states: $A\Big(\varphi_1^{(h_1)}\varphi_2^{(h_2)}\widetilde{\phi}_{\texttt{j}}\Big)\Big|^{(I_1\cdots I_{\texttt{2j}})}$ and $A\Big(\varphi_1^{(h_1-1)}\varphi_2^{(h_2-1)}\phi_{\texttt{j}}\Big)\Big|^{(I_1\cdots I_{\texttt{2j}})}$. From these amplitudes, we get 
\begin{equation}
	g_{h_1,h_2,\texttt{j}}=\mathfrak{g}_{h_1, h_2,\texttt{j}} 
\end{equation}
and this implies that \begin{equation}
	\frac{g_{h_1, h_2,\texttt{j}}g_{h_3,h_4,\texttt{j} }}{g_{h_1,h_2-1,\texttt{j}+1}g_{h_3,h_4-1, \texttt{j}+1}}=1
\end{equation}
\vspace{5pt}
 
\subsection{Four point function of the Hypermultiplet fermions}
In the case of $\mathcal{N}=2$ theories in $3+1 $ dimensions, the massless hypermultiplet is special because it is self-conjugate under parity. This provides us with a simple model to study four point function of self-conjugate massless multiplet in a supersymmetric theory. We devote this section to that. The three-point functions of two hyper-multiplet fields and one massive multiplet fields come from the following super-amplitude
\begin{equation}
	\mathcal{A}(\Sigma^{(\frac{1}{2})},\Sigma^{(\frac{1}{2})},\Phi_{\texttt{j}}) 
\qquad,\qquad 
\texttt{j}\in 2\mathbb{Z}+1
\end{equation}
We focus on the coupling constant of two massless fermions to massive spin $\texttt{j}$ particles. In \eqref{ntwosuperfield5.1}, we can see that there are five particles, and hence in a non-supersymmetric theory, there are ten allowed coupling constants. However, supersymmetry sets 5 of them to be zero and determines the other 5 in terms of a single coupling constant of the super amplitude $\mathfrak{g}_{\texttt{f,f,j}}$.: the particles that appear in the middle component (i.e. $\xi_{\texttt{j+1}}$, $\xi_{\texttt{j}}$ and $\xi_{\texttt{j-1}}$) couples to two fermions only through the minimal coupling given in \eqref{ntwothreepoint3}, \eqref{ntwothreepoint4.1} and \eqref{ntwothreepoint10.1}. 

Let's consider the four-point function of four fermions. All other amplitudes are related to this by supersymmetry. Furthermore, we focus only on the $s$ channel poles because other channels are related to $s$ channel by an exchange. We already know that the amplitude is non-zero only when two particles have positive helicity and two of them have negative helicity. Let's first consider $(1^+2^+3^-4^-)$. In this case, only the non-minimal couplings contribute. The amplitude is given by 
\begin{equation}
	s^{\texttt{j}}\, [12]\langle34\rangle\, \frac{(g_{\texttt{ff(j)}}^{(1)})^2}{m^{2\texttt{j}}}\,  
\widetilde{\mathcal{N}}_{\texttt{j};0,0}\, \jacobi_{\texttt{j}}^{(0,0)}(z)	
\label{ntwofourpoint1}
\end{equation}
For the $(1^+2^-3^+4^-)$ amplitude, the minimal couplings contribute. The total contribution is 
\begin{equation}
\begin{split}
	[13]\langle24\rangle\, \frac{(g_{\texttt{ff(j+1)}}^{(0)})^2}{m^{2\texttt{j}}}\,  \, s^{\texttt{j}-1 } \Bigg[& \left(\frac{s}{m^2}\right)^2 \widetilde{\mathcal{N}}_{\texttt{j+1};1,1}\jacobi_{\texttt{j}}^{(0,2)}(z)
+\left(\frac{s}{m^2}\right)\frac{(g_{\texttt{ff(j)}}^{(0)})^2}{(g_{\texttt{ff(j+1)}}^{(0)})^2}
\widetilde{\mathcal{N}}_{\texttt{j};1,1}\jacobi_{\texttt{j}-1}^{(0,2)}(z)
\\
&+\frac{(g_{\texttt{ff(j-1)}}^{(0)})^2}{(g_{\texttt{ff(j+1)}}^{(0)})^2}\widetilde{\mathcal{N}}_{\texttt{j-1};1,1}\jacobi_{\texttt{j}-2}^{(0,2)}(z)
\Bigg]	
\end{split}
\label{ntwofourpoint2}
\end{equation}
From \eqref{ntwothreepoint20} we see that 
\begin{equation}
\frac{(g_{\texttt{ff(j)}}^{(0)})^2}{(g_{\texttt{ff(j+1)}}^{(0)})^2}
=\frac{\texttt{j}}{ (\texttt{j}+1)}
\qquad,\qquad
\frac{(g_{\texttt{ff(j-1)}}^{(0)})^2}{(g_{\texttt{ff(j+1)}}^{(0)})^2}	
=\frac{(\texttt{j}-1)^2}{ (4\texttt{j}^2-1)}
\label{ntwofourpoint3}
\end{equation}
Then we can use the identity \eqref{fermionjacobiidentity} to write it as 
\begin{equation}
	[13]\langle24\rangle\, s^{\texttt{j}} \frac{(g_{\texttt{ff(j+1)}}^{(0)})^2}{m^{2\texttt{j}}}\,   \widetilde{\mathcal{N}}_{\texttt{j};0,0}\jacobi_{\texttt{j}}^{(0,0)}(z)
\label{ntwofourpoint4}
\end{equation}
Moreover, from eqn \eqref{ntwothreepoint4} we get $ (g_{\texttt{ff(j+1)}}^{(0)})^2 = (g_{\texttt{ff(j)}}^{(1)})^2$; this implies that the form factor for $(1^+2^+3^-4^-)$ (given in \eqref{ntwofourpoint1} is same as the form factor for the $(1^+2^-3^+4^-)$. In the Lorentz notation, the amplitude is 
\begin{equation}
	\mathcal{T}^{(\mathcal{N}=2)}_{\texttt{ffff}} \mathcal{F}(s,t)
\label{ntwofourpoint5}
\end{equation}
The tensor factor has the following expression in terms of the spinor helicity variables
\begin{equation}
	\mathcal{T}^{(\mathcal{N}=2)}_{\texttt{ffff}}= \Bigg[[12]\langle34\rangle\delta_{h_1+h_2,1}\delta_{h_3+h_4,-1}
	+ 2\longleftrightarrow 3+ 2\longleftrightarrow 4
	+\textrm{Parity conjugate}\Bigg]
\label{ntwofourpoint11}
\end{equation}
And the corresponding form is 
\begin{equation}
\mathcal{F}(s,t)=\frac{(g_{\texttt{ff(j+1)}}^{(0)})^2}{m^{2\texttt{j}}}\Bigg[ \frac{s^{\texttt{j}+1}\widetilde{\mathcal{N}}_{\texttt{j};0,0} 	\jacobi_{\texttt{j}}^{(0,0)}(z)}{s-m_{\texttt{j}}^2}+s\leftrightarrow t 
\Bigg]
\label{ntwofourpoint12}
\end{equation}
For the $(1^+2^+3^-4^-)$ configuration, this is a contribution from a single exchange, whereas, for the $(1^+2^-3^+4^-)$ configuration, this is a sum over three exchanges. 
 
\subsection{Photon amplitudes}
Now we focus on the four-photon/gluon amplitudes in $\mathcal{N}=2$ theories in 3+1 dimensions. In this case, we need to consider two different super-amplitudes
\begin{equation}
\begin{split}
&A(\Sigma^{(+1)},\Sigma^{(0)},\Phi_{\texttt{j}}) 
\quad,\quad	\texttt{j} \in \mathbb{Z}
\\
&A(\Sigma^{(+1)},\Sigma^{(+1)},\Phi_{\texttt{j}}) 
\quad,\quad	\texttt{j} \in 2\mathbb{Z}
\end{split}
\label{ntwofourpoint101}
\end{equation}
In the second case, $\texttt{j}$ must be an even integer from the exchange symmetry of bosons. In a parity invariant theory, we obtain  $A(\Sigma^{(0)},\Sigma^{(0)},\Phi_{\texttt{j}}) $ from the first super-amplitude using parity transformation. 

Let's start with first super-amplitude $A(\Sigma^{(+1)},\Sigma^{(0)},\Phi_{\texttt{j}})$; in this case, the two photons only have minimal coupling to the fields appearing in the middle component of the expansion in \eqref{ntwosuperfield4}: $\xi_{\texttt{j}+1}$, $\xi_{\texttt{j}}$ and $\xi_{\texttt{j}-1}$. The coupling constants satisfy the relations given in \eqref{ntwothreepoint20}. We denote the coupling constants of this super-amplitude as
\begin{equation}
	\mathfrak{g}_{1, 0,\texttt{j}}\equiv \mathfrak{g}_{\texttt{p}, \texttt{p},\texttt{j}}^{(0)}
\label{ntwofourpoint102}
\end{equation}
The super-script $(0)$ is to denote this has only minimal couplings of two photons. 

The other three-point super-amplitude is 
\begin{equation}
	\mathcal{A}(\Sigma^{(+1)},\Sigma^{(+1)},\Phi_{\texttt{j}}) 
\qquad,\qquad
\mathcal{A}(\Sigma^{(0)},\Sigma^{(0)},\Phi_{\texttt{j}}) 	
\label{ntwofourpoint103}
\end{equation}
These super-amplitudes have only non-minimal coupling of photons; From the first one, we get that two positive helicity photon couple to ${\widetilde  \phi}_{\texttt{j}}$ and in the second super-amplitude two negative helicity photons couple only to $ { \phi}_{\texttt{j}}$. We denote the coupling constant of this super-amplitude as 
\begin{equation}
	\mathfrak{g}_{1,1,\texttt{j}}=\mathfrak{g}_{0,0,\texttt{j}}\equiv \mathfrak{g}_{\texttt{p}, \texttt{p},\texttt{j}}^{(1)}
\label{ntwofourpoint104}
\end{equation}
The super-script ${(1)}$ denotes the non-minimal nature of the two photons coupling to the higher spin particles.

In the Lorentz covariant basis, the three point function of two photons and one higher spin particle in a $\mathcal{N}=1$ theory is given in \eqref{nonethreepoint41}. Using the above result, we can write it in $\mathcal{N}=2$ theories
\begin{equation}
\begin{split}
\widehat{\mathcal{A}}_{\texttt{j}}{}^{(0)}\Big|_{\mathcal{N}=2} &= \bigg(\widehat{\mathcal{A}}_{\texttt{j}+1}{}^{(0)} + \sqrt{2}\sqrt{\frac{(\texttt{j}-1)^2}{2\texttt{j}(\texttt{j}+1)}}\widehat{\mathcal{A}}_{\texttt{j}}{}^{(0)}+\sqrt{4}\sqrt{\frac{(\texttt{j}-2)^2(\texttt{j}-1)^2}{(2\texttt{j})^2(2\texttt{j}+1)(2\texttt{j}-1)}}\widehat{\mathcal{A}}_{\texttt{j}-1}{}^{(0)}\bigg)
\qquad,\qquad\texttt{j}\in 2\mathbb{Z}\\
\end{split}
\label{ntwofourpoint104.1}
\end{equation}
We omitted the superscript $\texttt{ppj}$ for sake of brevity.

Lets now discuss the four point function due to the exchange of super-multiplet. 
\begin{itemize}
	\item From $\mathcal{N}=1$ supersymmetry we already know that 
\begin{equation}
\mathcal{N}\ne 0\implies 	\mathcal{M}(1^+2^+3^+4^+)=0= \mathcal{M}(1^+2^+3^+4^-)
\label{ntwofourpoint105}
\end{equation}
In terms of the form factor, this implies
\begin{eqnarray}
	\mathcal{F}_4=\mathcal{F}_5=\mathcal{F}_6=\mathcal{F}_7=0
\end{eqnarray}
At the level of four point function, only thing to consider is $\mathcal{M}(1^+2^+3^-4^-)$ (permutations amongst particles $2,3,4$); so the only non-zero form factors are $\mathcal{F}_1$, $\mathcal{F}_2$ and $\mathcal{F}_3$. For $\mathcal{N}=1$ theories the final result can be found in \eqref{nonephotonfourpoint21} (see \eqref{nonephotonfourpoint22} for four gluon).  

We focus only on the $s$-channel contributions because the contribution in other channels can be obtained from the $s$-channel contributions. 

	\item The $s$-channel contribution to $\mathcal{M}(1^+2^+3^-4^-)$ comes only due to the non-minimal coupling in \eqref{ntwofourpoint103}. We know that in terms of the component amplitudes, the non-minimal coupling is there only for $\phi_{\texttt{j}}$ and ${\widetilde  \phi}_{\texttt{j}}$. Since there is only one contribution, this amplitude is the same as that of $\mathcal{N}=0,1$ theories. 
\begin{equation}
		\mathcal{F}_1\left(s,t,u|\mathcal{N}=2,\texttt{j}\in \mathbb{Z}\right)
	=\frac{(g_{\texttt{pp(j)}}^{(1)})^2}{m^{2\texttt{j}+2}}\frac{s^{\texttt{j}}\widetilde{\mathcal{N}}_{\texttt{j};0,0}\,  \jacobi^{(0,0)}_{\texttt{j}} \left(\frac{t-u}{s}\right) }{s-m^2} 
\end{equation}
	
	\item The $s$-channel contribution to $\mathcal{M}(1^+2^-3^+4^-)$ comes only due to the minimal couplings. So in $\mathcal{N}=2$ theories, there are three different contributions (all coming from the middle component of the superfield). The total contribution is given by  
\begin{equation}
\begin{split}
	([13]\langle24\rangle)^2\, \frac{(g_{\texttt{pp(j+1)}}^{(0)})^2}{m^{2\texttt{j}-4}}\, s^{\texttt{j}-3} \Bigg[&  \left(\frac{s}{m^2}\right)^2\widetilde{\mathcal{N}}_{\texttt{j+1};2,2}\jacobi_{\texttt{j}-1}^{(0,4)}(z)
+\left(\frac{s}{m^2}\right)\frac{(g_{\texttt{pp(j)}}^{(0)})^2}{(g_{\texttt{pp(j+1)}}^{(0)})^2}
\widetilde{\mathcal{N}}_{\texttt{j};2,2}\jacobi_{\texttt{j}-2}^{(0,4)}(z)
\\
&+\frac{(g_{\texttt{pp(j-1)}}^{(0)})^2}{(g_{\texttt{pp(j+1)}}^{(0)})^2}\widetilde{\mathcal{N}}_{\texttt{j}-1;2,2}\jacobi_{\texttt{j}-3}^{(0,4)}(z)
\Bigg]		
\end{split}
\label{ntwofourpoint106}
\end{equation}
From \eqref{ntwothreepoint20} we get the following ratio of coupling constant 
\begin{equation}
\frac{(g_{\texttt{pp(j)}}^{(0)})^2}{(g_{\texttt{pp(j+1)}}^{(0)})^2}
= \frac{(\texttt{j}-1)^2}{(\texttt{j})(\texttt{j}+1)}
\qquad,\qquad
\frac{(g_{\texttt{pp(j-1)}}^{(0)})^2}{(g_{\texttt{pp(j+1)}}^{(0)})^2}
= 	 \frac{(\texttt{j}-1)^2}{(\texttt{j}^2)(4\texttt{j}^2-1)}
\end{equation}
For the residue at the pole, we can set $s=m^2$. Using these results, we can simplify the four photons to be 
\begin{equation}
	([13]\langle24\rangle)^2\, \frac{(g_{\texttt{pp(j+1)}}^{(0)})^2 }{m^{2\texttt{j}}}
	\, s^{\texttt{j}-1} \frac{\widetilde{\mathcal{N}}_{\texttt{j};1,1}\jacobi_{\texttt{j}-1}^{(0,2)}(z)}{s-m^2}
\end{equation}
\end{itemize}
If we add up all the channels, then the four-photon amplitude is given by 
\begin{equation}
	\frac{(g_{\texttt{pp(j)}}^{(1)})^2}{m^{2\texttt{j}+2}}\widetilde{\mathcal{N}}_{\texttt{j};0,0}\,
\begin{pmatrix}
\frac{s^{\texttt{j}}  \jacobi^{(0,0)}_{\texttt{j}} \left(\frac{t-u}{s}\right) }{s-m^2}
\\
\\
\frac{t^{\texttt{j}}  \jacobi^{(0,0)}_{\texttt{j}} \left(\frac{s-u}{t}\right) }{t-m^2}
\\
\\
\frac{u^{\texttt{j}}   \jacobi^{(0,0)}_{\texttt{j}} \left(\frac{t-s}{t}\right) }{u-m^2}	
\end{pmatrix}
+ 
\frac{(g_{\texttt{pp(j+1)}}^{(0)})^2}{m^{2\texttt{j}}}\widetilde{\mathcal{N}}_{\texttt{j};1,1 }
\begin{pmatrix}
\frac{t^{\texttt{j}-1}\jacobi_{\texttt{j}-1}^{(0,2)}\left(\frac{s-u}{t}\right)}{t-m^2}
+\frac{u^{\texttt{j}-1}\jacobi_{\texttt{j}-1}^{(0,2)}\left(\frac{s-t}{u}\right)}{u-m^2}
\\
\\
\frac{s^{\texttt{j}-1}\jacobi_{\texttt{j}-1}^{(0,2)}\left(\frac{u-t}{s}\right)}{s-m^2}+
\frac{t^{\texttt{j}-1}\jacobi_{\texttt{j}-1}^{(0,2)}\left(\frac{s-u}{t}\right)}{t-m^2}

\\
\\
\frac{u^{\texttt{j}-1}\jacobi_{\texttt{j}-1}^{(0,2)}\left(\frac{s-t}{u}\right)}{u-m^2}
+\frac{s^{\texttt{j}-1}\jacobi_{\texttt{j}-1}^{(0,2)}\left(\frac{u-t}{s}\right)}{s-m^2}
\end{pmatrix} 
\end{equation}
In case of colour ordered four gluons, the $u$ channels poles will be absent. 
\begin{equation}
	\frac{(g_{\texttt{pp(j)}}^{(1)})^2}{m^{2\texttt{j}+2}}\widetilde{\mathcal{N}}_{\texttt{j};0,0}\,
\begin{pmatrix}
\frac{s^{\texttt{j}}  \jacobi^{(0,0)}_{\texttt{j}} \left(\frac{t-u}{s}\right) }{s-m^2}
\\
\\
0
\\
\\
\frac{t^{\texttt{j}}  \jacobi^{(0,0)}_{\texttt{j}} \left(\frac{s-u}{t}\right) }{t-m^2}
\end{pmatrix}
+ 
\frac{(g_{\texttt{pp(j+1)}}^{(0)})^2}{m^{2\texttt{j}}}\widetilde{\mathcal{N}}_{\texttt{j};1,1 }
\begin{pmatrix}
\frac{t^{\texttt{j}-1}\jacobi_{\texttt{j}-1}^{(0,2)}\left(\frac{s-u}{t}\right)}{t-m^2}
\\
\\
\frac{s^{\texttt{j}-1}\jacobi_{\texttt{j}-1}^{(0,2)}\left(\frac{u-t}{s}\right)}{s-m^2}+
\frac{t^{\texttt{j}-1}\jacobi_{\texttt{j}-1}^{(0,2)}\left(\frac{s-u}{t}\right)}{t-m^2}

\\
\\
\frac{s^{\texttt{j}-1}\jacobi_{\texttt{j}-1}^{(0,2)}\left(\frac{u-t}{s}\right)}{s-m^2}
\end{pmatrix} 
\end{equation}

\section{$\mathcal{N}=4$ supersymmetry} 
\label{sec:bcrsnfour}
This is the maximal supersymmetry in a non-gravitational theory. The four superspace coordinates are given by $\eta^A$ ($A=1,\cdots,4$). The photon/gluon multiplet is CPT self-conjugate. It is given by
\begin{equation}
	\Sigma^{(1)}= p^{(+1)}+\eta^A f_A^{(1/2)}+\frac{1}{2}\eta^{A}\eta^{B} \phi_{[AB]}^{(0)}+ \eta^A\eta^B\eta^C \bar f_{ABC}^{(-1/2)}+ \frac{1}{4!}\epsilon_{ABCD}\eta^A\eta^B\eta^C\eta^D  p^{(-1)}
\label{nfoursuperfield1}
\end{equation} 
In this the $R$ symmetry is $SU(4)$\footnote{From the perspective of dimensional reduction, this follows from the $\mathbb{T}^6$ reduction of $D=10,\, \mathcal{N}=1$ photon/gluon multiplet. $SO(6)\simeq SU(4)$ is the tangent space symmetry of the internal $T^6$. }. There are two helicities of the photon, four fermions (4 anti-fermion) and six scalars in the multiplet. This multiplet is self-conjugate under parity. We show this explicitly  
\begin{equation}
	\mathcal{P}\Sigma^{(1)}\mathcal{P}^{-1}= \zeta_{(1)}\Bigg[ p^{(-1)}+\iimg \eta^\dagger_A (f^A)^{(-1/2)}-\frac{1}{2}\eta^\dagger_A\eta^\dagger_A (\phi^{[AB]})^{(0)}-\iimg  \eta^\dagger_A\eta^\dagger_B\eta^\dagger_C (\bar f^{ABC})^{(1/2)}- \frac{1}{4!}\epsilon^{ABCD}\eta^\dagger_A\eta^\dagger_B\eta^\dagger_C\eta^\dagger_D  p^{(1)}\Bigg]
\label{nfoursuperfield2}
\end{equation} 
Let's now consider a massive supermultiplet. A massive super-multiplet in $\mathcal{N}=4$ theories have $256(2\texttt{j}+1)$ states; here $\texttt{j}$ is the spin of Clifford vacuum. A massive super multiplet is given by 
\begin{equation}
\begin{split}
\Phi_{\texttt{j}}^{(I_1\cdots I_\texttt{2j})}=& 		\phi_{\texttt{j}}^{(I_1\cdots I_\texttt{2j})}
+
\eta^A_J\psi_{A}^{J(I_1\cdots I_\texttt{2j})}
		+\eta^{A}_{J_1}\eta_{J_2}^{B}\xi_{AB}^{J_1J_2(I_1\cdots I_\texttt{2j})}
+
\\
&+ \eta^A_{J_1}\eta^B_{J_2}\eta^C_{J_3} \chi_{ABC}^{J_1J_2J_3(I_1\cdots I_\texttt{2j})}
%\\
%& 
+	\eta^A_{J_1}\eta^B_{J_2}\eta^C_{J_3}\eta^D_{J_4} \omega_{ABCD}^{J_1J_2J_3J_4(I_1\cdots I_\texttt{2j})}
\\
&+\cdots 
+\eta^8 {\widetilde  \phi}_{\texttt{j}}^{(I_1\cdots I_\texttt{2j})}
\end{split}	
\label{nfoursuperfield3}
\end{equation}
The details of the field content can be found in Table \ref{tab:nfourfieldcontentone}. $	\eta^8$ is the short-form of 
\begin{equation}
	\eta^8=\frac{1}{(4!)^2 (2!)^4}	\epsilon_{I_1I_5}	\epsilon_{I_2I_6}\epsilon_{I_3I_7}\epsilon_{I_4I_8}\epsilon^{ABCD}\epsilon^{EFGH}\eta^{I_1}_{A}\eta^{I_2}_B\eta^{I_3}_{C}\eta^{I_4}_{D}
	\eta^{I_5}_{E}\eta^{I_6}_F\eta^{I_7}_{G}\eta^{I_8}_{H}
\label{nfoursuperfield4}
\end{equation}
\begin{center}
	\begin{table}
		\tiny
		\begin{subtable}{0.4\textheight}
			\begin{tabular}{|c|c|c|}
				\hline
				Level & $R$-symmetry & Spin \\
				\hline
				&&\\
				0 & $\odot$ & 
				\begin{ytableau}
					1 & 2 & 3 & 4 & \dots  & 2\texttt{j} 
				\end{ytableau} \\
				\hline
				&&\\
				\multirow{2}{*}{1} &  \multirow{2}{*}{
					\ydiagram[*(yellow)]{1} } & 
				\begin{ytableau}
					1 & 2 & 3 & 4 & \dots  & 2\texttt{j} & *(green)
				\end{ytableau} \\
				& & 
				\begin{ytableau}
					1 & 2 & 3 & 4 & \dots  & \scriptstyle 2\texttt{j}-1 
				\end{ytableau} \\
				\hline
				&&\\
				\multirow{4}{*}{2} & \multirow{3}{*}{\ydiagram[*(yellow)]{1, 1}} & \begin{ytableau}
					1 & 2 & 3 & 4 & \dots  & 2\texttt{j} & *(green) & *(green)
				\end{ytableau}\\
				& & \begin{ytableau}
					1 & 2 & 3 & 4 & \dots  & 2\texttt{j} 
				\end{ytableau} \\
				& & \begin{ytableau}
					1 & 2 & 3 & 4 & \dots  & \scriptstyle 2\texttt{j}-2 
				\end{ytableau} \\
				\cline{2-3}
				&&\\
				& \ydiagram[*(yellow)]{2} & \begin{ytableau}
					1 & 2 & 3 & 4 & \dots  & 2\texttt{j} 
				\end{ytableau} \\
				\hline
				&&\\
				\multirow{6}{*}{3} & \multirow{4}{*}{\ydiagram[*(yellow)]{1,1,1}} & \begin{ytableau}
					1 & 2 & 3 & 4 & \dots  & 2\texttt{j} & *(green) & *(green) & *(green)
				\end{ytableau} \\
				& &\begin{ytableau}
					1 & 2 & 3 & 4 & \dots  & 2\texttt{j} & *(green) 
				\end{ytableau} \\
				& & \begin{ytableau}
					1 & 2 & 3 & 4 & \dots  & \scriptstyle 2\texttt{j}-1 
				\end{ytableau} \\
				& & \begin{ytableau}
					1 & 2 & 3 & 4 & \dots  & \scriptstyle 2\texttt{j}-3 
				\end{ytableau} \\
				\cline{2-3}
				&&\\
				&  \multirow{2}{*}{\ydiagram[*(yellow)]{2,1}} & \begin{ytableau}
					1 & 2 & 3 & 4 & \dots  & 2\texttt{j} & *(green) 
				\end{ytableau} \\
				& & \begin{ytableau}
					1 & 2 & 3 & 4 & \dots  & \scriptstyle 2\texttt{j}-1  
				\end{ytableau} \\
				&&\\ && \\
				\hline
				&&\\
				\multirow{9}{*}{4} & \multirow{5}{*}{$\odot$} & \begin{ytableau}
					1 & 2 & 3 & 4 & \dots  & 2\texttt{j} & *(green) & *(green) & *(green) & *(green)
				\end{ytableau} \\
				& & \begin{ytableau}
					1 & 2 & 3 & 4 & \dots  & 2\texttt{j} & *(green) & *(green) 
				\end{ytableau} \\
				& &\begin{ytableau}
					1 & 2 & 3 & 4 & \dots  & 2\texttt{j}
				\end{ytableau} \\
				& &\begin{ytableau}
					1 & 2 & 3 & 4 & \dots  & \scriptstyle 2\texttt{j}-2
				\end{ytableau} \\
				& & \begin{ytableau}
					1 & 2 & 3 & 4 & \dots  & \scriptstyle 2\texttt{j}-4
				\end{ytableau} \\
				\cline{2-3}
				&&\\
				& \multirow{3}{*}{\ydiagram[*(yellow)]{2,1,1}} & \begin{ytableau}
					1 & 2 & 3 & 4 & \dots  & 2\texttt{j} & *(green) & *(green) 
				\end{ytableau}  \\
				& &\begin{ytableau}
					1 & 2 & 3 & 4 & \dots  & 2\texttt{j} 
				\end{ytableau}  \\
				& & \begin{ytableau}
					1 & 2 & 3 & 4 & \dots  & \scriptstyle 2\texttt{j}-2  
				\end{ytableau}  \\
				&&\\
				\cline{2-3}
				&&\\
				& \ydiagram[*(yellow)]{2,2} &  \begin{ytableau}
					1 & 2 & 3 & 4 & \dots  & 2\texttt{j} 
				\end{ytableau}  \\
				&&\\
				\hline
			\end{tabular}
		\end{subtable}
		\begin{subtable}{0.48\textwidth}
			\begin{tabular}{|c|c|c|}
				\hline
				Level & $R$-symmetry & Spin \\
				\hline
				&&\\
				\multirow{6}{*}{5} & \multirow{4}{*}{\ydiagram[*(yellow)]{1}} & \begin{ytableau}
					1 & 2 & 3 & 4 & \dots  & 2\texttt{j} & *(green) & *(green) & *(green)
				\end{ytableau} \\
				& & \begin{ytableau}
					1 & 2 & 3 & 4 & \dots  & 2\texttt{j} & *(green) 
				\end{ytableau} \\
				& & \begin{ytableau}
					1 & 2 & 3 & 4 & \dots  & \scriptstyle 2\texttt{j}-1 
				\end{ytableau} \\
				& & \begin{ytableau}
					1 & 2 & 3 & 4 & \dots  & \scriptstyle 2\texttt{j}-3 
				\end{ytableau} \\
				\cline{2-3}
				&&\\
				& \multirow{2}{*}{\ydiagram[*(yellow)]{2,2,1}} & \begin{ytableau}
					1 & 2 & 3 & 4 & \dots  & 2\texttt{j} & *(green) 
				\end{ytableau} \\
				& & \begin{ytableau}
					1 & 2 & 3 & 4 & \dots  & \scriptstyle 2\texttt{j}-1 
				\end{ytableau} \\
				&&\\ &&\\ &&\\ && \\
				\hline
				&&\\
				\multirow{4}{*}{6} & \multirow{3}{*}{\ydiagram[*(yellow)]{1,1}} & \begin{ytableau}
					1 & 2 & 3 & 4 & \dots  & 2\texttt{j} & *(green) & *(green)
				\end{ytableau} \\
				& & \begin{ytableau}
					1 & 2 & 3 & 4 & \dots  & 2\texttt{j} 
				\end{ytableau} \\
				& & \begin{ytableau}
					1 & 2 & 3 & 4 & \dots  & \scriptstyle 2\texttt{j}-2 
				\end{ytableau} \\
				\cline{2-3}
				&&\\
				& \ydiagram[*(yellow)]{2,2,2} & \begin{ytableau}
					1 & 2 & 3 & 4 & \dots  & 2\texttt{j} 
				\end{ytableau} \\
				&&\\
				\hline
				&&\\
				\multirow{2}{*}{7} & \multirow{2}{*}{\ydiagram[*(yellow)]{1,1,1}} & 
				\begin{ytableau}
					1 & 2 & 3 & 4 & \dots  & 2\texttt{j} & *(green)
				\end{ytableau} \\
				& & 
				\begin{ytableau}
					1 & 2 & 3 & 4 & \dots  & \scriptstyle 2\texttt{j}-1 
				\end{ytableau} \\
			&& \\ && \\ && \\ && \\
				\hline
				&&\\
				8 & $\odot$ & 
				\begin{ytableau}
					1 & 2 & 3 & 4 & \dots  & 2\texttt{j} 
				\end{ytableau}  \\
				\hline
			\end{tabular}
		\end{subtable}
\caption{Field content of massive $\mathcal{N}=4$ multiplet  }
\label{tab:nfourfieldcontentone}
	\end{table}
\end{center}
The expression for $\texttt{j}=0$ can be found in \cite{Engelbrecht:2022aao}. The third term's decomposition in terms of irreducible representation can be found in \eqref{ntwosuperfield4} respectively. The decomposition of the fourth term has a lot of term terms. $\eta^{A}_I$ transforms as $\underline 4$ of $SU(4)_R$ and $\underline 2$ of little group $SU(2)$ and two $\eta$ s anti-commute. Here we demonstrate how to decompose the product of two $\eta$s into irrep of $SU(4)_R\otimes SU(2) $. 
\begin{equation}
	\eta^{A}_{J_1}\eta^{B}_{J_2}= \eta^{(A}_{[J_1}\eta^{B)}_{J_2]}+ \eta^{[A}_{(J_1}\eta^{B]}_{J_2)}
\label{nfoursuperfield5}
\end{equation}
The first term transforms as symmetric rank 2 (i.e the $10$ dimensional) representation of $SU(4)$ and trivial representation of $SU(2)$, and the second term transforms as $6$ of $SU(4)$ and $3$ of $SU(2)$. 

Let's now consider the product of 4 $\eta$s
\begin{equation}
\begin{split}
\eta^{A}_{J_1}\eta^{B}_{J_2}\eta^{C}_{J_3}\eta^{D}_{J_4}&= \eta^{[A}_{J_1}\eta^{B}_{J_2}\eta^{C}_{J_3}\eta^{D]}_{J_4}+\cdots 
\end{split}	
\label{nfoursuperfield8}
\end{equation}
In \eqref{nfoursuperfield8}, only the first term transforms trivially under the $SU(R)$; other terms are charged under the $R$-symmetry. So the components of the superfields multiplying those terms are also charged under the $R$ symmetry. We focus on terms that are not neutral under the $R$ symmetry, and hence we consider the first term only. The first term transforms under the spin $2$ representation of the little group. The decomposition of the first term in terms of irreducible representations is given by, 
\begin{eqnarray}
 \frac{1}{4!}\epsilon_{ABCD}&&\left[\eta^{A}_{I}\eta^{B}_{J}\eta^{C}_{K}\eta^{D}_{L}\psi_{\texttt{j}+2}^{(I_{1}\ldots I_{2\texttt{j}})IJKL} + \sqrt{\frac{\texttt{j}}{\texttt{2}+\texttt{j}}}\eta^{A(I_{1}}\eta^{B}_{J}\eta^{C}_{K}\eta^{D}_{L}\psi_{\texttt{j}+1}^{I_{2}\ldots I_{2\texttt{j}})JKL} \right. \nonumber\\
&& \left. + \sqrt{\frac{(2\texttt{j}-1)\texttt{j}}{(2\texttt{j}+3)(\texttt{j}+1)}}\eta^{A(I_{1}}\eta^{BI_{2}}\eta^{C}_{K}\eta^{D}_{L}\psi_{\texttt{j}}^{I_{3}\ldots I_{2\texttt{j}})KL} + \sqrt{\frac{(2\texttt{j}-1)(\texttt{j}-1)}{(2\texttt{j}+1)(\texttt{j}+1)}}\eta^{A(I_{1}}\eta^{BI_{2}}\eta^{CI_{3}}\eta^{D}_{L}\psi_{\texttt{j}-1}^{I_{4}\ldots I_{2\texttt{j}})L} \right. \nonumber\\
&& \left. +  \sqrt{\frac{2\texttt{j}-3}{2\texttt{j}+1}}\eta^{A(I_{1}}\eta^{BI_{2}}\eta^{CI_{3}}\eta^{DI_{5}}\psi_{\texttt{j}-2}^{I_{5}\ldots I_{2\texttt{j}})}\right] 
\label{nfoursuperfield11}
\end{eqnarray}
In terms of $SU(2)$ representation theory, this is simply
\begin{equation}
	2\otimes \texttt{j}= \texttt{j}+2\oplus \texttt{j}+1\oplus \texttt{j}\oplus \texttt{j}-1\oplus \texttt{j}-2
\label{nfoursuperfield12}
\end{equation}
The numbers in the decomposition are simply Clebsh-Gordon coefficients. So field content of the $R$ symmetry neutral sector is 
\begin{equation}
\phi_{\texttt{j}}\quad ,\quad {\widetilde  \phi}_{\texttt{j}}\quad, \quad 
\omega_{\texttt{j}+2}\quad, \quad 
\omega_{\texttt{j}+1}\quad, \quad 
\omega_{\texttt{j}}\quad, \quad 
\omega_{\texttt{j}-1}\quad, \quad 
\omega_{\texttt{j}-2} 	
\label{nfoursuperfield13}
\end{equation}
All other components in the supermultiplet are charged under the $R$ symmetry. Various components can be obtained from the following projection
\begin{equation}
\begin{split}
\phi_{\texttt{j}}^{(I_1\cdots I_\texttt{2j})}
&= 
\Phi_{\texttt{j}}^{(I_1\cdots I_\texttt{2j})}\Big|_{\eta \rightarrow 0 }
\\
\omega_{\texttt{j}+2}^{(I_1\cdots I_\texttt{2j+4})}&=
\frac{1}{4!} 
\epsilon^{ABCD}\partial_A^{(I_1}\partial_B^{I_2}\partial_C^{I_3}\partial_D^{I_4}\Phi_{\texttt{j}}^{I_5\cdots I_\texttt{2j+4})}\Big|_{\eta \rightarrow 0 }
\\
\omega_{\texttt{j}+1}^{(I_1\cdots I_\texttt{2j+2})}&=	
\frac{1}{4!} 
\sqrt{\frac{\texttt{j}}{(\texttt{j}+2)}}
\epsilon^{ABCD}\partial_{A,J}\partial_B^{(I_1}\partial_C^{I_2}\partial_D^{I_3}\Phi_{\texttt{j}}^{JI_4\cdots I_\texttt{2j+2})}\Big|_{\eta \rightarrow 0 }
\\
\omega_{\texttt{j}}^{(I_1\cdots I_\texttt{2j})}&=
\frac{1}{4!} \sqrt{\frac{(\texttt{j})(2\texttt{j}-1)}{(2\texttt{j}+3)(2\texttt{j}+1)}}
\epsilon^{ABCD}\partial_{A,J_1}\partial_{B,J_2}\partial_C^{(I_1}\partial_D^{I_2}\Phi_{\texttt{j}}^{J_1J_2I_3\cdots I_\texttt{2j})}\Big|_{\eta \rightarrow 0 }
\\
\omega_{\texttt{j}-1}^{(I_1\cdots I_\texttt{2j-2})}&=	
\frac{1}{4!} \sqrt{\frac{(2\texttt{j}-1)(\texttt{j}-1)}{(2\texttt{j}+1)(\texttt{j}+1) }}
\epsilon^{ABCD}\partial_{A,J_1}\partial_{B,J_2}\partial_{C,J_3}\partial_D^{(I_1}\Phi_{\texttt{j}}^{J_1J_2J_3I_2\cdots I_\texttt{2j-2})}\Big|_{\eta \rightarrow 0 }
\\
\omega_{\texttt{j}-2}^{(I_1\cdots I_\texttt{2j-4})}&=
\frac{1}{4!} \sqrt{\frac{ (2\texttt{j}-3)}{(2\texttt{j}+1)}}
\epsilon^{ABCD}\partial_{A,J_1}\partial_{B, J_2}\partial_{C, J_3}\partial_{D, J_4}\Phi_{\texttt{j}}^{(J_1J_2J_3J_4I_1\cdots I_\texttt{2j-4})}\Big|_{\eta \rightarrow 0 }
\\
{\widetilde  \phi}_{\texttt{j}}^{(I_1\cdots I_\texttt{2j})}
&= 
 \epsilon_{J_1J_2}\epsilon_{J_3J_4}   \epsilon_{J_5J_6}\epsilon_{J_7J_8}  \epsilon^{A_1B_1C_1D_1}\epsilon^{A_2B_2C_2D_2} \frac{\partial^8}{\partial\eta_{J_1}^{A_1}\partial\eta_{J_2}^{B_1}\partial\eta_{J_3}^{C_1}\partial\eta_{J_4}^{D_1}\partial\eta_{J_5}^{A_2}\partial\eta_{J_6}^{B_2}\partial\eta_{J_7}^{C_2}\partial\eta_{J_8}^{D_2}}
\Phi_{\texttt{j}}^{(I_1\cdots I_\texttt{2j})}\Big|_{\eta \rightarrow 0 }
\end{split}	
\label{nfoursuperfield16}
\end{equation}

\paragraph{Parity}
The action of parity on the massless supermultiplet is simple. It just flips the helicity. The photon multiplet in $\mathcal{N}=4$ theories is self-conjugate. The massive representations are always parity invariant. The self-conjugacy of the supermultiplet imposes the following parity transformation rules for the components 
\begin{equation}
	(\phi_{\texttt{j}}, {\widetilde  \phi}_{\texttt{j}} ,  
\omega_{\texttt{j}+2}, 
\omega_{\texttt{j}+1} ,  
\omega_{\texttt{j}} , 
\omega_{\texttt{j}-1} , 
\omega_{\texttt{j}-2})
\longrightarrow 
	({\widetilde  \phi}_{\texttt{j}} ,  \phi_{\texttt{j}}, 
\omega_{\texttt{j}+2}, 
\omega_{\texttt{j}+1} ,  
\omega_{\texttt{j}} , 
\omega_{\texttt{j}-1} , 
\omega_{\texttt{j}-2})
\label{nfourparity}
\end{equation}

\subsection{Three point functions}
In the $\mathcal{N}=4$ theories, let's consider the three-point function of a two-photon supermultiplet and one massive spin $\texttt{j}$ supermultiplet $\Phi_\texttt{j}$. 
\begin{equation} 
\texttt{j}\in \mathbb{Z}+\frac{1}{2} \implies A(\Sigma^{(+1)},\Sigma^{(+1)},\Phi_{\texttt{j}})=0 
\label{nfourthreepoint1}
\end{equation}
Furthermore, if the massless spin of one particle doesn't carry any other charges (i.e. it's abelian gauge theory), then from the spin-statistics theorem, it follows that the massive multiplet must have even spin. 
\begin{equation} 
  A(\Sigma^{(+1)},\Sigma^{(+1)},\Phi_{\texttt{j}})\ne 0 \implies \texttt{j}\in 2\mathbb{Z}
\label{nfourthreepoint2}
\end{equation}
For the rest of the section, we consider two general massless multiplets with helicity $h_1$ and $h_2$. In the end, we will specialize in the case of $h_1=h_2=1$. The three-point function of two massless multiplets and one massive multiplet $\mathcal{A}(\Sigma_1^{(h_1)}\Sigma_2^{(h_2)}\Phi_{\texttt{j}})	\Big|^{(I_1\cdots I_{\texttt{2j}})}$   is given by 
\begin{equation}
\frac{\mathfrak{g}_{h_1,h_2,\texttt{j}}}{m^{3\texttt{j}+h_1+h_2+3}}\delta^{(8)}(\mathcal{Q}_{(3)}^\dagger) 
	\Big([12]^{\texttt{j}+ h_1+h_2} 
	\langle 13^{(I_1} \rangle \cdots \langle 13^{I_{\texttt{j}+h_2 -h_1}} \rangle \langle 23^{\texttt{j}+h_2 -h_1+1} \rangle\cdots \langle 23^{I_{2\texttt{j}})} \rangle 
\Big)	
\label{nfourthreepoint3}
\end{equation} 
The delta function imposes conservation of super-charges, and for $\mathcal{N}=4$ theories, it is given by 
\begin{equation}
	\delta^{(8)}(\mathcal{Q}_{(3)}^\dagger)=  \prod_{A=1}^4\Bigg( \langle 12\rangle \eta_1^A\eta_2^A + \langle 13^J\rangle \eta_1^A\eta_{3,J}^A + \langle 23^J\rangle \eta_2^A\eta_{3,J}^A + \frac{m_3}{2} \eta_{3,J}^A\eta_{3}^{A,J}\Bigg)
\label{nfourthreepoint4}
\end{equation}
The subscript in $\mathcal{Q}_{(3)}$ denotes that it is a delta function for a three-point function. We want to determine the couplings to the component amplitudes
\begin{equation}
	g_{h_1,h_2,\texttt{j}\pm 2}
\quad,\quad
g_{h_1,h_2,\texttt{j}\pm 1}	
\quad,\quad
g_{h_1,h_2,\texttt{j}}	
\label{nfourthreepoint11}
\end{equation}
In particular, we want to restrict to case of top and bottom components of the massless superfields. All other components are charged under the $R$ symmetry. We can use the projection defined in \eqref{nfoursuperfield16} to determine these coupling constants. We are not presenting the details here. The method is very similar to the analysis of $\mathcal{N}=1$ and $\mathcal{N}=2$ theories. The final result is as follows:  
\begin{equation}
\begin{split}
	g_{h_1,h_2-2,\texttt{j+2}}&=  \mathfrak{g}_{h_1,h_2,\texttt{j}}
\\	
	g_{h_1,h_2-2,\texttt{j+1}}&= (\texttt{j}+h_2-h_1) \sqrt{\frac{1}{\texttt{j}(\texttt{j}+2)}} \mathfrak{g}_{h_1,h_2,\texttt{j}}
\\	
	g_{h_1,h_2-2,\texttt{j}}&=(\texttt{j}+h_2-h_1)(\texttt{j}+h_2-h_1-1)\sqrt{\frac{3}{(2\texttt{j}+3)(2\texttt{j}+2)2\texttt{j}(2\texttt{j}-1)}} \mathfrak{g}_{h_1,h_2,\texttt{j}}
\\	
	g_{h_1,h_2-2,\texttt{j-1}}&= \frac{(\texttt{j}+h_2-h_1)!}{(\texttt{j}+h_2-h_1-3)!}
	\sqrt{\frac{4}{(2\texttt{j}+2)(2\texttt{j}+1)(2\texttt{j})^2(2\texttt{j}-1)(2\texttt{j}-2)}} 
\mathfrak{g}_{h_1,h_2,\texttt{j}}	
\\	
	g_{h_1,h_2-2,\texttt{j-2}}&= \frac{(\texttt{j}+h_2-h_1)!}{(\texttt{j}+h_2-h_1-4)!}
	\frac{1}{\sqrt{(2\texttt{j}+1)(2\texttt{j})^2 (2\texttt{j}-1)^2(2\texttt{j}-2)^2(2\texttt{j}-3)}} 
\mathfrak{g}_{h_1,h_2,\texttt{j}}	
\\	
	g_{h_1,h_2,\texttt{j}}&= \mathfrak{g}_{h_1,h_2,\texttt{j}}
	\\	
	g_{h_1-2,h_2-2,\texttt{j}}&= \mathfrak{g}_{h_1,h_2,\texttt{j}}
\end{split}	
\label{nfourthreepoint12}
\end{equation}
These on-shell results can be used to write down the answer in Lorentz covariant basis.
 \begin{equation}
\begin{split}
\widehat{\mathcal{A}}_{\texttt{j}}{}^{(0)} \Big|_{\mathcal{N}=4}&= \bigg[\sqrt{2}\,\widehat{\mathcal{A}}_{\texttt{j}+2}{}^{(0)} +\sqrt{4} \sqrt{\Big(\frac{\texttt{j}}{\texttt{j}+2}\Big)}\widehat{\mathcal{A}}_{\texttt{j}+1}{}^{(0)} + \sqrt{8}\sqrt{\Big(\frac{3 (\texttt{j}-1)^2 \texttt{j}}{2 \left(4 (\texttt{j}+2) \texttt{j}^2+\texttt{j}-3\right)}\Big)}\widehat{\mathcal{A}}_{\texttt{j}}{}^{(0)}\\
&+\sqrt{16}\sqrt{\Big(\frac{(\texttt{j}-2)^2 (\texttt{j}-1)}{4 (2 \texttt{j}+1) \left(2 \texttt{j}^2+\texttt{j}-1\right)}\Big)}\widehat{\mathcal{A}}_{\texttt{j}-1}{}^{(0)} +\sqrt{32} \sqrt{\Big(\frac{\left(\texttt{j}^2-5 \texttt{j}+6\right)^2}{16 (2 \texttt{j}-3) (2 \texttt{j}-1)^2 (2 \texttt{j}+1)}\Big)}\widehat{\mathcal{A}}_{\texttt{j}-2}{}^{(0)}\bigg]
\end{split}	
\end{equation}	  
$\widehat{\mathcal{A}}_{\texttt{j}}{}^{(0)} $ is defined in \eqref{photonreview73}. We omitted the superscript $\texttt{ppj}$ for sake of brevity.  In the next section, we will see the consequence of these relations on four-photon/gluon amplitude.

\subsection{Four photon/gluon amplitudes}
Let's consider the four gluon tree level amplitude in $\mathcal{N}=4$ theories. In \cite{Alday:2007hr} it was argued that the tensor structure is unique for four gluon amplitudes in $\mathcal{N}=4$ theories, and it is given by $t_8\mathcal{B}^4$ 
\begin{eqnarray}
	t_8\mathcal{B}^4=\frac{1}{2}\times \Bigg\{2^4 \textrm{tr}\left(\mathcal{B}_1\mathcal{B}_2\mathcal{B}_3\mathcal{B}_4\right)-2^2\Big( \textrm{tr}\left(\mathcal{B}_1\mathcal{B}_2\right)\textrm{tr}\left(\mathcal{B}_3\mathcal{B}_4\right)+\textrm{perm.}\Big)\Bigg\}
\label{t8ten8}
\end{eqnarray}
Here we demonstrate it for tree level amplitude  due to the exchange of higher spin particles. 

For the moment, we focus only on $s$-channel exchange since other channels can be generated by the exchange of $1,3$ and $1,4$. If we consider the $(1^+2^+3^-4^-)$ configuration, the amplitude gets only contribution from the top/bottom component of the superfield. $(1^+2^+)$ couples only to $\widetilde{\phi}$ and $(3^-4^-)$ couples only to $\phi$. Due to supersymmetry, the propagator non-zero propagator is $\phi - \widetilde{\phi}$ and it gives rise to the following contribution
\begin{equation}
	\mathcal{F}_1\left(s,t,u|\mathcal{N}=4,\texttt{j}\in \mathbb{Z}\right)
	=\frac{(g_{\texttt{pp(j)}}^{(1)})^2}{m^{2\texttt{j}+2}}\frac{s^{\texttt{j}}\widetilde{\mathcal{N}}_{\texttt{j};0,0}\,  \jacobi^{(0,0)}_{\texttt{j}} \left(\frac{t-u}{s}\right) }{s-m^2}
\label{nfourfourpoint1}
\end{equation}
Let's now consider the $(1^+2^-3^+4^-)$ configuration. It gets contributions from five fields (all of these appear in the middle component of the superfield). The total contribution is given by 
\begin{equation}
\begin{split}
	([13]\langle24\rangle)^2\, \frac{(g_{\texttt{pp(j+2)}}^{(0)})^2}{m^{2\texttt{j}-6}}\,s^{\texttt{j}-4} \Bigg[&\left(\frac{s}{m^2}\right)^{4}  \widetilde{\mathcal{N}}_{\texttt{j+2};2,2}\jacobi_{\texttt{j}}^{(0,4)}(z_s)
+\left(\frac{s}{m^2}\right)^3\frac{(g_{\texttt{pp(j+1)}}^{(0)})^2}{(g_{\texttt{pp(j+2)}}^{(0)})^2}
\widetilde{\mathcal{N}}_{\texttt{j+1};2,2}\jacobi_{\texttt{j}-1}^{(0,4)}(z_s)
\\
&
+\left(\frac{s}{m^2}\right)^2
\frac{(g_{\texttt{pp(j)}}^{(0)})^2}{(g_{\texttt{pp(j+2)}}^{(0)})^2}\widetilde{\mathcal{N}}_{\texttt{j};2,2}\jacobi_{\texttt{j}-2}^{(0,4)}(z_s)
+\left(\frac{s}{m^2}\right)
\frac{(g_{\texttt{pp(j-1)}}^{(0)})^2}{(g_{\texttt{pp(j+2)}}^{(0)})^2}\widetilde{\mathcal{N}}_{\texttt{j}-1;2,2}\jacobi_{\texttt{j}-3}^{(0,4)}(z_s)
\\
&
+
\frac{(g_{\texttt{pp(j-2)}}^{(0)})^2}{(g_{\texttt{pp(j+2)}}^{(0)})^2}\widetilde{\mathcal{N}}_{\texttt{j}-2;2,2}\jacobi_{\texttt{j}-4}^{(0,4)}(z_s)
\Bigg]		
\end{split}
\label{nfourfourpoint2}
\end{equation}
Here $z_s=\frac{t-u}{s}$.  Now we use \eqref{nfourthreepoint12} for $h_1=h_2=1$, along with \eqref{photonreview71} to get  
\begin{equation}
\begin{split}
\frac{(g_{\texttt{pp(j+1)}}^{(0)})^2}{(g_{\texttt{pp(j+2)}}^{(0)})^2}
=\frac{2\texttt{j}}{\texttt{j}+2}
\qquad,\qquad
\frac{(g_{\texttt{ppj}}^{(0)})^2}{(g_{\texttt{pp(j+2)}}^{(0)})^2}	
=\frac{12\texttt{j}(\texttt{j}-1)^2}{2(\texttt{j}+1)(2\texttt{j}-1)(2\texttt{j}+3)}
\\
\frac{(g_{\texttt{pp(j-1)}}^{(0)})^2}{(g_{\texttt{pp(j+2)}}^{(0)})^2}
=\frac{8(\texttt{j}-2)^2(\texttt{j}-1)}{4(\texttt{j}+1)(4\texttt{j}^2-1)}
\qquad,\qquad
\frac{(g_{\texttt{pp(j-2)}}^{(0)})^2}{(g_{\texttt{pp(j+2)}}^{(0)})^2}
=\frac{16(\texttt{j}-2)^2(\texttt{j}-3)^2}{16(2\texttt{j}-1)^2(2\texttt{j}-3)(2\texttt{j}+1)}
\end{split}
\label{nfourfourpoint3}
\end{equation}
We can use identity \eqref{photonjacobiidentity} to simplify the answer to 
\begin{equation}
	([13]\langle24\rangle)^2\, \frac{(g_{\texttt{pp(j+2)}}^{(0)})^2}{m^{2\texttt{j}+2}}\, s^{\texttt{j}}
	\widetilde{\mathcal{N}}_{\texttt{j};0,0}\jacobi_{\texttt{j}}^{(0,0)}\left(\frac{t-u}{s} \right)
\label{nfourfourpoint4}
\end{equation}
Similarly, we can compute the $s$-channel contribution to $(1^+2^-3^-4^+)$ by exchanging 3 and 4. From \eqref{nfourfourpoint1} and \eqref{nfourfourpoint4} we can see that the contribution to $s$-channel amplitude from both configurations is the same. So the $s$-channel contribution is 
\begin{equation}
	\mathcal{F}\Big|_{s-\textrm{channel}}
	=
	\frac{(g_{\texttt{pp(j)}}^{(1)})^2}{m^{2\texttt{j}+2}}\widetilde{\mathcal{N}}_{\texttt{j};0,0}\,
\begin{pmatrix}
1
\\
\\
1
\\
\\
1
\end{pmatrix} 
\left(
\frac{s^{\texttt{j}}  \jacobi^{(0,0)}_{\texttt{j}} \left(\frac{t-u}{s}\right) }{s-m^2}
\right)
\label{nfourfourpoint5}
\end{equation}
If we add all the channels, then the four-photon amplitude is given by 
\begin{equation}
	\frac{(g_{\texttt{pp(j)}}^{(1)})^2}{m^{2\texttt{j}+2}}\widetilde{\mathcal{N}}_{\texttt{j};0,0}\,
\begin{pmatrix}
1
\\
\\
1
\\
\\
1
\end{pmatrix} 
\left[
\frac{s^{\texttt{j}}  \jacobi^{(0,0)}_{\texttt{j}} \left(\frac{t-u}{s}\right) }{s-m^2}
+\frac{t^{\texttt{j}}  \jacobi^{(0,0)}_{\texttt{j}} \left(\frac{s-u}{t}\right) }{t-m^2}
+\frac{u^{\texttt{j}}   \jacobi^{(0,0)}_{\texttt{j}} \left(\frac{t-s}{t}\right) }{u-m^2}	
\right]
\label{nfourfourpoint11}
\end{equation}
In the case of four gluons, we consider colour-ordered amplitude. In this case, $u$ channel poles are absent. 
\begin{equation}
	\frac{(g_{\texttt{pp(j)}}^{(1)})^2}{m^{2\texttt{j}+2}}\widetilde{\mathcal{N}}_{\texttt{j};0,0}\,
\begin{pmatrix}
1
\\
\\
1
\\
\\
1
\end{pmatrix} 
\left[
\frac{s^{\texttt{j}}  \jacobi^{(0,0)}_{\texttt{j}} \left(\frac{t-u}{s}\right) }{s-m^2}
+\frac{t^{\texttt{j}}  \jacobi^{(0,0)}_{\texttt{j}} \left(\frac{s-u}{t}\right) }{t-m^2}
\right]
\label{nfourfourpoint12}
\end{equation}
Since the coefficient of all the tensor structures are the same, the tensor structure for the gluon/photon amplitude is
\begin{equation}
\mathcal{T}_1+\mathcal{T}_2+\mathcal{T}_3\equiv 	t_8\mathcal{B}^4 
\label{nfourfourpoint13}
\end{equation}
The expression for $	t_8\mathcal{B}^4 $ can be found in \eqref{t8ten8}. The four-photon amplitude due to massive spin $\texttt{j}$ superfield exchange is 
\begin{equation}
	\mathfrak{g}^2\widetilde{\mathcal{N}}_{\texttt{j},0,0}\, t_8\mathcal{B}^4 \frac{\legendre_{\texttt{j}}}{s-m^2} 
\label{nfourfourpoint14}
\end{equation}
This result is very suggestive of the fact that the four-photon amplitude in $D$ spacetime dimension with 16 super-symmetries is given by 
\begin{equation}
		t_8\mathcal{B}^4 \Bigg[  s^{\texttt{j}}\frac{\gegen^{(\frac{D-3}{2})}_{\texttt{j}}(z)}{s-m_{\texttt{j}}^2}+\textrm{two more channels}\Bigg]
\qquad,\qquad
	\texttt{j}\in 2\mathbb{Z}^+
\end{equation}
 
 \section{Conclusion and future direction}
\label{sec:bcrsconclusion}
Our long-term goal is to construct a perturbative interacting theory of massive higher spin particles. As a baby step towards that direction, in \cite{Balasubramanian:2021act} the expression for tree-level scattering amplitudes for massless external particles due to higher spin exchange were written down. The key method was to write down the massless spinning amplitude as a derivative operator acting on the four scalar amplitude.  The expression for four gravitons is very tedious and hence difficult to use in other contexts. In this work, we used supersymmetry to explore the structures of tree-level amplitudes with higher spin exchanges. We systematically derived the constraints from supersymmetry on three-point functions. We derived the form factor and tensor factor due to different amounts of supersymmetry. We found that the form factor for maximum supersymmetry is proportional to Legendre polynomial, which is also the form factor for four scalar amplitude. In absence of supersymmetry, a massless irreducible representations in $3+1$ dimensions are not parity self-conjugate unless it is a scalar representation. This feature remains true in the presence of supersymmetry. The only self-conjugate massless multiplets are $\mathcal{N}=2$ hyper-multiplet,  $\mathcal{N}=4$ vector multiplet, $\mathcal{N}=8$ gravity multiplet. In all these cases, the amplitude is very simple. The form factor turns out to be 
\begin{equation}
\jacobi^{(0,0)}_{\texttt{j}}(z)=  \legendre_{\texttt{j}}(z)
\end{equation}
$\texttt{j}$ is the spin of Clifford vacuum. This is already known for four scalar amplitudes. One conceptual way is to note that  the tensor factor for maximally symmetric theory in any number of spacetime dimensions is unique due to supersymmetry \cite{Alday:2007hr}. The two self-conjugate massless (short) multiplets combine to give a massive representation (long) multiplet with scalar Clifford vacuum; this implies that the three-point amplitude of two self-conjugate massless multiplets and one massive supermultiplet is unique in any number of spacetime dimensions \footnote{In $3+1$ dimensions, the amplitude of two massless particles and one massive particle is also unique even without supersymmetry \cite{Arkani-Hamed:2017jhn}.}. Hence in a supersymmetric theory, the four scattering of massless self-conjugate multiplet can be thought of as the two-point function of a massive multiplet. This makes the tensor factor unique. So there is only one form factor for the scattering of massless self-conjugate multiplet in supersymmetric theories. In $3+1$ dimensions, the form factor is the Legendre polynomial. This probably hints that the form factor for maximal supersymmetric theory in any number of dimensions is Gegenbauer polynomial. It would be nice to demonstrate this explicitly.  

In a unitary theory, the residue of a pole can be written as a positive sum over physical exchanges. String theory is known to be unitary due to the no-ghost theorems. So one can try to verify the unitarity of string theory by checking the residue at poles. In \cite{Maity:2021obe}, Veneziano amplitude was analyzed in 3+1 dimension; it is an amplitude of four scalars, and hence the residue at a pole can be written as a positive sum of Legendre polynomial. A complete analysis was done in \cite{Arkani-Hamed:2022gsa}. In this work, the authors also considered four gluon amplitude in the type I string theory (which is a theory with 16 super-charges) and showed that the residue could be decomposed into a positive sum of Gegenbauer polynomial. Our work suggests that the Gegenbauer polynomial should be interpreted as the contribution from a massive super-multiplet, and then the co-efficient can be interpreted as the square of the coupling constant of two photon (multiplet)s to that super-multiplet.  

The loop amplitudes in supersymmetric theory are very simple. The loop amplitudes are important to learn about the low energy EFTs (for example see \cite{Bern:2021ppb, Bern:2022yes}). One could attempt to compute the loop amplitudes in the supersymmetric theories. In maximally supersymmetric theory, the loop contribution can be computed from loops with external scalars only, and as a result, the computation simplifies. The non-maximal cases are difficult to compute. 

In recent times, it has been shown that flat space scattering amplitudes with massless external legs can be written in terms of the correlation function of a non-unitary CFT living on the celestial sphere \cite{Pasterski:2016qvg, Pasterski:2020pdk, Pasterski:2021raf}. For a complete understanding of the celestial CFT, one would like to know if there is a massive higher spin exchange then, what it corresponds to in the celestial CFT. 

In \cite{Caron-Huot:2016icg} the authors have shown that the amplitude of four external scalars resembles that of string theory in the large $s$ and large $t$ limit. An important part of that analysis relies on the positivity of the residues at the pole. Again in maximally supersymmetric theory, the proof can be straightforwardly extended because the amplitude is essentially proportional to the four scalar amplitude. In the case of spinning amplitudes without supersymmetry, the positivity of the residue is not obvious. For example, consider the process with the configuration $(1^+2^+3^+4^-)$; in that case, the residue is to proportional to the product of two different coupling constants, and hence it is not necessarily positive definite. We have seen that $\mathcal{N}=1$ rules out a non-trivial scattering for this configuration. So it would be interesting to explore the uniqueness of four-photon amplitude for $\mathcal{N}=1,2$ supersymmetric theories. At every mass level, these theories have two different coupling constants (one for minimal and one for non-minimal). In string theory, these two coupling constants are related to each other. We would like to check whether such a relation is more generic by doing an analysis following  \cite{Caron-Huot:2016icg}. 

Another interesting direction to explore could be amplitudes with massive external legs. In this paper, we focussed on the process with only massless external legs. Compton amplitude has two massive external legs. Compton amplitude is extremely important in black hole physics. In general, Compton amplitude is important to know what is a good basis to write down scattering amplitudes with massive external legs. 

This method can be extended to scattering amplitudes in $\mathcal{N}=8$ in 3+1 dimensions. In particular, one can compute the four graviton amplitude in $\mathcal{N}=8$ theories. The tensor factor for four graviton amplitude in $\mathcal{N}=8$ is known to be unique. From this analysis, we expect the form factor to be just Legendre polynomial. It would be nice to show it explicitly. Any discussion of gluon and graviton amplitude is incomplete without referring to a double copy. In certain theories, the graviton amplitudes can be written as two copies of gluon amplitudes. This is known as double copy. The first known example is the amplitudes in open string theory and closed string theory \cite{Kawai:1985xq}. In recent time a lot of examples were found \cite{Bern:2008qj, Bern:2019prr, Carrasco:2020ywq} . Higher spin amplitudes can be another place to explore the applicability of double copy. In \cite{Engelbrecht:2022aao} authors analyzed double copy at the level of three-point function in the presence of supersymmetry. It would be nice to check it at the level of the four-point function with massive higher spin exchanges (and higher point functions). Our preliminary investigation seems to suggest that the double copy relation is true only if there is some restriction on the three-point functions and on the spectrum. We hope to report a comprehensive analysis in the near future.  
 
\paragraph{Acknowledgement} We thank Md. Abhishek, Sourav Ballav, Chandan Jana and Amey Yeole for many useful discussions. We are grateful to Subramanya Hegde, Diksha Jain, Dileep Jatkar, Aakash Kumar, Alok Laddha, Manoj Mandal, especially Joydeep Chakravarty, Raj Patil and Sourav Ballav for their comment on the draft. MKNB is grateful to CSIR for the fellowship. AR is grateful to ICTP, Trieste, for hospitality during this work.  AR would like to thank the organizers of the Regional Strings Meeting in NISER, Bhubaneswar where preliminay version of this work was presented. APS would like to thank HRI, Allahabad, for the hospitality during this work. We are grateful to the people of India for their generous funding for research in basic sciences.

\newpage
\appendix

\section{Notation and convention}
\label{app:bcrsnotation}

\begin{subequations}    
\begin{eqnarray} 
\textrm{Metric} 
\quad \quad &&  \eta_{\mu \nu }=
\textrm{diag }(-1,1,\cdots,1)
\\
\textrm{Lorentz indices}\quad \quad && \mu,\nu 
\\
\textrm{Spinor indices}\quad \quad &&  a, b, \dot a, \dot b  
\\
\textrm{Little group indices}\quad \quad &&  I,J  
\\
\textrm{$R$ symmetry indices}\quad \quad &&  A,B,C,D  
\\
\textrm{External Momentum}\quad \quad &&  k_\mu 
%\\
%\textrm{Lorentz generators}\quad \quad &&  ??????
%\mathcal{J}_{\mu \nu} 
%\\
%\textrm{$n$-point amplitude}\quad \quad &&  \mathcal{A}^n(\epsilon_a,k_a)
%\\
%\textrm{Helicity amplitude}\quad \quad &&  \mathcal{A}^{j_1\cdots j_n}_{m_1\cdots m_n}
\\
\textrm{Particle labels }\quad \quad &&  i,l
\\
\textrm{Momentum difference}\quad \quad &&  k_{ab}=k_a-k_b
\\
\textrm{Mandestam variables}\quad \quad &&  s,t,u
\\
\textrm{Bosonic polarizations}\quad \quad &&  \zeta^{(\texttt{j})}_{\mu_1\cdots \mu_\texttt{j} }, \epsilon_\mu
\\
\textrm{Form factor}\quad \quad &&  \mathcal{F} (s,t,u)
\\
\textrm{Tensor factor}\quad \quad &&  \mathcal{T} 
\\
\textrm{Wigner (small-)$d$ matrix}\quad \quad &&  d_{m^\prime m}^{(\texttt{j})}( \beta ) 
\\
\textrm{Legendre polynomial}\quad \quad && \legendre_\texttt{\texttt{j } }(z)
\\
\textrm{Gegenbauer polynomial}\quad \quad && \gegen^{(\beta)}_n(z)
\\
\textrm{Jacobi polynomial}\quad \quad && \jacobi^{\alpha,\beta}_{\texttt{j }}(z)
\\
\textrm{Masses of particles}\quad \quad &&  m, m_a, m_\texttt{j } 
\\
\textrm{Spin}\quad \quad &&  \texttt{j}
\\
\textrm{Helicity}\quad \quad &&  h
\\
\textrm{Linearized Maxwell field strength}\quad \quad &&  \mathcal{B}_{\mu \nu}
\\
\textrm{Linearized Riemann tensor}\quad \quad &&  \mathcal{R}_{\mu \nu \rho \sigma}
\\
\textrm{Super-space co-ordinate}\quad \quad &&  \eta
\\
\textrm{Number of supersymmetry in $3+1$ dimension}\quad \quad &&  \mathcal{N}
\end{eqnarray} 
\end{subequations} 

We follow the following convention for the Mandelstam variables,
\begin{eqnarray}
s=-(k_1+k_2)^2
&\hspace{40pt} &
%\nonumber\\
t=-(k_1+k_4)^2
\nonumber\\
&u=-(k_1+k_3)^2&
%    \nonumber\\
\label{hscomp22} 
\end{eqnarray}
This is same as convention in Green-Schwarz-Witten \cite{Green:2012oqa} \footnote{vol.1 page 373, 378} but different from Polchinski \cite{Polchinski:1998rq}. We also follow the convention such that all the {\it external particles} are  {\it outgoing}. 

\section{Spinor-Helicity conventions}
\label{app:bcrsshconevtions}

In the literature, there are more than one conventions for the spinor-helicity conventions. In this appendix, we spell it out explicitly. Our convention is consistent with the book \cite{Elvang:2015rqa}. 
\begin{enumerate}
	\item We define {\it square bracket} with $a$ indices and {\it angle bracket } with $\dot a$ indices. 
\begin{equation}
\lambda_{a}\, =|\lambda]_{a}\,
\qquad,\qquad
 \tilde \lambda_{\dot a} = \langle  \tilde \lambda|_{\dot a }	
 \label{bcrsshconv1}
\end{equation}
The inner product in the $SU(2)$ indices are as follows
\begin{equation}
		\langle il\rangle =  \tilde \lambda_{i\dot a }\, 	\tilde \lambda_l^{\dot a }
\qquad,\qquad
[il]=\lambda_i^a\,  \lambda_{la}
 \label{bcrsshconv2}
\end{equation}
	\item Moreover, we also follow mostly positive sign convention for the metric. So  the momenta of a massless particle is given by 
\begin{equation}
	p_{i\, a \dot a}=-\lambda_{i\, a}\, \tilde \lambda_{i\, \dot a } =-|p]_{i\, a}\, \langle p|_{i\, \dot a }
 \label{bcrsshconv3}
\end{equation}
The action of the little group for helicity $h$ particle is 
\begin{equation}
	(|p], |p\rangle) \longrightarrow (t |p], t^{-1}|p\rangle)
 \label{bcrsshconv4}
\end{equation}
$t$ is a phase. Under little group scaling the scattering amplitudes 
\begin{equation}
	\mathcal{A}(h_1,\cdots, h_n)\longrightarrow \prod_{i=1}^n\, (t_{i})^{2h_i}\mathcal{A}(h_1,\cdots, h_n)
 \label{bcrsshconv5}
\end{equation}

	\item For massless particles, all the scattering amplitudes are written in terms of gauge-invariant quantities.  
	
The Maxwell field strength is given by 	
\begin{equation}
\begin{split}
	\mathcal{B}_{\mu\nu}\longrightarrow &\mathcal{B}_{a \dot a b\dot b }=  \mathcal{B}^{(+)}_{ab }\epsilon_{\dot a\dot b }+\mathcal{B}^{(-)}_{\dot a\dot b  }\epsilon_{ab}
\\
&\mathcal{B}^{(+)}_{ab }= \sqrt{2}\, \lambda_{a }  \lambda_{b}
\qquad,\qquad
\mathcal{B}^{(-)}_{\dot a\dot b  }= \sqrt{2}\, \tilde \lambda_{\dot a }\tilde \lambda_{\dot b}	
\end{split}
 \label{bcrsshconv11}
\end{equation}
The gauge invariant expression for massless spin 2 fields involve Riemann tensor 
\begin{equation}
	\mathcal{R}_{\mu \nu \rho \sigma}\longrightarrow \mathcal{R}_{a \dot ab \dot bc \dot cd \dot d  }
=\epsilon_{ab}\epsilon_{cd } \mathcal{R}^{-}_{ \dot a \dot b\dot c\dot d  }	
+\epsilon_{\dot a \dot b}\epsilon_{\dot c \dot d} \mathcal{R}^{+}_{abcd   }	
 \label{bcrsshconv12}
\end{equation}

 \item For massive particles, the momenta can be written as
 \begin{equation}
	p_{i\, a\dot a}=-\lambda_{i\,I, a}\, \tilde \lambda^I_{i\,  \dot a} =-| p_I]_{i\,a}\, \langle  p|^I_{i\, \dot a }
 \label{bcrsshconv13}
\end{equation}

\begin{equation}
	\langle p^I p^J\rangle = m\, \epsilon^{IJ}
\qquad,
\qquad	
[ p^I p^J] =- m\, \epsilon^{IJ} 
 \label{bcrsshconv14}
\end{equation}
We list down a few more relations that is useful for us. First we write down the on-shell conditions
\begin{equation}
\begin{split}
&	p\,|p^I]=-m |p^I\rangle 
\qquad,\qquad
	p\, |p^I\rangle =-m\,  |p^I]	
\\
&[p^I|\, p =m\, \langle p^I|	
\qquad,\qquad	
\langle p^I|\, p =m [p^I|	
\end{split}
 \label{bcrsshconv15}
\end{equation}   
The completeness relation of spinors is given by 
\begin{equation}
	|p_I]_a[p^I|^b=m\, \delta_a^b 
\qquad,
\qquad
	|p_I\rangle_{\dot a}\langle p^I|^{\dot b}=-m\, \delta_{\dot a}^{\dot b}
 \label{bcrsshconv21}
\end{equation}
 
	\item The polarization of massive spin 1 particle is given by 
\begin{equation}
	\epsilon_{a\dot a}^{(I_1I_2)}= \sqrt{2}\frac{\lambda_a^{(I_1}\tilde \lambda_{\dot a}^{I_2)}}{m}= \sqrt{2} \frac{|p]_a^{(I_1}\langle p|_{\dot a}^{I_2)}}{m}
 \label{bcrsshconv22}
\end{equation}  
Polarisation of a massive particle satisfies transversality condition $p^\mu \epsilon_\mu=0$. We now check it in the spinor helicity language 
\begin{equation}
	p^{\dot a b}\epsilon_{b \dot a}^{(I_1I_2)}= -\left(|p_I\rangle^{\dot a}[p^I|^b\right) \left(\frac{|p]_b^{(I_1}\langle p|_{\dot a}^{I_2)}}{m}\right)
	=|p_I\rangle^{\dot a}\epsilon^{I(I_1}\langle p|_{\dot a}^{I_2)}=m \epsilon^{(I_1I_2)}=0
 \label{bcrsshconv23}
\end{equation}
The factor $\sqrt{2}$ is important to ensure orthonormality of polarisation. For a spin $\texttt{j}$ particle the polarisation \cite{Guevara:2018wpp} is given by 
\begin{equation}
	\epsilon_{a_1\cdots a_\texttt{j}\dot a\cdots \dot a_\texttt{j}}^{(I_1\cdots I_{2\texttt{j}} )}=\frac{2^{\frac{\texttt{j}}{2}}}{m^\texttt{j}}
	\lambda_{a_1}^{(I_1}\cdots \lambda_{a_\texttt{j}}^{I_\texttt{j}}\tilde \lambda_{\dot a_{1}}^{I_{\texttt{j}+1}}\cdots \tilde \lambda_{\dot a_{\texttt{j}}}^{I_{2\texttt{j}})}
 \label{bcrsshconv24}
\end{equation}
 
\end{enumerate}

\subsection{Spinors}
Consider a  massless particle with the following 4-momenta   
 \begin{eqnarray}
    p^\mu =E(1, \sin \theta \cos \phi , \sin \theta \sin \phi, \cos \theta )
\quad,\quad  0\leq \theta \leq \pi   
\quad,\quad 0\leq \phi \leq 2\pi 
\label{spinorconv1}
\end{eqnarray}
From this expression we get
\begin{eqnarray}
p_{\alpha \dot \alpha}    = 2E 
    \left(
\begin{matrix}
   -\sin^2 \frac{\theta}{2} & \sin \frac{\theta}{2}\cos \frac{\theta}{2} e^{-\iimg \phi}\\
  \sin \frac{\theta}{2}\cos \frac{\theta}{2}e^{\iimg \phi} & -\cos^2 \frac{\theta}{2} 
\end{matrix}
\right)
\label{spinorconv2}
\end{eqnarray}
This implies 
\begin{eqnarray}
    |p]_{ a}
    &= \sqrt{2E} 
        \left(
\begin{matrix}
   \sin \frac{\theta}{2}  e^{-\iimg\frac{\phi}{2}}\\
  -\cos \frac{\theta}{2}e^{\iimg \frac{\phi}{2}}  
\end{matrix}
\right)
\qquad\qquad
\langle p|_{\dot a}
=
\sqrt{2E} 
        \left(
\begin{matrix}
   \sin \frac{\theta}{2}  e^{\iimg \frac{\phi}{2}} &
  -\cos \frac{\theta}{2}e^{-\iimg\frac{\phi}{2}}  
\end{matrix}
\right)
\label{spinorconv3}
\\
    |p\rangle^{\dot a}
    &= \sqrt{2E} 
        \left(
\begin{matrix}
   \cos \frac{\theta}{2}  e^{-\iimg \frac{\phi}{2}}\\
  \sin \frac{\theta}{2}e^{\iimg \frac{\phi}{2}}  
\end{matrix}
\right)
\qquad\qquad
[ p|^{a}
=
\sqrt{2E} 
        \left(
\begin{matrix}
 \cos \frac{\theta}{2}e^{\iimg\frac{\phi}{2}} &
   \sin \frac{\theta}{2}  e^{-\iimg \frac{\phi}{2}}  
\end{matrix}
\right)
\label{spinorconv4}
\end{eqnarray}
For future purposes we set $\phi=0$. One more formula that is useful for us is
\begin{equation}
	\langle ij\rangle = 2\sqrt{E_iE_j}\sin\left(\frac{\theta_j-\theta_i}{2} \right)
\label{spinorconv5}
\end{equation}

\subsection{Center of mass frame}
The choice for center of mass frame is given by 
\begin{equation}
	\begin{split}
		p_1&= E(-1,0,0,1)
\\		
		p_2&= E(-1,0,0,-1)
\\
		p_3&= E(1,\sin\theta,0,\cos\theta )
\\
		p_4&= E(1,-\sin\theta,0,-\cos\theta )
	\end{split}
\label{fourmasslessconv1}
\end{equation}
For this choice of Mandelstam variables are
\begin{equation}
	s=4E^2\qquad,\qquad t= -2E^2(1-\cos\theta) 
	\qquad,\qquad u= -2E^2(1+ \cos\theta) 
\label{fourmasslessconv2}
\end{equation}
The spinor helicity variables follow from \eqref{spinorconv3}
\begin{equation}	
	|3\rangle =  \sqrt{2E} 
        \left(
\begin{matrix}
   \cos \frac{\theta}{2}   \\
  \sin \frac{\theta}{2}   
\end{matrix}
\right)
\qquad,\qquad	
	|4\rangle =\sqrt{2E} 
        \left(
\begin{matrix}
   \sin \frac{\theta}{2}   \\
   -\cos \frac{\theta}{2}  
\end{matrix}
\right)
\label{fourmasslessconv3}
\end{equation}
In order to get spinors for $p_1$ and $p_2$ we do $E\longrightarrow -E$ and $\theta \longrightarrow\pi- \theta$
\begin{equation}	
	|1\rangle =  \iimg \sqrt{2E} 
        \left(
\begin{matrix}
   0   \\
  1   
\end{matrix}
\right)
\qquad,\qquad	
	|2\rangle = \iimg  \sqrt{2E} 
      \left(
\begin{matrix}
   1  \\
   0 
\end{matrix}
\right)
\label{fourmasslessconv4}
\end{equation}
Now we can use \eqref{spinorconv5}
\begin{equation}
	\langle 12\rangle =2E=-\langle 34\rangle 
\quad,\quad 
\langle 14\rangle 	=-\iimg 2E \sin \frac{\theta}{2} =-\langle 23\rangle 
\quad,\quad 
\langle 13\rangle 	=-\iimg 2E \cos \frac{\theta}{2} =\langle 24\rangle 
\label{fourmasslessconv5}
\end{equation}
In order to get the box brackets we use 
\begin{equation}
	[ij]=\langle ji\rangle^\star =-\langle ij\rangle^\star
\label{fourmasslessconv6}
\end{equation}

\subsection{Grassmann Fourier transformation}
\label{seubsec:grassmannfouriertransform}
Any function of grassmann variables can be written in two ways; either in terms of $\eta$s or in terms of $\eta^\dagger$s. These expansions are related by an integral transformation which is often referred as the grassmann Fourier transformation. It is defined in the following way
\begin{equation}
	f(\eta)=\int d\eta^\dagger \tilde e^{\eta^\dagger \eta} f(\eta^\dagger)
\end{equation}
Since any function of grassmann variable can only be one the two kinds: 1) constant function 2) delta function, the Fourier transformation is straightforward and it inter-charges two function just like in bosonic variable.
\begin{equation}
\int d\eta^\dagger  e^{\eta^\dagger \eta} 
\left\{
\begin{matrix}
	1\\
	\\
	\delta(\eta^\dagger)(\equiv\eta^\dagger)   
\end{matrix}
\right.
= 
\left\{
\begin{matrix}
	\delta(\eta)(\equiv\eta)\\
	\\
	1
\end{matrix}
\right.
\end{equation}
For example, we can take fourier transformation of massless superfields
\begin{equation}
	\widetilde{\Sigma}^{+1}=\int d \eta\, e^{\eta \eta^\dagger}\,  \Sigma^{(1)}= p^{(1)}\eta^\dagger + f^{(\frac{1}{2})}
\end{equation}
Similarly we can also take Fourier transformation of massive super-field 
\begin{equation}
\begin{split}
	\widetilde{\Phi}^{I_1\cdots I_{\texttt{2j}}}
&	= 
\int d^2\eta \exp[\eta_{I}\eta^\dagger_I ] 	\Phi^{I_1\cdots I_{\texttt{2j}}}
\\
&={\widetilde  \phi}_{\texttt{j}}^{(I_1\cdots I_\texttt{2j})}+\eta^\dagger_J\left(\phi_{\texttt{j}+\frac{1}{2}}^{(JI_1\cdots I_\texttt{2j})}+\sqrt{\frac{2\texttt{j} }{2\texttt{j}+1}}\sum_{k=1}^\texttt{2j}\epsilon^{J I_k}\phi_{\texttt{j}-\frac{1}{2}}^{(I_1\cdots I_{k-1}I_{k+1}\cdots I_\texttt{j})} \right)-\frac{1}{2}\eta^\dagger_J(\eta^\dagger)^J\phi_{\texttt{j}}^{(I_1\cdots I_\texttt{2j})}
\end{split}
\end{equation}

Here we have used two important formula
\begin{equation}
	\int [d^2\eta] \eta_I\eta_J=-\epsilon_{IJ} 
\qquad,\qquad
\int [d^2\eta] \eta_I\eta^I=2 	
\end{equation}
In this paper, it has been useful at least in two cases: 1) to determine the action of parity, 2) to determine various co-efficient in the super-field expansion.
\begin{equation}
	d^2\eta = \frac{1}{2}\epsilon^{IJ}d\eta_I d\eta_J
\end{equation}

\subsection{Jacobi identities}
The basic identity that we use repeatedly in this paper is 
\begin{equation}
	\jacobi^{(0,\beta)}_{\texttt{j}-\beta}(z)= \frac{\texttt{j}+1}{2\texttt{j}-\beta+1}\jacobi^{(0,\beta+1)}_{\texttt{j}-\beta}(z)+\frac{\texttt{j}-\beta}{2\texttt{j}-\beta+1}\jacobi^{(0,\beta+1)}_{\texttt{j}-\beta-1}(z)
\label{basicidentity}	
\end{equation}	
In particular, there are three applications of this: 
\begin{enumerate}
	\item Identity for $\mathcal{N}=2$ hypermultiplet amplitude
\begin{equation}
		\jacobi^{(0,0)}_{\texttt{j}}(z)= 
		\frac{(\texttt{j}+2)}{2(2\texttt{j}+1)}\jacobi^{(0,2)}_{\texttt{j}}(z)
		+\frac{1}{2}\jacobi^{(0,2)}_{\texttt{j}-1}(z)
				+\frac{(\texttt{j}-1)}{2(2\texttt{j}+1)}\jacobi^{(0,2)}_{\texttt{j}-2}(z)
\label{fermionjacobiidentity}				
\end{equation}

	\item Identity for $\mathcal{N}=4$ Vector multiplet amplitude
\begin{equation}
\begin{split}
\jacobi^{(0,0)}_{\texttt{j}}(z)= 
&\,\, \frac{(\texttt{j}+3)(\texttt{j}+4)}{4(2\texttt{j}+1)(2\texttt{j}+3)}\jacobi^{(0,4)}_{\texttt{j}}(z)
+\frac{(\texttt{j}+3)}{2(2\texttt{j}+1)}\jacobi^{(0,4)}_{\texttt{j}-1}(z)
+\frac{3(\texttt{j}-1)(\texttt{j}+2)}{2(2\texttt{j}+3)(2\texttt{j}-1)}\jacobi^{(0,4)}_{\texttt{j}-2}(z)
\\
&+\frac{(\texttt{j}-2)}{2(2\texttt{j}+1)}\jacobi^{(0,4)}_{\texttt{j}-3}(z)
+\frac{(\texttt{j}-2)(\texttt{j}-3)}{4(2\texttt{j}-1)(2\texttt{j}+1)}\jacobi^{(0,4)}_{\texttt{j}-4}(z)
\end{split}
\label{photonjacobiidentity}				
\end{equation}

\end{enumerate}

\section{Tree level amplitude in massive spinor helicity formalism}
\label{app:bcrstreelevelshamp}

In this section we summarise the computation of tree level amplitudes of massless external states due to the exchange of massive higher spin states. The key steps of computation can be found in \cite{Arkani-Hamed:2017jhn}. 

\paragraph{Propagator}
In the $SU(2)$ language the spin $\texttt{j}$ propagator has the following form
\begin{equation}
	\langle \Phi^{I_1\cdots I_{2\texttt{j}}}\Phi^{J_1\cdots J_{2\texttt{j}}}\rangle
	=\frac{\epsilon^{I_1J_1}\cdots \epsilon^{I_{2\texttt{j}}J_{2\texttt{j}}} }{p^2+M^2-\iimg \varepsilon} \Bigg|_{\textrm{symmetric in } Is}
\end{equation}
We have to glue three point functions, we use the following identity 
\begin{equation}
	A_3(h_1,h_2,P_{1\cdots 2\texttt{j}})A_3(h_3,h_4,-P^{1\cdots 2\texttt{j}})
\end{equation}
Our prescription for analytic continuation is 
\begin{equation}
	|-P^I\rangle = |P^I\rangle
\qquad,\qquad	
	|-P^I]= -|P^I]
\end{equation}
When we contract the left hand side and right hand side, we use
\begin{equation}
	|P_I\rangle_{\dot a} \langle P^I|^{\dot b}=-m{\delta_{\dot a }}^{\dot b} 
\end{equation}
So when we do the contraction we get 
\begin{equation}
\begin{split}
	&\frac{(-m)^{\texttt{2j}}}{m^{4\texttt{j}+h_1+h_2+h_3+h_4-2}}
[12]^{\texttt{j}+h_1+h_2}
[34]^{\texttt{j}+h_3+h_4}	
\\
&\sum_a	\mathfrak{C}(\texttt{j},a)\langle 13\rangle^a \langle 14\rangle^{\texttt{j}+h_2-h_1-a}\langle 24\rangle^{h_1+h_3-h_2-h_4+a}\langle 23\rangle^{\texttt{j}+h_4-h_3-a} 	
\end{split}
\end{equation}
Now we have to determine the combinatorial factor $\mathfrak{C}(\texttt{j},a)$. We the computation in steps. 
\begin{enumerate}
	\item We start with the combinatorial factor for $\langle 13\rangle^a$
\begin{equation}
	\frac{(\texttt{j}+h_2-h_1)! }{(\texttt{j}+h_2-h_1-a)! a!}
	\frac{(\texttt{j}+h_4-h_3)! }{(\texttt{j}+h_4-h_3-a)! a!}	a!
\end{equation}

	\item Then we compute the combinatorial factor for $\langle 24\rangle^{h_1+h_3-h_2-h_4+a}$
\begin{equation}
	\frac{(\texttt{j}+h_1-h_2)! }{(\texttt{j}+h_2-h_1-a)! (h_1+h_3-h_2-h_4+a)!}
	\frac{(\texttt{j}+h_4-h_3)! }{(\texttt{j}+h_3-h_4-a)!  } 
\end{equation}
	
	\item Then we compute for $\langle 14\rangle^{\texttt{j}+h_2-h_1-a}$
\begin{equation}
	(\texttt{j}+h_2-h_1-a)!
\end{equation}
	\item Then we compute for $\langle 23\rangle^{\texttt{j}+h_4-h_3-a} $
\begin{equation}
	(\texttt{j}+h_4-h_3-a)!
\end{equation}
\end{enumerate}
Putting al together we get 
\begin{equation}
	\mathfrak{C}(\texttt{j},a)
= \frac{1}{(\texttt{2j})!}	 
	\frac{(\texttt{j}+h_2-h_1)! (\texttt{j}+h_4-h_3)! (\texttt{j}+h_1-h_2)!( \texttt{j}+h_4-h_3)! }{a! (\texttt{j}+h_2-h_1-a)!(\texttt{j}+h_4-h_3-a)! (h_1+h_3-h_2-h_4+a)!}
\end{equation}

\subsection{Angular distribution in the com frame}
Now we compute the angular distribution in the com frame. The com frame spinors are given by 
\begin{equation}
	\langle 12\rangle =2E=-\langle 34\rangle 
\quad,\quad 
\langle 14\rangle 	=-\iimg 2E \sin \frac{\theta}{2} =-\langle 23\rangle 
\quad,\quad 
\langle 13\rangle 	=-\iimg 2E \cos \frac{\theta}{2} =\langle 24\rangle 
\end{equation}
Putting it back in the above equation we get 
\begin{equation}
\begin{split}
	&\frac{(-m)^{\texttt{2j}}}{m^{4\texttt{j}+h_1+h_2+h_3+h_4-2}}
s^{\texttt{2j}+\frac{h_1+h_2+h_3+h_4}{2}}	(-1)^{\texttt{j}+h_3+h_4}	
\\
&(\iimg)^{2\texttt{j}}\sum_a (-1)^{a+h_3-h_4+\texttt{j}}	\mathfrak{C}(\texttt{j},a) \left(\cos\frac{\theta}{2}\right)^{2a+h+h^\prime} \left(\sin\frac{\theta}{2} \right)^{2\texttt{j}-2a-h-h^\prime} 
\end{split}
\end{equation}
We define $h=h_1-h_2$, $h^\prime=h_3-h_4$. We simplify it (and since we are looking for the residue at the pole, we use $s=m^2$)
\begin{equation}
\begin{split}
	&
s^{\texttt{j}+1}	(-1)^{5\texttt{j}+h_3+h_4}	
\sum_a  (-1)^{a+h^\prime}	\mathfrak{C}(\texttt{j},a) \left(\cos\frac{\theta}{2}\right)^{2a+h+h^\prime} \left(\sin\frac{\theta}{2} \right)^{2\texttt{j}-2a-h-h^\prime} 
\end{split}
\end{equation}
First we note that 
\begin{equation}
	\mathfrak{C}(\texttt{j},a)= \bar{\mathcal{N}}_{\texttt{j};h,h^\prime}\frac{\sqrt{(\texttt{j}-h)! (\texttt{j}-h^\prime)! (\texttt{j}+h)!( \texttt{j}+h^\prime)!} }{a! (\texttt{j}-h-a)!(\texttt{j}-h^\prime-a)! (h+h^\prime+a)!}
\end{equation}
Then the summation part can be written as 
\begin{equation}
\begin{split}
&	\sum_a 	(-1)^{a+h^\prime} 
	\frac{\sqrt{(\texttt{j}-h)! (\texttt{j}-h^\prime)! (\texttt{j}+h)!( \texttt{j}+h^\prime)!} }{a! (\texttt{j}-h-a)!(\texttt{j}-h^\prime-a)! (h+h^\prime+a)!}\left(\cos\frac{\theta}{2}\right)^{2a+h+h^\prime} \left(\sin\frac{\theta}{2} \right)^{2\texttt{j}-2a-h-h^\prime} 
\\
=&	\sum_a 	(-1)^{a+h^\prime} 
	\frac{\sqrt{(\texttt{j}-h)! (\texttt{j}-h^\prime)! (\texttt{j}+h)!( \texttt{j}+h^\prime)!} }{a! (\texttt{j}-h-a)!(\texttt{j}-h^\prime-a)! (h+h^\prime+a)!}
	\left(\sin\frac{\pi -\theta}{2}\right)^{2a+h+h^\prime} 
	\left(\cos\frac{\pi-\theta}{2}\right)^{2\texttt{j}-2a-h-h^\prime} 
\end{split}
\end{equation}
This agrees with standard formula for wigner matrix  (for example, check \href{https://en.wikipedia.org/wiki/Wigner_D-matrix}{wiki})
\begin{equation}
(-1)^{-h  }d^{(\texttt{j})}_{h,-h^\prime}(\pi - \theta)
=(-1)^{\texttt{j}} d^{(\texttt{j})}_{h,h^\prime}(\theta)  
\end{equation}
So the full answer is 
\begin{equation}
	s^{\texttt{j}+1}	(-1)^{6\texttt{j}+h_3+h_4}	d^{(\texttt{j})}_{h,h^\prime}(\theta)  =s^{\texttt{j}+1}	(-1)^{\texttt{j}}	d^{(\texttt{j})}_{h,h^\prime}(\theta)  
\end{equation}

\section{Four photon tensor structures}
\label{app:bcrspppptsturtcures}

\subsection{Redundancy in the basis element in $3+1$ dimensions }
\label{subapp:bcrspppptsturtcuresone}

In $ 2\rightarrow 2 $ scattering in $D$-dimensions, the plane of action is a plane spanned by the three of the momenta. We can resolve the Polarization of the particles along the plane and orthogonal to the plane.
\begin{equation}
\epsilon{}^\mu= \epsilon^{\parallel}{}^\mu+\epsilon^{\perp}{}^\mu
\label{redundancy1}
\end{equation}
For four photons, we can constrain the parallel component using the transversality condition,
\begin{equation}
\epsilon_i^{\parallel}{}^\mu = \beta_i\, k_i{}^\mu + \alpha_i\bigg[\frac{k_{i+1}{}^\mu}{s}-\frac{k_{i-1}{}^\mu}{t}\bigg]
\label{redundancy2}
\end{equation}
Any scattering amplitude can be described in the language of $ (\epsilon^\perp_i,\alpha_i) $.
The $ \alpha_i $'s are given by,
\begin{equation}
\alpha_i = \frac{2}{s+t}\times\bigg(k_{i-1}\cdot \mathcal{B}_i\cdot k_{i+1}\bigg)
\label{redundancy3}
\end{equation}
We can also determine $ \epsilon^\perp_i $,
\begin{equation}
\begin{split}
\epsilon^\perp_1{}^\mu &= \frac{2}{t}\Bigg[\mathcal{B}^{\mu\nu}_1k_{4\,\nu}-\bigg(\frac{k_4\cdot \mathcal{B}_1\cdot k_2}{s(s+t)}\bigg)\Big[(s+t)k^\mu_1+t k^\mu_2 -s k^\mu_4\Big]\Bigg]\\
\epsilon^\perp_2{}^\mu &= \frac{2}{t}\Bigg[\mathcal{B}^{\mu\nu}_2 k_{3\,\nu}-\bigg(\frac{k_3\cdot \mathcal{B}_2\cdot k_1}{s(s+t)}\bigg)\Big[(s+t)k^\mu_2+t k^\mu_1 -s k^\mu_3\Big]\Bigg]\\
\epsilon^\perp_3{}^\mu &= \frac{2}{t}\Bigg[\mathcal{B}^{\mu\nu}_3k_{2\,\nu}-\bigg(\frac{k_2\cdot \mathcal{B}_3\cdot k_4}{s(s+t)}\bigg)\Big[(s+t)k^\mu_3+t k^\mu_4 -s k^\mu_2\Big]\Bigg]\\
\epsilon^\perp_4{}^\mu &= \frac{2}{t}\Bigg[\mathcal{B}^{\mu\nu}_4k_{1\,\nu}-\bigg(\frac{k_1\cdot \mathcal{B}_4\cdot k_3}{s(s+t)}\bigg)\Big[(s+t)k^\mu_4+t k^\mu_3 -s k^\mu_1\Big]\Bigg]
\end{split}
\end{equation}
In $ D=3+1 $, the $ \epsilon^\perp $ are numbers and hence,
\begin{equation}
(\epsilon^\perp_1\cdot\epsilon^\perp_2)(\epsilon^\perp_3\cdot \epsilon^\perp_4) = (\epsilon^\perp_1\cdot\epsilon^\perp_4)(\epsilon^\perp_2\cdot \epsilon^\perp_3) = (\epsilon^\perp_1\cdot\epsilon^\perp_3)(\epsilon^\perp_2\cdot \epsilon^\perp_4)
\label{redundancy5}
\end{equation}
We use \eqref{redundancy3} to detect ``gauge invariance" in the basis for tensor structures. The above quantities can be written in our basis for tensor structures given in \eqref{}. 
\begin{equation}
\begin{split}
(\epsilon^\perp_1\cdot\epsilon^\perp_2)(\epsilon^\perp_3\cdot \epsilon^\perp_4)&=\sum_{\alpha=1}^7 \mathcal{T}_\alpha f^{(1234)}_\alpha 
\\	
(\epsilon^\perp_1\cdot\epsilon^\perp_3)(\epsilon^\perp_2\cdot \epsilon^\perp_4)&=\sum_{\alpha=1}^7 \mathcal{T}_\alpha f^{(1324)}_\alpha
\\
(\epsilon^\perp_1\cdot\epsilon^\perp_4)(\epsilon^\perp_2\cdot \epsilon^\perp_3)&=\sum_{\alpha=1}^7 \mathcal{T}_\alpha f^{(1423)}_\alpha
\end{split}	
\end{equation}
$f^{(1234)}$(,$f^{(1324)}$ and $f^{(1423)}$) can be thought as a column vector with 7 components. 
\begin{equation}
\begin{split}
f^{(1234)}
&=
% \frac{2}{s^2 (z+1)^2(z-1)^2}
\mathcal{C}  \begin{pmatrix}
 	(z^2-1)^2\\
	4(z-1)^2 \\
	4(z+1)^2\\
	-(z^2-1)^2\\
	- 4 (z+2)(z-1)^2 \\
	4(z-2)(z+1)^2 \\
	\frac{8}{s}  (z^2-1)
\end{pmatrix}
%\end{split}
%\end{equation}
%\begin{equation}
%\begin{split}
\quad,\quad
f^{(1423)}
%(\epsilon^\perp_1\cdot\epsilon^\perp_4)(\epsilon^\perp_2\cdot \epsilon^\perp_3)
= 
% \frac{2}{s^2 (z+1)^2(z-1)^2}
\mathcal{C} \begin{pmatrix}
	(z^2-1)^2 \\
	4(z-1)^2 \\
	4(z+1)^2\\
	-(z-5)(z-1)(z+1)^2\\
	- 4 (3z+1)(z-1) \\
	-4 (z+1)^2 \\
	\frac{8}{s}(z^2-1)  
\end{pmatrix}
\\
%\end{split}
%\end{equation}
%
%\begin{equation}
%\begin{split}
f^{(1324)}
%(\epsilon^\perp_1\cdot\epsilon^\perp_4)(\epsilon^\perp_2\cdot \epsilon^\perp_3)
&= 
% \frac{2}{s^2 (z+1)^2(z-1)^2}
\mathcal{C} \begin{pmatrix}
	(z^2-1)^2 \\
	4(z-1)^2 \\
	4(z+1)^2\\
	-2(z+5)(z+1)(z-1)^2\\
	-4 (z-1)^2 \\
	 -4 (3z-1)(z+1) \\	
	\frac{8}{s}(z^2-1)  
\end{pmatrix}
\quad,\quad 
\mathcal{C}=\frac{2}{s^2 (z+1)^2(z-1)^2}
\end{split}
\end{equation}
From \eqref{redundancy5}, it follows that if we take the difference between any two of the above quantities, then we should get zero. Note that the first three entries and the last entries of the above three column vectors are the same. So when we take the difference, the first three entries and the seventh entry become zero. The differences are given by 
\begin{equation}
u = \begin{pmatrix}
 0 \\
0 \\
0 \\
z+1 \\
z-3 \\
-z-1 \\
0 \\
\end{pmatrix}\quad,\quad v =\begin{pmatrix}
 0 \\
0 \\
0 \\
-z \\
1 \\
-1 \\
0 \\
\end{pmatrix}
\end{equation}
These three elements generate ``gauge invariance"(redundancy) of four-photon S-matrices. If two form factors differ in the following way
\begin{equation}
	\mathcal{F}(s,z)\longrightarrow \mathcal{F}(s,z)+ u f_1(s,z)+ vf_2(s,z) 
\end{equation}
Here $\mathcal{F}(s,z)$, $u$ and $v$ are seven dimensional column vectors and $f_1(s,z)$ and $f_2(s,z)$ are two arbitrary functions of $s$ and $z$.

\subsection{Scattering amplitude and dimensional analysis}
\label{subapp:bcrspppptsturtcurestwo}

Here we give a short note for dimensional analysis of scattering amplitudes. We start with the quantum fields. 
\begin{itemize}
	\item From the kinetic term, it follows that any bosonic field has mass dimension 1. 
	\item We fourier transform the fields and write in terms of creation and annihilation operators. 
	\item The creation and annihilation have mass dimension $1$ in $3+1$ dimensions. And this implies that the polarization does not have any mass dimension. 
	\item Any $n$-point bosonic scattering amplitude is the expectation value of $n$-creation operators, and it has mass dimension $-n $. 
	\item In computing the mass dimension of momentum space amplitude, one has to consider the mass dimension of the delta function, which we do not write explicitly. 
\end{itemize}
Let's do a few examples. The scalar-scalar-spin $\texttt{j}$ amplitude is 
\begin{equation}
	\frac{g}{M^{\texttt{j}-1}}(\epsilon_3\cdot k_{12})^\texttt{j}\delta^{(4)}(k_1+k_2+k_3+k_4) 
\end{equation}
This amplitude has mass $\texttt{j}-4-(\texttt{j}-1)=-3$. Similarly, the photon-photon-spin $\texttt{j}$ amplitude is given by
\begin{equation}
		\frac{g}{M^{\texttt{j}-1}}(\mathcal{W}_{(12)}^{\mu\nu}\epsilon_{3\mu}\epsilon_{3\nu})(\epsilon_3\cdot k_{12})^\texttt{j-2}\delta^{(4)}(k_1+k_2+k_3+k_4) 
\end{equation}
The non-minimal amplitude is given by
\begin{equation}
		\frac{g}{M^{\texttt{j}-3}}(\mathcal{W}_{(12)})(\epsilon_3\cdot k_{12})^\texttt{j}\delta^{(4)}(k_1+k_2+k_3+k_4) 
\end{equation}

\addcontentsline{toc}{section}{References}
\bibliographystyle{utphys}
\bibliography{susyhspin}

\providecommand{\href}[2]{#2}\begingroup\raggedright\begin{thebibliography}{10}

\bibitem{Vasiliev:1990en}
M.~A. Vasiliev, ``{Consistent equation for interacting gauge fields of all
  spins in (3+1)-dimensions},''
  \href{http://dx.doi.org/10.1016/0370-2693(90)91400-6}{{\em Phys. Lett. B}
  {\bfseries 243} (1990) 378--382}.

\bibitem{Camanho:2014apa}
X.~O. Camanho, J.~D. Edelstein, J.~Maldacena, and A.~Zhiboedov, ``{Causality
  Constraints on Corrections to the Graviton Three-Point Coupling},''
  \href{http://dx.doi.org/10.1007/JHEP02(2016)020}{{\em JHEP} {\bfseries 02}
  (2016) 020}, \href{http://arxiv.org/abs/1407.5597}{{\ttfamily arXiv:1407.5597
  [hep-th]}}.

\bibitem{Maity:2021obe}
P.~Maity, ``{Positivity of the Veneziano amplitude in D = 4},''
  \href{http://dx.doi.org/10.1007/JHEP04(2022)064}{{\em JHEP} {\bfseries 04}
  (2022) 064}, \href{http://arxiv.org/abs/2110.01578}{{\ttfamily
  arXiv:2110.01578 [hep-th]}}.

\bibitem{Arkani-Hamed:2022gsa}
N.~Arkani-Hamed, L.~Eberhardt, Y.-t. Huang, and S.~Mizera, ``{On unitarity of
  tree-level string amplitudes},''
  \href{http://dx.doi.org/10.1007/JHEP02(2022)197}{{\em JHEP} {\bfseries 02}
  (2022) 197}, \href{http://arxiv.org/abs/2201.11575}{{\ttfamily
  arXiv:2201.11575 [hep-th]}}.

\bibitem{Chakraborty:2020rxf}
S.~Chakraborty, S.~D. Chowdhury, T.~Gopalka, S.~Kundu, S.~Minwalla, and
  A.~Mishra, ``{Classification of all 3 particle S-matrices quadratic in
  photons or gravitons},''
  \href{http://dx.doi.org/10.1007/JHEP04(2020)110}{{\em JHEP} {\bfseries 04}
  (2020) 110}, \href{http://arxiv.org/abs/2001.07117}{{\ttfamily
  arXiv:2001.07117 [hep-th]}}.

\bibitem{Balasubramanian:2021act}
M.~K.~N. Balasubramanian, R.~Patil, and A.~Rudra, ``{Spinning amplitudes from
  scalar amplitudes},'' \href{http://dx.doi.org/10.1007/JHEP11(2021)151}{{\em
  JHEP} {\bfseries 11} (2021) 151},
  \href{http://arxiv.org/abs/2106.05301}{{\ttfamily arXiv:2106.05301
  [hep-th]}}.

\bibitem{Coleman:1967ad}
S.~R. Coleman and J.~Mandula, ``{All Possible Symmetries of the S Matrix},''
  \href{http://dx.doi.org/10.1103/PhysRev.159.1251}{{\em Phys. Rev.} {\bfseries
  159} (1967) 1251--1256}.

\bibitem{Herderschee:2019ofc}
A.~Herderschee, S.~Koren, and T.~Trott, ``{Massive On-Shell Supersymmetric
  Scattering Amplitudes},''
  \href{http://dx.doi.org/10.1007/JHEP10(2019)092}{{\em JHEP} {\bfseries 10}
  (2019) 092}, \href{http://arxiv.org/abs/1902.07204}{{\ttfamily
  arXiv:1902.07204 [hep-th]}}.

\bibitem{Herderschee:2019dmc}
A.~Herderschee, S.~Koren, and T.~Trott, ``{Constructing $ \mathcal{N} $ = 4
  Coulomb branch superamplitudes},''
  \href{http://dx.doi.org/10.1007/JHEP08(2019)107}{{\em JHEP} {\bfseries 08}
  (2019) 107}, \href{http://arxiv.org/abs/1902.07205}{{\ttfamily
  arXiv:1902.07205 [hep-th]}}.

\bibitem{Engelbrecht:2022aao}
L.~Engelbrecht, C.~R.~T. Jones, and S.~Paranjape, ``{Supersymmetric Massive
  Gravity},'' \href{http://arxiv.org/abs/2205.12982}{{\ttfamily
  arXiv:2205.12982 [hep-th]}}.

\bibitem{Chiodaroli:2022ssi}
M.~Chiodaroli, M.~Gunaydin, H.~Johansson, and R.~Roiban, ``{Spinor-helicity
  formalism for massive and massless amplitudes in five dimensions},''
  \href{http://arxiv.org/abs/2202.08257}{{\ttfamily arXiv:2202.08257
  [hep-th]}}.

\bibitem{Abhishek:2022nqv}
M.~Abhishek, S.~Hegde, D.~P. Jatkar, and A.~P. Saha, ``{Scattering Amplitudes
  and BCFW in $\mathcal{N}=2^{\ast}$ Theory},''
  \href{http://arxiv.org/abs/2202.12204}{{\ttfamily arXiv:2202.12204
  [hep-th]}}.

\bibitem{Liu:2020fgu}
J.-Y. Liu and Z.-M. You, ``{The supersymmetric spinning polynomial},''
  \href{http://arxiv.org/abs/2011.11299}{{\ttfamily arXiv:2011.11299
  [hep-th]}}.

\bibitem{Alday:2007hr}
L.~F. Alday and J.~M. Maldacena, ``{Gluon scattering amplitudes at strong
  coupling},'' \href{http://dx.doi.org/10.1088/1126-6708/2007/06/064}{{\em
  JHEP} {\bfseries 06} (2007) 064},
  \href{http://arxiv.org/abs/0705.0303}{{\ttfamily arXiv:0705.0303 [hep-th]}}.

\bibitem{Arkani-Hamed:2017jhn}
N.~Arkani-Hamed, T.-C. Huang, and Y.-t. Huang, ``{Scattering Amplitudes For All
  Masses and Spins},''
\href{http://arxiv.org/abs/1709.04891}{{\ttfamily arXiv:1709.04891 [hep-th]}}.
%%CITATION = ARXIV:1709.04891;%%.

\bibitem{Boels:2012if}
R.~H. Boels, ``{Three particle superstring amplitudes with massive legs},''
  \href{http://dx.doi.org/10.1007/JHEP06(2012)026}{{\em JHEP} {\bfseries 06}
  (2012) 026}, \href{http://arxiv.org/abs/1201.2655}{{\ttfamily arXiv:1201.2655
  [hep-th]}}.

\bibitem{Conde:2016vxs}
E.~Conde and A.~Marzolla, ``{Lorentz Constraints on Massive Three-Point
  Amplitudes},'' \href{http://dx.doi.org/10.1007/JHEP09(2016)041}{{\em JHEP}
  {\bfseries 09} (2016) 041}, \href{http://arxiv.org/abs/1601.08113}{{\ttfamily
  arXiv:1601.08113 [hep-th]}}.

\bibitem{Conde:2016izb}
E.~Conde, E.~Joung, and K.~Mkrtchyan, ``{Spinor-Helicity Three-Point Amplitudes
  from Local Cubic Interactions},''
  \href{http://dx.doi.org/10.1007/JHEP08(2016)040}{{\em JHEP} {\bfseries 08}
  (2016) 040}, \href{http://arxiv.org/abs/1605.07402}{{\ttfamily
  arXiv:1605.07402 [hep-th]}}.

\bibitem{Chowdhury:2019kaq}
S.~D. Chowdhury, A.~Gadde, T.~Gopalka, I.~Halder, L.~Janagal, and S.~Minwalla,
  ``{Classifying and constraining local four photon and four graviton
  S-matrices},''
\href{http://arxiv.org/abs/1910.14392}{{\ttfamily arXiv:1910.14392 [hep-th]}}.
%%CITATION = ARXIV:1910.14392;%%.

\bibitem{DeAngelis:2022qco}
S.~De~Angelis, ``{Amplitude bases in generic EFTs},''
  \href{http://dx.doi.org/10.1007/JHEP08(2022)299}{{\em JHEP} {\bfseries 08}
  (2022) 299}, \href{http://arxiv.org/abs/2202.02681}{{\ttfamily
  arXiv:2202.02681 [hep-th]}}.

\bibitem{Salam:1976ib}
A.~Salam and J.~A. Strathdee, ``{Supersymmetry and Superfields},''
  \href{http://dx.doi.org/10.1002/prop.19780260202}{{\em Fortsch. Phys.}
  {\bfseries 26} (1978) 57}.

\bibitem{Bern:2021ppb}
Z.~Bern, D.~Kosmopoulos, and A.~Zhiboedov, ``{Gravitational effective field
  theory islands, low-spin dominance, and the four-graviton amplitude},''
  \href{http://dx.doi.org/10.1088/1751-8121/ac0e51}{{\em J. Phys. A} {\bfseries
  54} no.~34, (2021) 344002}, \href{http://arxiv.org/abs/2103.12728}{{\ttfamily
  arXiv:2103.12728 [hep-th]}}.

\bibitem{Bern:2022yes}
Z.~Bern, E.~Herrmann, D.~Kosmopoulos, and R.~Roiban, ``{Effective Field Theory
  Islands from Perturbative and Nonperturbative Four-Graviton Amplitudes},''
  \href{http://arxiv.org/abs/2205.01655}{{\ttfamily arXiv:2205.01655
  [hep-th]}}.

\bibitem{Pasterski:2016qvg}
S.~Pasterski, S.-H. Shao, and A.~Strominger, ``{Flat Space Amplitudes and
  Conformal Symmetry of the Celestial Sphere},''
  \href{http://dx.doi.org/10.1103/PhysRevD.96.065026}{{\em Phys. Rev. D}
  {\bfseries 96} no.~6, (2017) 065026},
  \href{http://arxiv.org/abs/1701.00049}{{\ttfamily arXiv:1701.00049
  [hep-th]}}.

\bibitem{Pasterski:2020pdk}
S.~Pasterski and A.~Puhm, ``{Shifting spin on the celestial sphere},''
  \href{http://dx.doi.org/10.1103/PhysRevD.104.086020}{{\em Phys. Rev. D}
  {\bfseries 104} no.~8, (2021) 086020},
  \href{http://arxiv.org/abs/2012.15694}{{\ttfamily arXiv:2012.15694
  [hep-th]}}.

\bibitem{Pasterski:2021raf}
S.~Pasterski, M.~Pate, and A.-M. Raclariu, ``{Celestial Holography},'' in {\em
  {2022 Snowmass Summer Study}}.
\newblock 11, 2021.
\newblock \href{http://arxiv.org/abs/2111.11392}{{\ttfamily arXiv:2111.11392
  [hep-th]}}.

\bibitem{Caron-Huot:2016icg}
S.~Caron-Huot, Z.~Komargodski, A.~Sever, and A.~Zhiboedov, ``{Strings from
  Massive Higher Spins: The Asymptotic Uniqueness of the Veneziano
  Amplitude},'' \href{http://dx.doi.org/10.1007/JHEP10(2017)026}{{\em JHEP}
  {\bfseries 10} (2017) 026},
\href{http://arxiv.org/abs/1607.04253}{{\ttfamily arXiv:1607.04253 [hep-th]}}.
%%CITATION = ARXIV:1607.04253;%%.

\bibitem{Kawai:1985xq}
H.~Kawai, D.~C. Lewellen, and S.~H.~H. Tye, ``{A Relation Between Tree
  Amplitudes of Closed and Open Strings},''
\href{http://dx.doi.org/10.1016/0550-3213(86)90362-7}{{\em Nucl. Phys.}
  {\bfseries B269} (1986) 1--23}.
%%CITATION = NUPHA,B269,1;%%.

\bibitem{Bern:2008qj}
Z.~Bern, J.~J.~M. Carrasco, and H.~Johansson, ``{New Relations for Gauge-Theory
  Amplitudes},'' \href{http://dx.doi.org/10.1103/PhysRevD.78.085011}{{\em Phys.
  Rev. D} {\bfseries 78} (2008) 085011},
  \href{http://arxiv.org/abs/0805.3993}{{\ttfamily arXiv:0805.3993 [hep-ph]}}.

\bibitem{Bern:2019prr}
Z.~Bern, J.~J. Carrasco, M.~Chiodaroli, H.~Johansson, and R.~Roiban, ``{The
  Duality Between Color and Kinematics and its Applications},''
  \href{http://arxiv.org/abs/1909.01358}{{\ttfamily arXiv:1909.01358
  [hep-th]}}.

\bibitem{Carrasco:2020ywq}
J.~J.~M. Carrasco and I.~A. Vazquez-Holm, ``{Loop-Level Double-Copy for Massive
  Quantum Particles},''
  \href{http://dx.doi.org/10.1103/PhysRevD.103.045002}{{\em Phys. Rev. D}
  {\bfseries 103} no.~4, (2021) 045002},
  \href{http://arxiv.org/abs/2010.13435}{{\ttfamily arXiv:2010.13435
  [hep-th]}}.

\bibitem{Green:2012oqa}
M.~B. Green, J.~H. Schwarz, and E.~Witten,
  \href{http://dx.doi.org/10.1017/CBO9781139248563}{{\em {Superstring Theory
  Vol. 1}}}.
\newblock Cambridge Monographs on Mathematical Physics. Cambridge University
  Press, 2012.
\newblock
\url{http://www.cambridge.org/mw/academic/subjects/physics/theoretical-physics-and-mathematical-physics/superstring-theory-25th-anniversary-edition-volume-1?format=AR}.
\newblock
%%CITATION = INSPIRE-1384878;%%.

\bibitem{Polchinski:1998rq}
J.~Polchinski, \href{http://dx.doi.org/10.1017/CBO9780511816079}{{\em {String
  theory. Vol. 1: An introduction to the bosonic string}}}.
\newblock Cambridge Monographs on Mathematical Physics. Cambridge University
  Press, 12, 2007.

\bibitem{Elvang:2015rqa}
H.~Elvang and Y.-t. Huang, {\em {Scattering Amplitudes in Gauge Theory and
  Gravity}}.
\newblock Cambridge University Press, 4, 2015.

\bibitem{Guevara:2018wpp}
A.~Guevara, A.~Ochirov, and J.~Vines, ``{Scattering of Spinning Black Holes
  from Exponentiated Soft Factors},''
  \href{http://dx.doi.org/10.1007/JHEP09(2019)056}{{\em JHEP} {\bfseries 09}
  (2019) 056}, \href{http://arxiv.org/abs/1812.06895}{{\ttfamily
  arXiv:1812.06895 [hep-th]}}.

\end{thebibliography}\endgroup

\end{document}